\documentclass[reprint,amsmath,amssymb,aps,prd,showpacs,floatfix,numerical,preprintnumbers]{revtex4-1}

\usepackage{graphicx}
\usepackage{dcolumn}
\usepackage{bm}
\usepackage{atlasphysics}

\usepackage[colorlinks,breaklinks,pdftitle={Study of jets produced in association with a W boson in pp collisions at sqrt(s)=7 TeV with the ATLAS detector},pdfauthor={The ATLAS Collaboration},dvipdfm]{hyperref}
\hypersetup{linkcolor=blue,citecolor=blue,filecolor=black,urlcolor=blue}

\begin{document}

\preprint{CERN-PH-EP-2011-218}
\preprint{Submitted to Physical Review D}
\vspace*{0.1cm}

\title{Study of jets produced in association with a $W$ boson in
\boldmath $pp$ collisions \\ at $\sqrt{s}=7$\unboldmath~\TeV\ with the
ATLAS detector}

\author{The ATLAS Collaboration}
\altaffiliation{Full author list given at the end of the article.}

\date{January 24, 2012}

\begin{abstract}
We report a study of final states containing a $W$ boson and hadronic
jets, produced in proton-proton collisions at a center-of-mass energy
of 7 \TeV. The data were collected with the ATLAS detector at the CERN
LHC and comprise the full 2010 data sample of $36$~\ipb. Cross
sections are determined using both the electron and muon decay modes
of the $W$ boson and are presented as a function of inclusive jet
multiplicity, $N_{\rm{jet}}$, for up to five jets. At each
multiplicity, cross sections are presented as a function of jet
transverse momentum, the scalar sum of the transverse momenta of the
charged lepton, missing transverse momentum, and all jets, the
invariant mass spectra of jets, and the rapidity distributions of
various combinations of leptons and final-state jets. The results,
corrected for all detector effects and for all backgrounds such as
diboson and top quark pair production, are compared with
particle-level predictions from perturbative QCD. Leading-order
multiparton event generators, normalized to the NNLO total cross
section for inclusive $W$-boson production, describe the data
reasonably well for all measured inclusive jet multiplicities.
Next-to-leading-order calculations from {\sc MCFM}, studied here for
$N_{\rm{jet}} \le 2$, and {\sc BlackHat-Sherpa}, studied here for
$N_{\rm{jet}} \le 4$, are found to be mostly in good agreement with
the data.
\end{abstract}

\maketitle

\section{Introduction}
\par The study of massive vector boson production in association with
one or more jets is an important test of quantum chromodynamics
(QCD). These final states are also a significant background to studies
of Standard Model processes such as \ttbar, diboson, and single-top
production, as well as to searches for the Higgs boson and for physics
beyond the Standard Model. Thus, measurements of the cross section and
kinematic properties, and comparisons with theoretical predictions,
are of significant interest. Measurements of $W$+jets production in
proton-antiproton collisions at $\sqrt{s}=1.96$~\TeV\ have been
reported by the CDF and D0 Collaborations~\cite{CDFWjets:2008,
D0W:2011} and for $\sqrt{s}=7$~\TeV\ proton-proton collisions by the
CMS Collaboration~\cite{CMSWJets}. Measurements of jets produced in
association with a $Z$ boson were also performed using $p\bar{p}$
collisions at
$\sqrt{s}=1.96$~\TeV~\cite{Abazov:2009av,Abazov:2009pp,PhysRevLett.100.102001}
and $pp$ collisions at
$\sqrt{s}=7$~\TeV~\cite{ATLAS_Zjets_2011,CMSWJets}. The study
presented here is complementary to the measurement of the transverse
momentum distribution of $W$ bosons conducted by the ATLAS
Collaboration~\cite{W_Pt_ATLAS}.

\par This paper reports a measurement at the CERN Large Hadron
Collider (LHC) of the $W$+jets cross section for proton-proton ($pp$)
collisions at a center-of-mass energy ($\sqrt{s}$) of 7~\TeV, using
the ATLAS detector. The measurement is based on the full 2010 data
sample, corresponding to an integrated luminosity of approximately
$36$~\ipb. It is an extension of an earlier ATLAS measurement of both
the electron and muon decay modes of the $W$ boson based on
1.3~\ipb~\cite{WJets}. Compared to the earlier result, uncertainties
in both the jet energy scale and luminosity are reduced, acceptance
for the jets is expanded, and event reconstruction and simulation are
improved. The improved reconstruction brings better alignment of the
detector systems and reduction of backgrounds in the electron channel.

\par The results have been corrected for all known detector effects
and are quoted in a specific range of jet and lepton kinematics, fully
covered by the detector acceptance. This avoids model-dependent
extrapolations and facilitates comparisons with theoretical
predictions. Theoretical calculations at next-to-leading order (NLO)
in perturbative QCD (pQCD) have been computed inclusively for up to
four jets~\cite{EllisWjets:2009, BlackHat4jets} and are compared with
the data.

\section{The ATLAS detector}

\par ATLAS uses a right-handed coordinate system with its origin at
the nominal $pp$ interaction point (IP) in the center of the detector
and the $z$-axis along the beam pipe. The $x$-axis points from the IP
to the center of the LHC ring, and the $y$-axis points
upward. Cylindrical coordinates $(r,\phi)$ are used in the transverse
plane, $\phi$ being the azimuthal angle around the beam pipe. The
pseudorapidity is defined in terms of the polar angle $\theta$ as
$\eta=-\ln\tan(\theta/2)$ and the rapidity is defined as
$y=\ln[(E+p_{z})/(E-p_{z})] / 2$. The separation between final state
particles is defined as $\Delta R=\sqrt{(\Delta y)^2+(\Delta\phi)^2}$
and is Lorentz invariant under boosts along the $z$-axis.

\par The ATLAS detector \cite{AtlasDetector,CSCbook} consists of an
inner tracking system (inner detector, or ID) surrounded by a thin
superconducting solenoid providing a 2T magnetic field,
electromagnetic and hadronic calorimeters, and a muon spectrometer
(MS). The ID consists of pixel and silicon microstrip detectors,
surrounded by a transition radiation tracker. The electromagnetic
calorimeter is a liquid-argon and lead detector, split into barrel
($|\eta| < 1.475$) and endcap ($1.375 < |\eta| < 3.2$) regions. Hadron
calorimetry is based on two different detector technologies. The
barrel ($|\eta| < 0.8$) and extended barrel ($0.8 < |\eta| < 1.7$)
calorimeters are composed of scintillator and steel, while the
hadronic endcap calorimeters ($1.5 < |\eta| < 3.2$) utilize
liquid-argon and copper. The forward calorimeters ($3.1 < |\eta| <
4.9$) are instrumented with liquid-argon/copper and
liquid-argon/tungsten, providing electromagnetic and hadronic energy
measurements, respectively. The MS is based on three large
superconducting toroids arranged with an eight-fold azimuthal coil
symmetry around the calorimeters, and a system of three stations of
chambers for triggering and for precise track measurements.

\section{Data and online event selection}

\par The data for this analysis were collected during LHC operation in
2010 with proton-proton interactions at a center-of-mass energy of
7~\TeV. The collisions occurred within pairs of bunches of up to
$\sim1.1 \times 10^{11}$ protons per bunch. The bunches were
configured in trains with a time separation between bunches of 150~ns
and a longer separation between trains. Data were collected with up to
348 colliding bunch pairs per beam revolution. This configuration led
to a peak instantaneous luminosity of up to
$2.1\times10^{32}$~cm$^{-2}$s$^{-1}$ that corresponds to an average of
3.8 inelastic collisions per bunch crossing. Typical values were lower
as the luminosity degraded during the data taking fills which lasted
up to 20 hours. On average, the data contain 2.1 inelastic collisions
per bunch crossing.

\par Application of beam, detector, and data-quality requirements
resulted in a total integrated luminosity of 36~\ipb. The uncertainty
on the luminosity is 3.4\%~\cite{lumiv15,lumi1}. The integrated
luminosities for the data samples associated with the electron and
muon decay modes of the $W$ boson were calculated separately and
differ by 1.7\%.

\par Events were selected online if they satisfied either the electron
or muon criteria described below. Criteria for electron and muon
identification, as well as for event selection, followed closely those
of the previous 1.3~\ipb\ $W$+jets cross section
analysis~\cite{WJets}.

\par For this analysis, the following kinematic requirements were
imposed on events in order to enter the selected sample:
\begin{itemize}
\item $\pt^{\ell}>20$ \GeV\ ($\ell$ = electron or muon),
\item $|\eta^e|<2.47$ (except $1.37<|\eta^e|<1.52$) or $|\eta^\mu|<$2.4,
\item \met~$>25$~\GeV\ (missing transverse momentum),
\item $m_{\rm T}(W)>40$ \GeV,
\item $\pt^{\rm jet}>30$ \GeV,
\item $|y^{\rm jet}|<4.4$ and $\Delta R(\ell,\rm jet)>0.5$.
\end{itemize}
These selection criteria differ slightly from the fiducial acceptance
to which measured cross sections are finally corrected, which is
described in Section~\ref{sec:unfold}. The transverse momenta of the
leptons and neutrinos from \Wen\ and \Wmn\ decays are denoted as
$\pt^\ell$ and $\pt^\nu$, respectively. The transverse momentum of the
neutrino is determined as \met, the missing transverse momentum, from
the requirement that the total transverse momentum of all final-state
particles is a zero vector. The calculation of \met\ and the transverse
mass of the $W$, $m_{\rm{T}}(W)$, are discussed later in
Section~\ref{sec:met}.

\par All measured cross sections are corrected for any detection
losses within these regions. The lower bound $\pt^{\rm
jet}$~$>$~30~\GeV\ is chosen to facilitate comparisons with other
experiments and with next-to-leading-order QCD predictions.
Appendix~\ref{app:xsec_20GeV} shows analogous results with $\pt^{\rm
jet}$~$>$~20~\GeV\ in order to facilitate validation of the QCD
description in Monte Carlo generators and future theoretical
developments in this area.

\subsection{Electron selection}

\par In the electron channel, events were selected online using two
different triggers depending on the instantaneous luminosity. The
tighter trigger requirement corresponds to 99.1\% of the data and is a
subset of the looser one. It required the presence of at least one
electromagnetic cluster in the calorimeter with transverse energy
above 15~\GeV\ in the region of $|\eta| < 2.5$. The final selection
requirements were applied by the online event
filter~\cite{AtlasDetector} and the kinematic variables correspond
closely to those in the offline analysis described in
Section~\ref{sec:ele_offline}.

\par The impact of the trigger efficiency was small for electrons with
\ET~$>$~20~\GeV, as required in this analysis. The efficiency was
measured using \Zee\ decays identified in the experimental data. It
was found to be $99.0 \pm 0.5\%$ and constant over the full kinematic
region of this measurement~\cite{incWv16,ElePerf}.

\subsection{Muon selection}
\par In the muon channel, events were selected online using a trigger
that required the presence of a muon candidate reconstructed in both
the muon spectrometer and inner detector, consistent with having
originated from the interaction region. The candidate was required to
have \pT~$>$~10~\GeV\ or \pT~$>$~13~\GeV\ (depending on the
data-taking period) and $|\eta| < 2.4$. The higher threshold was used
to collect most of the data. As in the electron case, these
requirements were imposed in the online event filter and were less
stringent than those applied offline. The offline selection is
documented later in Section~\ref{sec:muo_offline}. The average trigger
efficiency was measured to be $\sim$85\% including the reduced
geometrical acceptance in the central region.

\begin{table*}[!htb]
\caption{Samples of simulated signal events used in this analysis. The
  $W$ samples are normalized to the inclusive NNLO cross section of
  10.46~nb calculated with FEWZ~\cite{FEWZ} using the MSTW2008 PDF
  set~\cite{MSTW2008}. For {\sc Pythia}, the inclusive $W$ sample is
  based on a $2\rightarrow 1$ matrix element merged with a
  $2\rightarrow 2$ matrix element and a leading-logarithmic parton
  shower. Details of PDF sets, final-state photon radiation, and
  underlying event tunes are given in the text.}
\begin{center}
\begin{tabular}{l|lr}
\hline\hline
\raisebox{-0.4ex}{Physics process} & \multicolumn{2}{c}{Generator} \\
\hline
$W$ inclusive ($W\rightarrow\ell\nu$; $\ell = e,\mu,\tau$) &{\sc Pythia} 6.4.21 &\cite{Pythia} \\
$W$ + jets ($W\rightarrow\ell\nu$; $\ell = e,\mu$; $0 \le N_{\rm parton}\le5$) &{\sc Alpgen} 2.13 &\cite{Alpgen} \\
$W$ + jets ($W\rightarrow\ell\nu$; $\ell = e,\mu$; $0 \le N_{\rm parton}\le5$) &{\sc Sherpa} 1.3.1 &\cite{Sherpa} \\
\hline\hline
\end{tabular}
\label{tab:MC}
\end{center}
\end{table*}

\begin{table*}[!htb]
\centering
\caption{Samples of simulated background events used in this analysis.
The $Z$+jets samples were normalized using the inclusive cross
sections from FEWZ~\cite{FEWZ} code that utilized MSTW2008 PDF
set~\cite{MSTW2008}. The $t\bar{t}$ cross section is given at
next-to-leading order (plus next-to-next-to-leading-log, NNLL). The
dijet cross sections are given at leading order (LO) in pQCD. For
these samples, the variable $\hat{p}_{\mathrm{T}}$ is the average \pT\
of the two outgoing partons from the hard-scattering process before
modification by initial- and final-state radiation and the underlying
event. Details of PDF sets, final-state photon radiation, and
underlying event tunes are given in the text.}
\begin{tabular}{l|lc|r r c}
\hline\hline
\raisebox{-0.4ex}{Physics process} & \multicolumn{2}{c|}{Generator}&
\multicolumn{3}{c}{\raisebox{-0.4ex}{$\sigma \cdot \rm{BR}$  (nb)}} \\
\hline
$Z$ + jets ($Z\rightarrow\ell\ell$; $\ell=e,\mu$; $m_{\ell\ell}>40$~\GeV; $0 \le N_{\rm parton} \le 5$) &{\sc Alpgen} 2.13 &\cite{Alpgen} & 1.07 &NNLO&\cite{FEWZ} \\
$Z\rightarrow\tau\tau$ ($m_{\ell\ell}>60$~\GeV) &{\sc Pythia} 6.4.21 &\cite{Pythia} & 0.989 &NNLO&\cite{FEWZ} \\
$t\bar{t}$ &  \multicolumn{2}{l|}{{\sc PowHeg}-HVQ}&  & & \\
           &  \hspace{0.5cm}v1.01 patch 4 &\cite{Powheg} & 0.165 &NLO+NNLL  & \cite{ttbar_th}\\
$t\bar{t}$ &{\sc AcerMC} 3.7&\cite{AcerMC} & 0.165 &NLO+NNLL  & \cite{ttbar_th}\\
Single-top $t\rightarrow \ell \nu q$ ($s$-channel) & {\sc Mc@Nlo} 3.3.1&\cite{MC_at_NLO,MC_at_NLO_st} & 4.3$\times 10^{-4}$ & NLO & \cite{MCFM} \\
Single-top $t\rightarrow \ell \nu q$ ($t$-channel) & {\sc Mc@Nlo} 3.3.1&\cite{MC_at_NLO,MC_at_NLO_st} & 6.34$\times 10^{-3}$ & NLO & \cite{MCFM} \\
Single-top ($Wt$) & {\sc Mc@Nlo} 3.3.1&\cite{MC_at_NLO,MC_at_NLO_Wt} & 13.1$\times 10^{-3}$ & NLO & \cite{MCFM} \\
$WW$ & {\sc Herwig} 6.510&\cite{Herwig} & 44.9$\times 10^{-3}$  & NLO & \cite{MCFM} \\
$WZ$ ($m_Z > 60 \GeV$) & {\sc Herwig} 6.510&\cite{Herwig} & 18.5$\times 10^{-3}$  & NLO & \cite{MCFM} \\
$ZZ$ ($m_Z > 60 \GeV$) & {\sc Herwig} 6.510&\cite{Herwig} & 5.96$\times 10^{-3}$  & NLO & \cite{MCFM} \\
Dijet ($\mu$ channel,     $\hat{p}_{\mathrm{T}}> 8$~\GeV, $\pT^\mu>$~8~\GeV)  &{\sc Pythia} 6.4.21&\cite{Pythia}  & 10.6$\times 10^6$  &LO&\cite{Pythia} \\
\hline\hline
\end{tabular}
\label{tab:MCbkg}
\end{table*}

\section{Simulated event samples}
\label{sec:mc}

\par Simulated event samples were used for most background estimates,
for the correction of the signal yield for detector effects and for
comparisons of results to theoretical expectations. The detector
simulation~\cite{AtlasSimulation} was performed using
GEANT4~\cite{Geant4}. The simulated event samples are summarized in
Table~\ref{tab:MC} for signal simulations and Table~\ref{tab:MCbkg}
for the background simulations. The {\sc Alpgen} and {\sc Mc@Nlo}
samples were interfaced to {\sc Herwig} for parton shower and
fragmentation processes and to {\sc Jimmy} v4.31~\cite{Jimmy} for
underlying event simulation. Similarly, {\sc Jimmy} was used for the
underlying event simulation in the diboson samples produced with {\sc
Herwig}. The {\sc AcerMC} \ttbar\ samples were showered with {\sc
Pythia} where the default settings for initial state radiation (ISR)
and final state radiation (FSR) were altered~\footnote{The ISR and FSR
rates were increased or decreased individually or simultaneously. To
decrease ISR, parameters PARP(67) and PARP(64) were adjusted from 4
and 1 to 0.5 and 4, respectively. They were set to 6 and 0.25 to
increase ISR. Similarly, parameters PARP(72) and PARJ(82) were
switched from 0.192~\GeV\ and 1~\GeV\ to 0.096~\GeV\ and 2~\GeV\ to
decrease FSR and to 0.384~\GeV\ and 0.5~\GeV\ to increase FSR.}. The
parameterization of the factorization scale used for the matrix
element (ME) calculation in the {\sc Alpgen} samples was chosen to be
$Q_0^2=m_V^2+\displaystyle\sum\limits_{\rm{partons}}^{}(\pt^2)$, where
$m_V$ is the mass of a $W$ or $Z$ boson and the decay products of the
boson are not included in the sum~\cite{Alpgen}. The parton-jet
matching was performed at \mbox{$\pT^{\rm{jet}}=20$}~\GeV\ with the
MLM matching scheme~\cite{MLMmatch} using jets from the cone
clustering algorithm with \mbox{$R=0.7$}. The default renormalization
and factorization scales were used in the {\sc Sherpa} samples and the
parton-jet matching was performed at \mbox{$\pT^{\rm{jet}}=30$}~\GeV\
using the CKKW matching scheme~\cite{CKKW1,CKKW2}. Parton density
functions (PDFs) were: CTEQ6L1~\cite{CTEQ6L1} for the {\sc Alpgen}
samples and the parton showering and underlying event in the {\sc
PowHeg} samples interfaced to {\sc Pythia}; MRST 2007
$\rm{LO}^*$~\cite{LO*} for {\sc Pythia}, {\sc AcerMC}, and the diboson
samples; and CTEQ6.6M~\cite{CTEQ6.6M} for {\sc Mc@Nlo}, {\sc Sherpa},
and the NLO matrix element calculations in {\sc PowHeg}. The radiation
of photons from charged leptons was treated in {\sc Herwig} and {\sc
Pythia} using {\sc Photos} v2.15.4~\cite{Photos}. {\sc Tauola}
v1.0.2~\cite{Tauola} was used for $\tau$ lepton decays. The underlying
event tunes were the ATLAS MC10 tunes: ATLAS underlying event tune \#1
(AUET1)~\cite{AUET1} for the {\sc Herwig}, {\sc Alpgen}, and {\sc
Mc@Nlo} samples; ATLAS minimum bias 1 (AMBT1)~\cite{AMBT1} for {\sc
Pythia}, {\sc AcerMC}, and {\sc PowHeg} samples. These two tunes were
derived using $pp$ collisions at $\sqrt{s}=7$~\TeV\ produced at the
LHC. The samples generated with {\sc Sherpa} used the default
underlying event tune determined from lower energy measurements and
$pp$ data from the LHC.

\par Samples were generated with minimum bias interactions overlaid on
the hard-scattering event to account for the multiple $pp$
interactions in the same beam crossing (pile-up). The minimum bias
interactions were simulated with {\sc Pythia} with the AMBT1
tune. These samples were then re-weighted so the distribution of the
number of primary vertices matched that of the data.

\section{Offline event analysis}

\par Events were selected if they satisfied the criteria described
above and had at least one interaction vertex with three or more
associated charged particle tracks, located within 200~mm in $z$ from
the center of the detector. For these data the luminous region had a
typical RMS size of $\sim60$~mm in $z$. The position resolution of
reconstructed vertices along $z$ was $\sim0.1$~mm for a vertex with 10
tracks. For the sample of events passing the single-lepton trigger the
mean number of interaction vertices was 2.1 per event. The primary
vertex was taken as the one with the largest $\Sigma\pT^2$ of
associated tracks. Events with significant noise in the calorimeters,
cosmic rays, and beam-induced background were rejected~\cite{JES}.

\subsection{Jet selection}
\par Jets were reconstructed from energy observed in the calorimeter
cells using the anti-$k_{\rm{t}}$ algorithm~\cite{Cacciari:2008gp}
with a radius parameter \mbox{$R=0.4$}~\cite{JES}. Since the volume of
individual cells is small compared to the volume of the
electromagnetic and hadronic energy showers, cells were grouped into
clusters depending on their signal size relative to
noise~\cite{topoclusters}. These clusters formed the input to the jet
reconstruction. Since a jet involves many clusters a mass can be
calculated and the jet rapidity rather than pseudorapidity was
determined.

\par To account for the difference in calorimeter response between
electrons and hadrons of the same energy, and to correct for other
experimental effects, a \pT\ and $\eta$-dependent factor, derived from
simulated events, was applied to each jet to provide an average energy
scale correction~\cite{JES}. Jets were required to have a rapidity
$|y| < 4.4$ and \pT~$>$~30~\GeV. To ensure a reliable energy
measurement all jets within $\Delta R < 0.5$ of an electron or muon
(that passed the lepton identification requirements) were explicitly
not considered, regardless of the jet \pT\ or rapidity, but the event
itself was retained. Jets consistent with detector noise, cosmic rays,
or beam halo were rejected~\cite{JES}. The jet rejection requirement 
was more stringent than that applied to events.

\par
\begin{figure}[htb]
\centering
\includegraphics[width=0.95\linewidth]{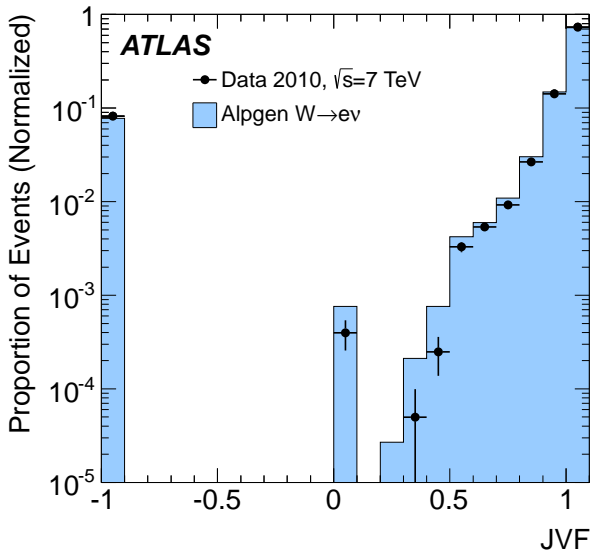}
\caption{Jet-vertex fraction (JVF) distribution for all jets in the
\Wen\ sample. The events at $-1$ correspond to jets where the JVF
could not be calculated, while the peak near 0 corresponds to jets
from a secondary vertex. For the data 99.1\% of the jets pass the
requirement that the absolute value of the JVF be greater than 0.75,
while for the Monte Carlo sample this rate is 98.8\%.}
\label{fig:JVF}     
\end{figure}

 To suppress jets arising from additional $pp$ interactions a
parameter called the jet-vertex fraction (JVF) was calculated for each
jet in the event. After associating tracks to jets by requiring
$\Delta R<0.4$ between tracks and a jet, the JVF was computed for each
jet as the scalar sum of \pT\ of all associated tracks from the
primary vertex divided by the total \pT\ associated with that jet from
all vertices. The JVF could not be calculated for jets which fell
outside the fiducial tracking region (\mbox{$|\eta|<2.5$}) or which
had no matching tracks so these were assigned a value of $-1$ for
accounting purposes. Only jets with the absolute value of the JVF
smaller than 0.75 were rejected so that jets with a JVF of $-1$ were
kept.  Figure~\ref{fig:JVF} shows the distribution of this parameter
for all jets in the \Wen\ data and Monte Carlo event samples. The
requirement on the JVF is most important for low \pT\ jets and for the
data with high instantaneous luminosity.

\par The pile-up collisions also add a uniform background of particles
to the events and slightly increase the measured jet energies. The jet
energy calibration factor described above contains a correction for
this effect.

\par No minimum separation $\Delta R$ was required between final state
jets, but the measured jet response changed for separations less than
\mbox{$\Delta R<0.5$}. This distortion in the response was corrected
by the event reconstruction efficiency calculation and residual
effects enter the estimated systematic uncertainties.

\par After the application of all jet requirements, the efficiency for
reconstructing jets was determined from simulation to be $\sim97\%$
for jets with \mbox{\pT~=~30}~\GeV, rising to close to 100\% for jets
above 80~\GeV. The uncertainties in the jet energy scale and jet
energy resolution were determined in separate studies~\cite{JES}. The
uncertainties in the jet energy scale were 2.5--14\%, and depended on
the $\eta$ and \pt\ of the jet. The uncertainty on the jet energy
resolution was $\sim10\%$ for each jet, relative to the nominal
resolution which also varied with $\eta$ and \pt.

\subsection{Missing transverse momentum and $m_{\rm{T}}(W)$}
\label{sec:met}

\par The calculation of missing transverse momentum (\met) and
transverse mass of $W$ bosons ($m_{\rm{T}}(W)$) followed the
prescription in Refs.~\cite{MetRefFinal} and~\cite{incWv16}. $m_{\rm
T}(W)$ was defined by the lepton and neutrino \pT\ and direction as
\mbox{$\ensuremath{m_{\mathrm{T}}}(W)=
\sqrt{2\ensuremath{p_{\mathrm{T}}}^{\ell}\ensuremath{p_{\mathrm{T}}}^{\nu}(1-\cos(\phi^{\ell}-\phi^{\nu}))}$},
where the $(x,y)$ components of the neutrino momentum were taken to be
the same as the corresponding \met\ components. \met\ was calculated
from the energy deposits in calorimeter cells inside three-dimensional
clusters~\cite{topoclusters}. These clusters were then corrected to
account for the different response to hadrons compared to electrons or
photons, as well as dead material and out-of-cluster energy
losses~\cite{lochadtopo}. Only clusters within $|\eta| < 4.5$ were
used. In the muon channel, \met\ was corrected for the muon momentum
and its energy deposit in the calorimeters. Events were required to
have \met~$>$~25 GeV and $m_{\rm{T}}(W)$~$>$~40~\GeV.

\subsection{\Wen\ + jets final state}
\label{sec:ele_offline}

\begin{figure*}[htb]
\centering
\includegraphics[width=0.32\linewidth]{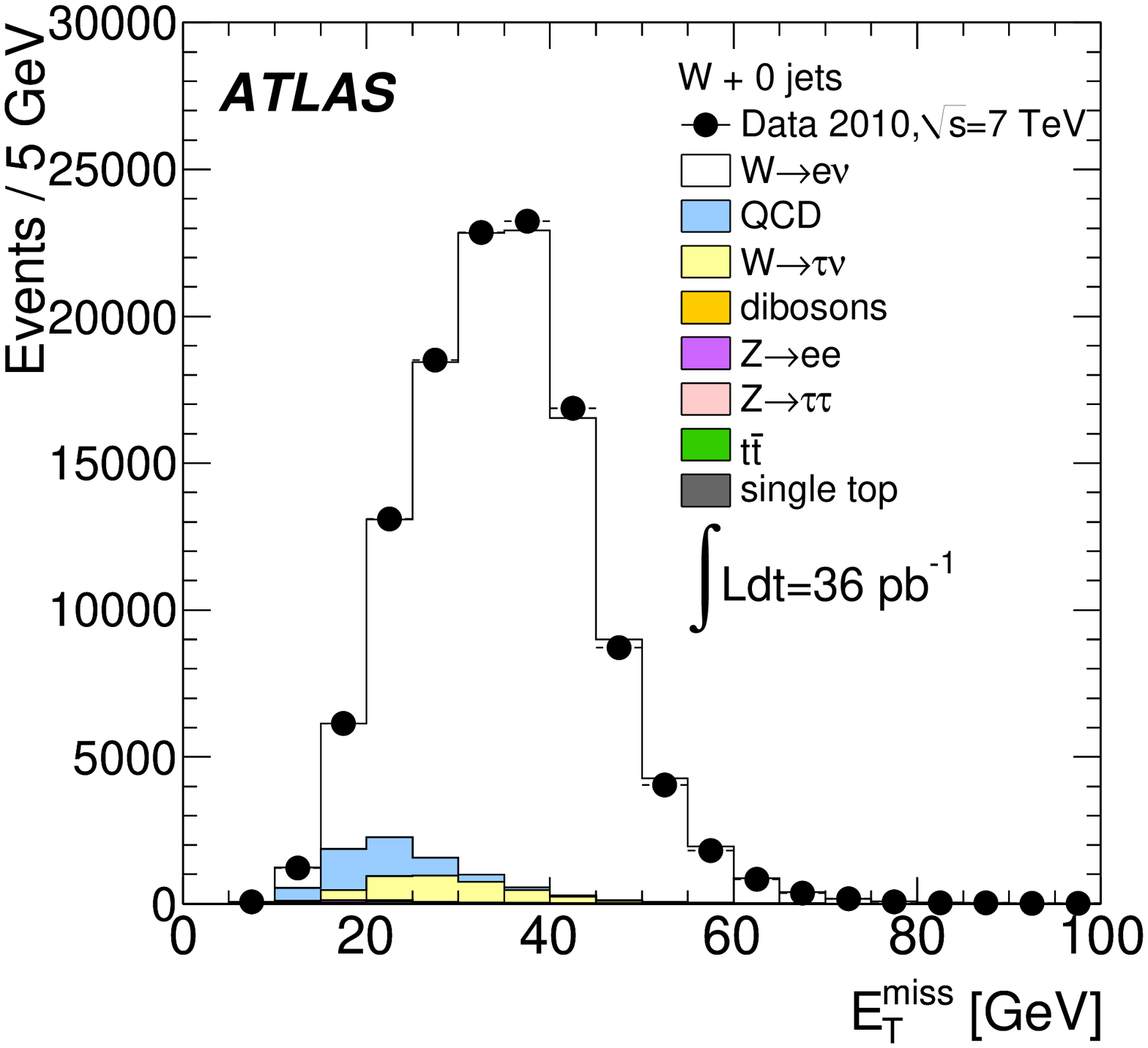}
\includegraphics[width=0.32\linewidth]{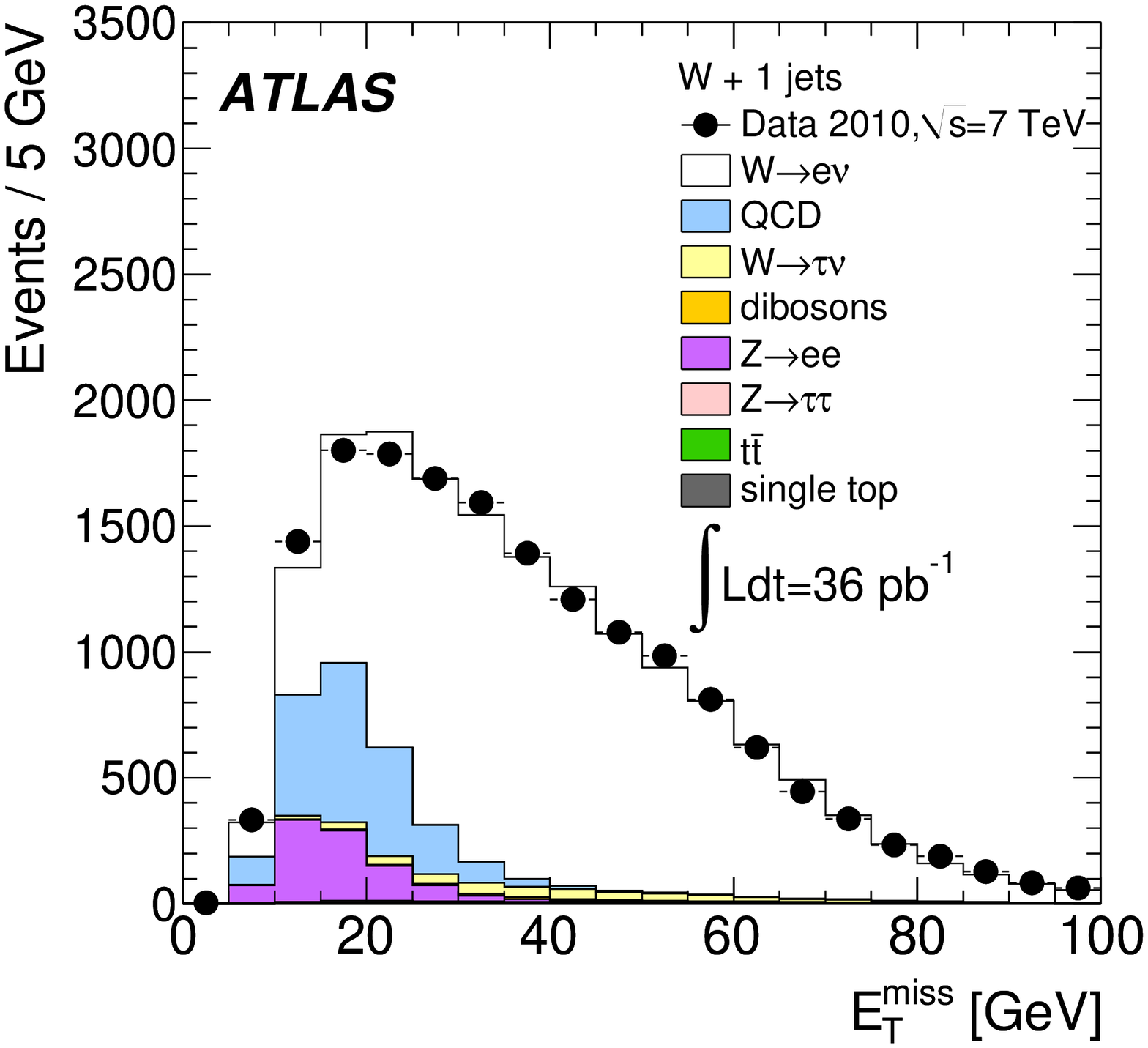}
\includegraphics[width=0.32\linewidth]{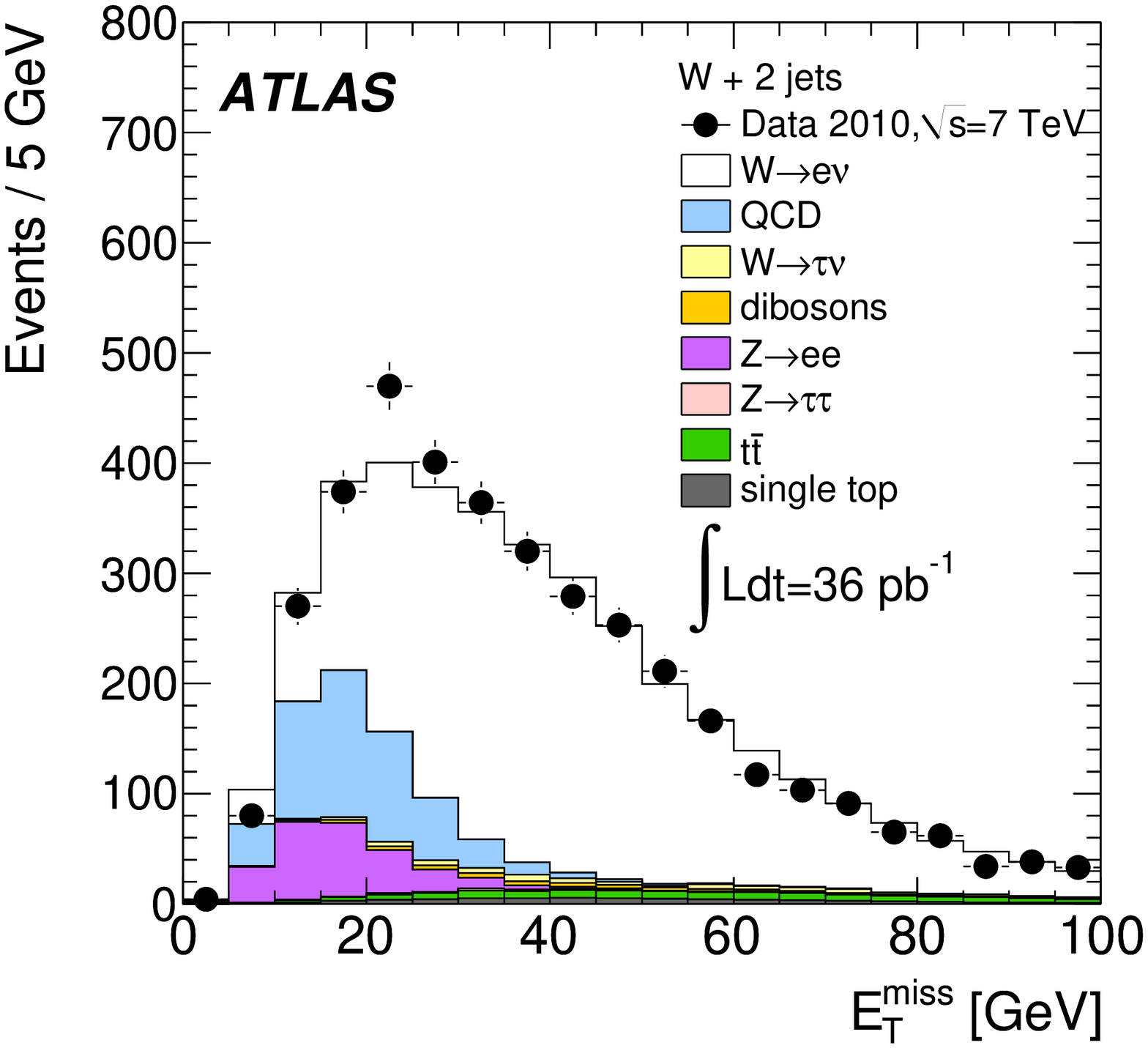}
\includegraphics[width=0.32\linewidth]{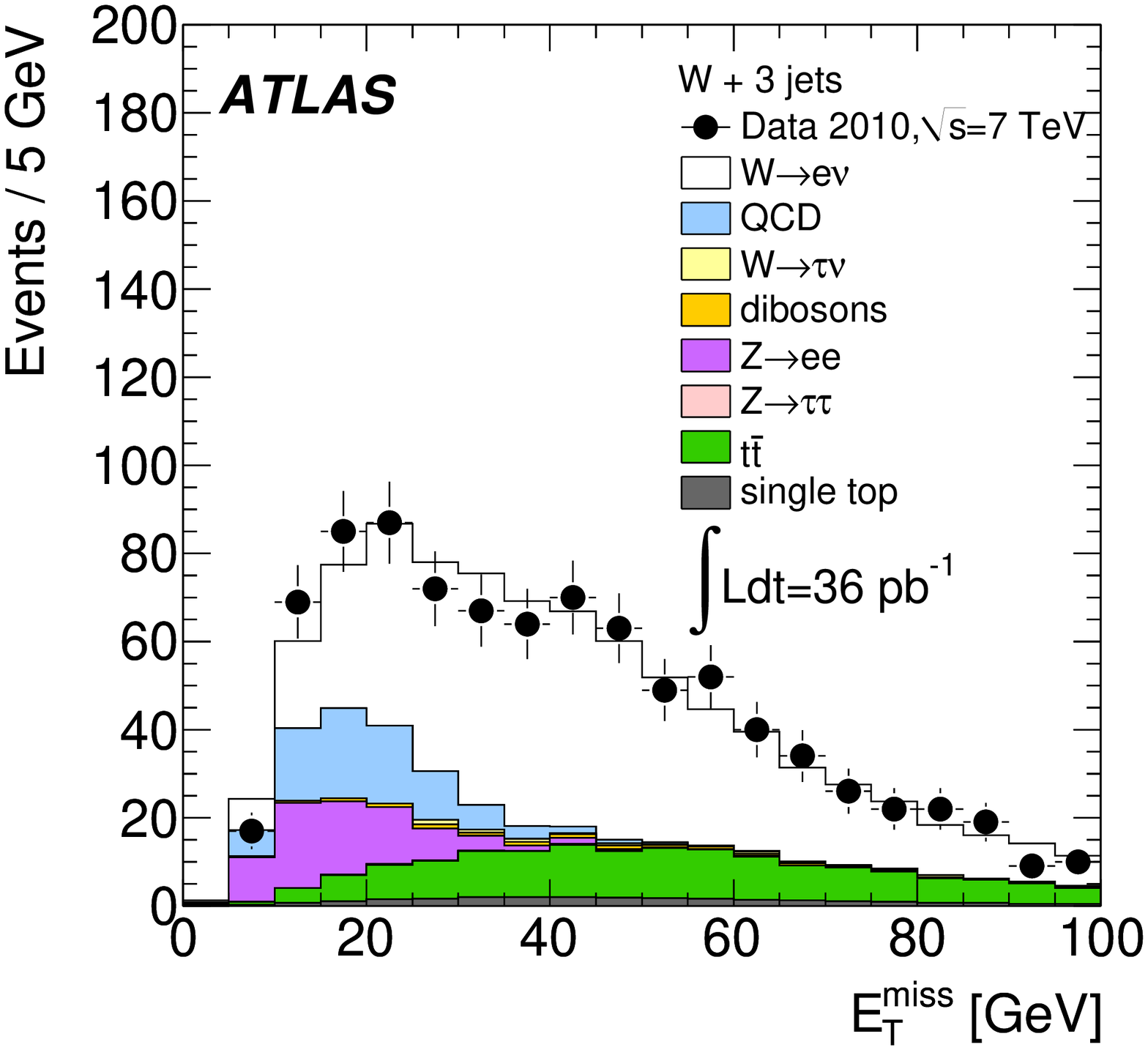}
\includegraphics[width=0.32\linewidth]{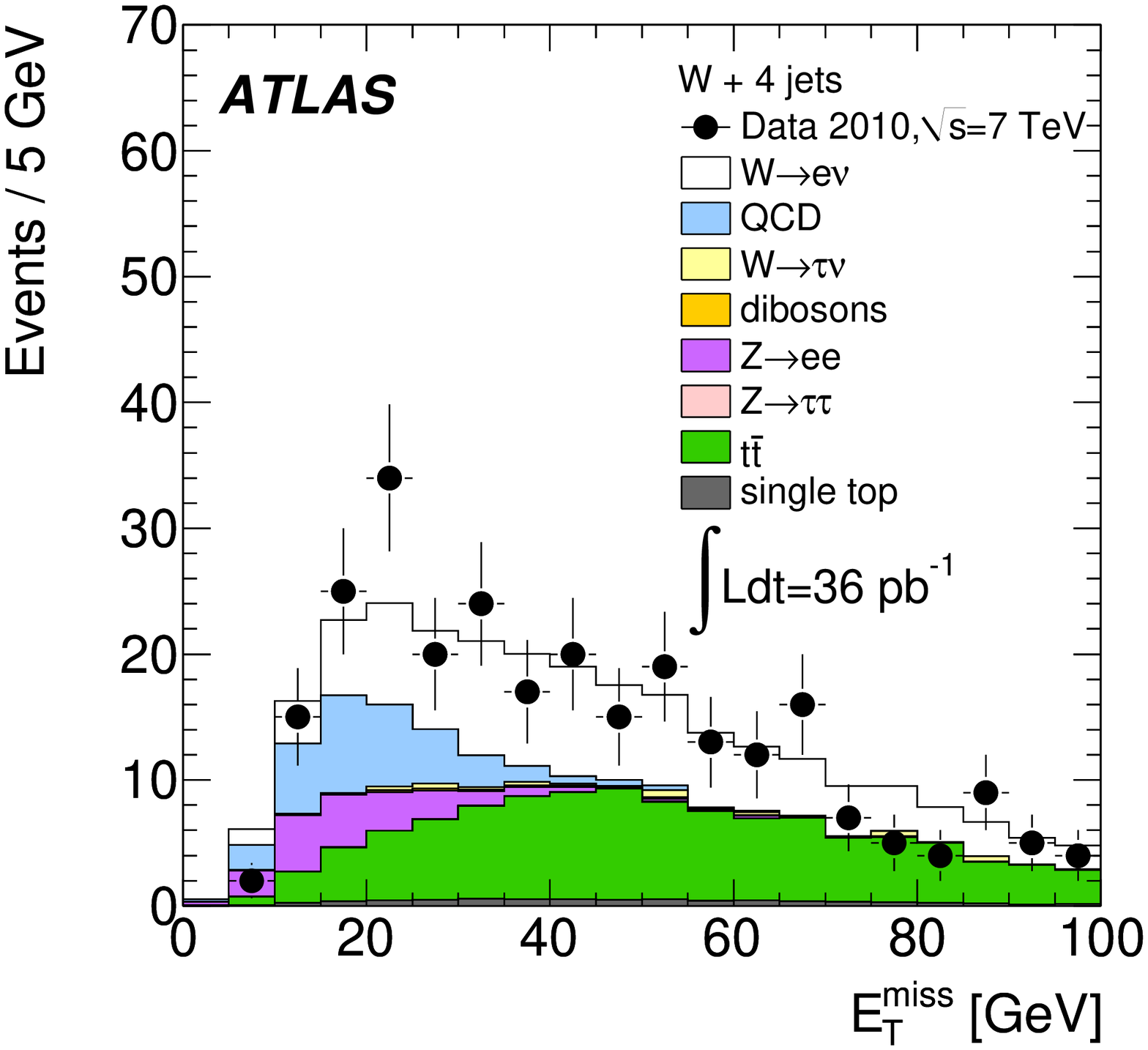}
\includegraphics[width=0.32\linewidth]{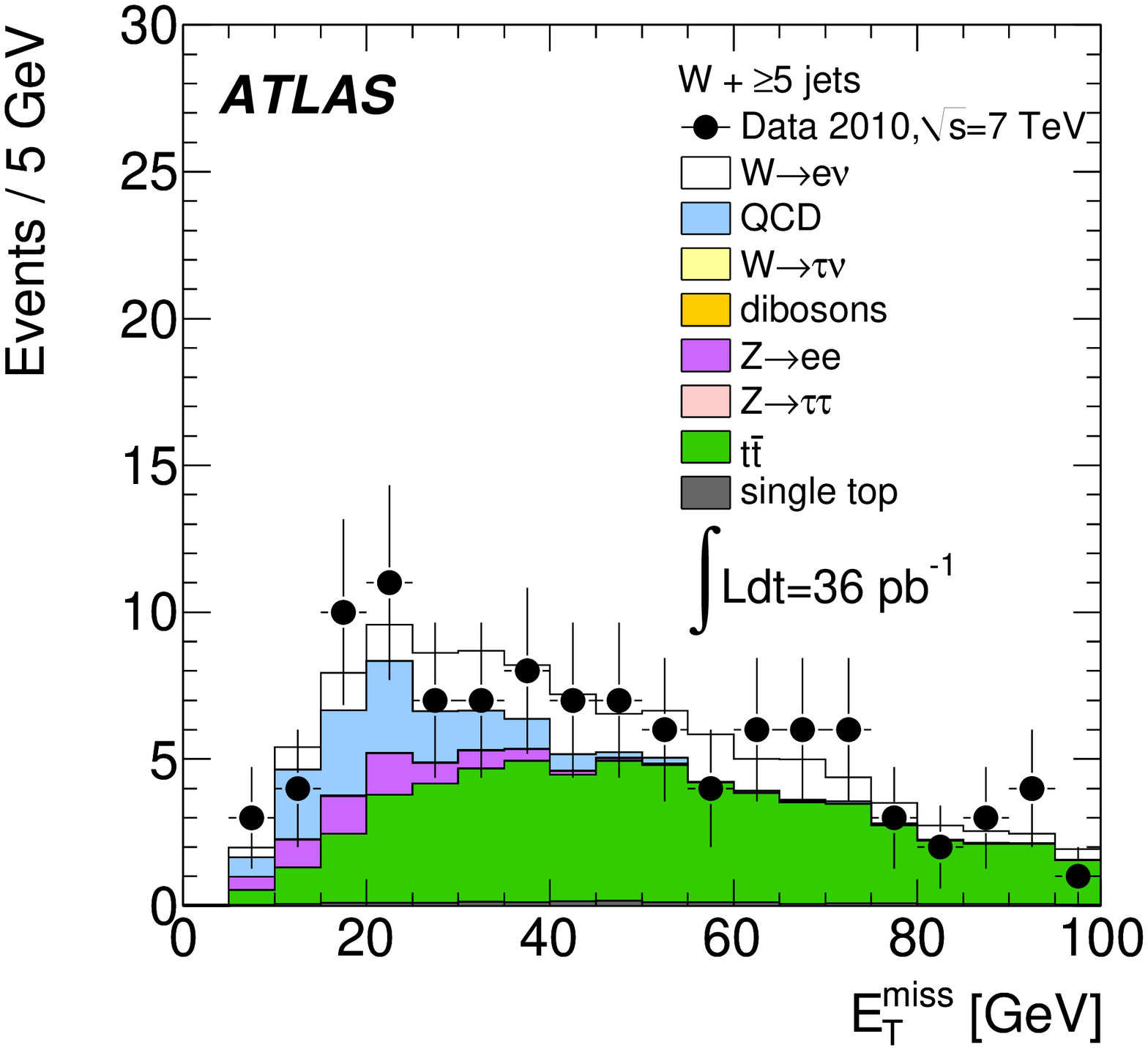}
\caption{Result of the \met\ template fits used to obtain an estimate
of the multijet background for \Wen\ events, in bins of exclusive jet
multiplicity. The data is shown with the statistical uncertainties
only. In this case the multijet template was obtained with relaxed
shower shape requirements, as described in the text. The data with
$\ge$ 5 jets are not used for measurements because of low event
multiplicity and a poor signal-to-background ratio. The event
multiplicity and the ratio were better for
$\pt^{\rm{jet}}$~$>$~20~\GeV. $W$ candidate events were required to
have \met~$>$~25~GeV.}
\label{fig:fitresultsI}     
\end{figure*}

\begin{figure*}[htb]
\centering
\includegraphics[width=0.32\linewidth]{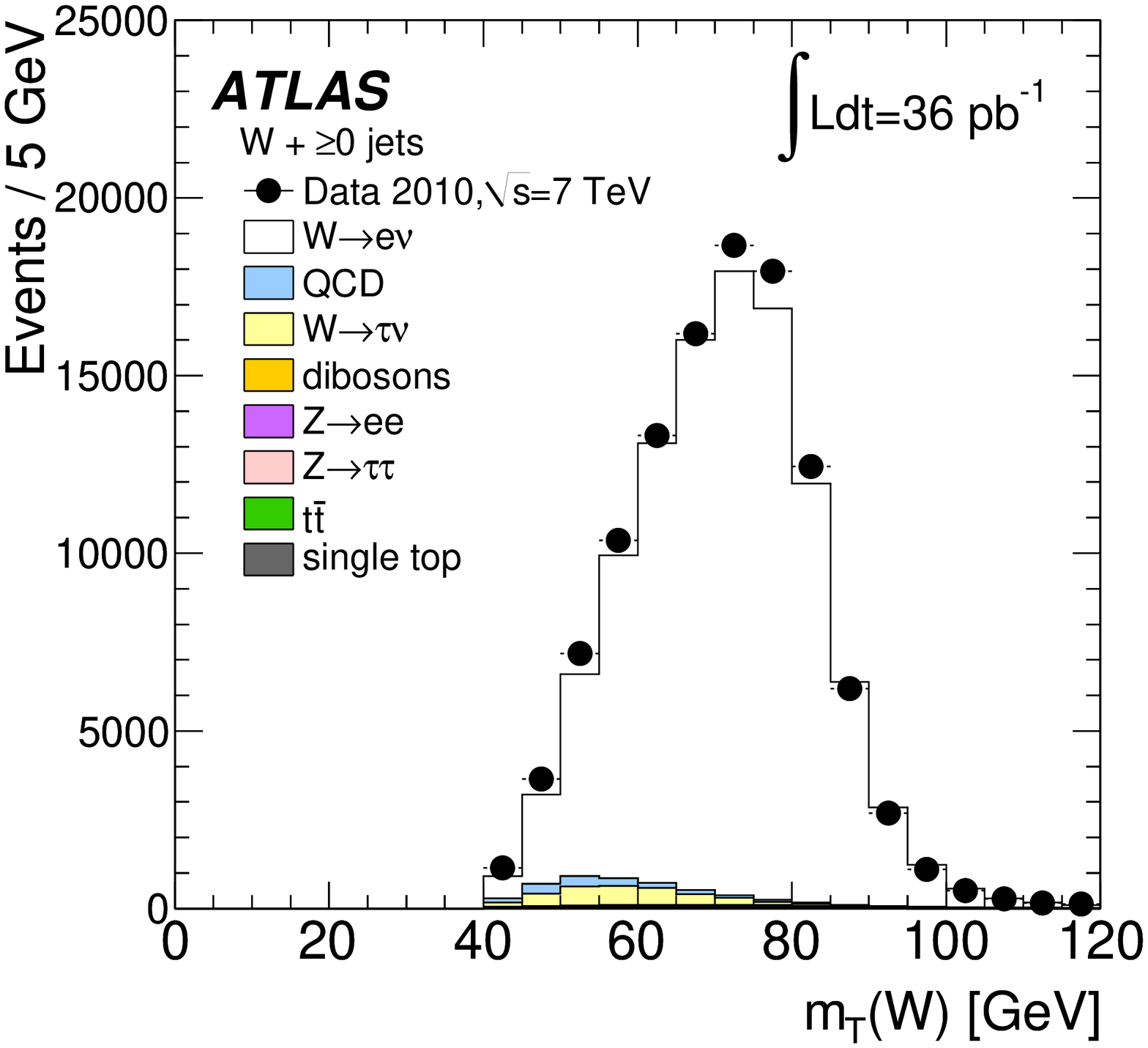}
\includegraphics[width=0.32\linewidth]{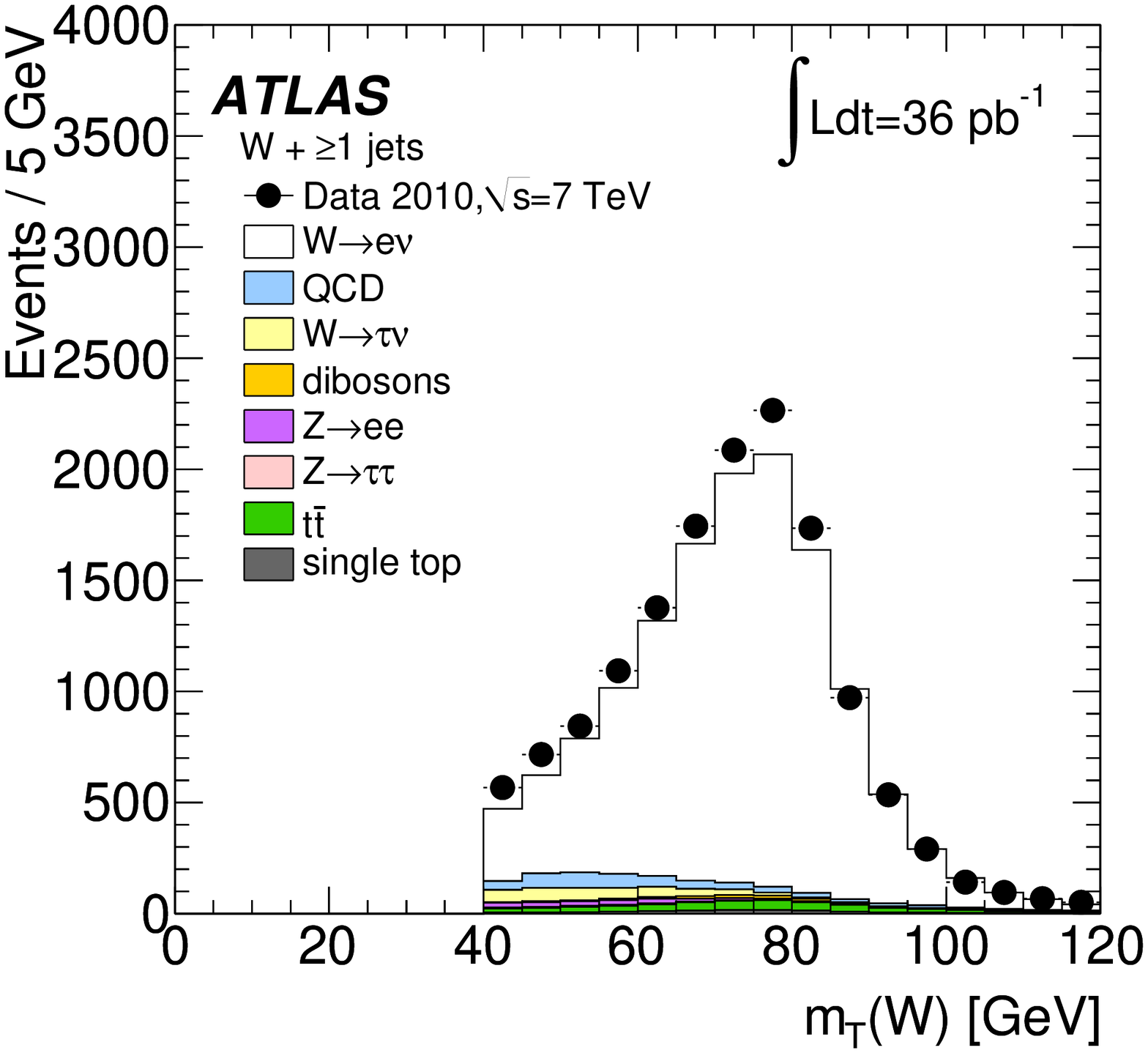}
\includegraphics[width=0.32\linewidth]{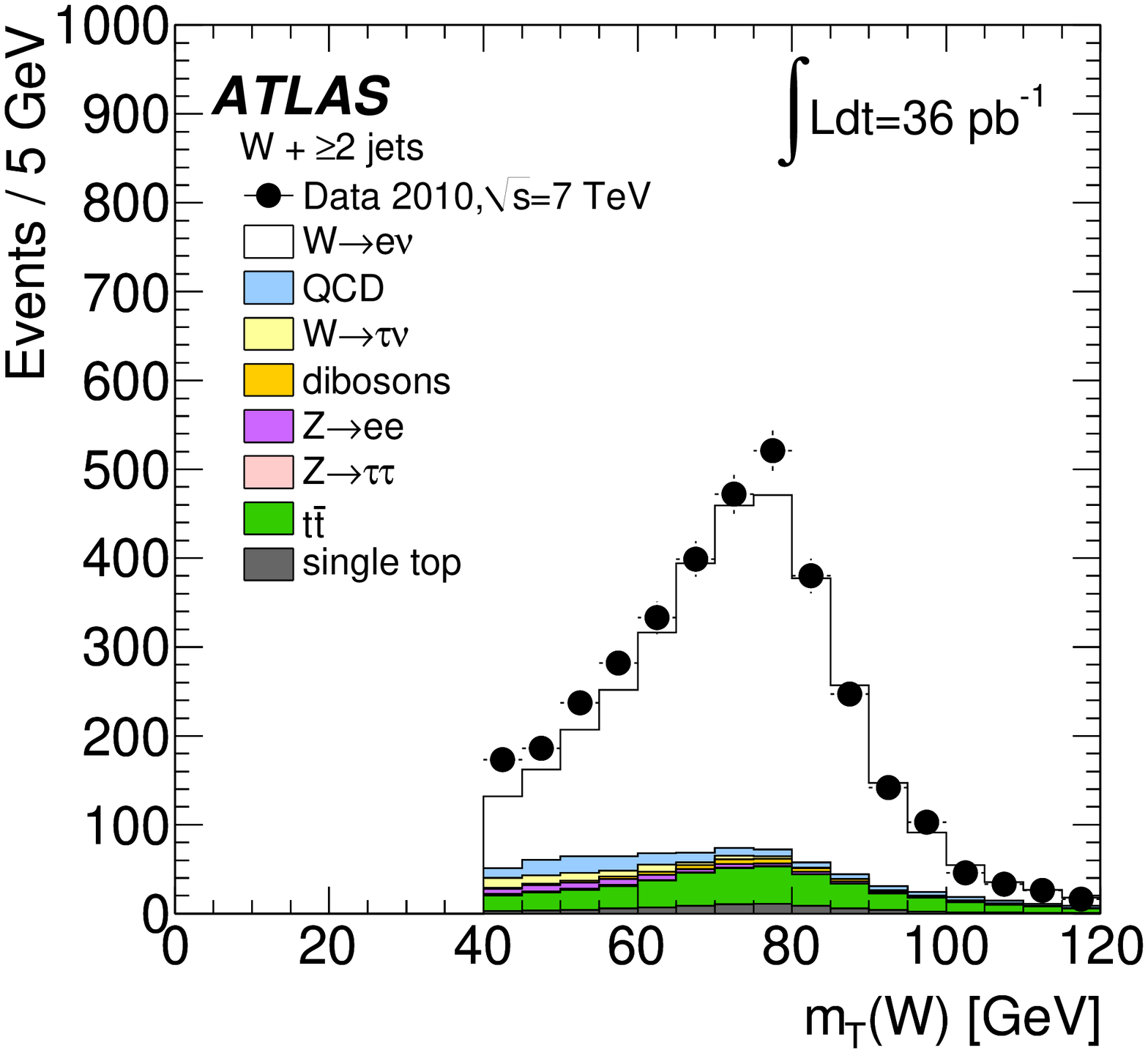}
\includegraphics[width=0.32\linewidth]{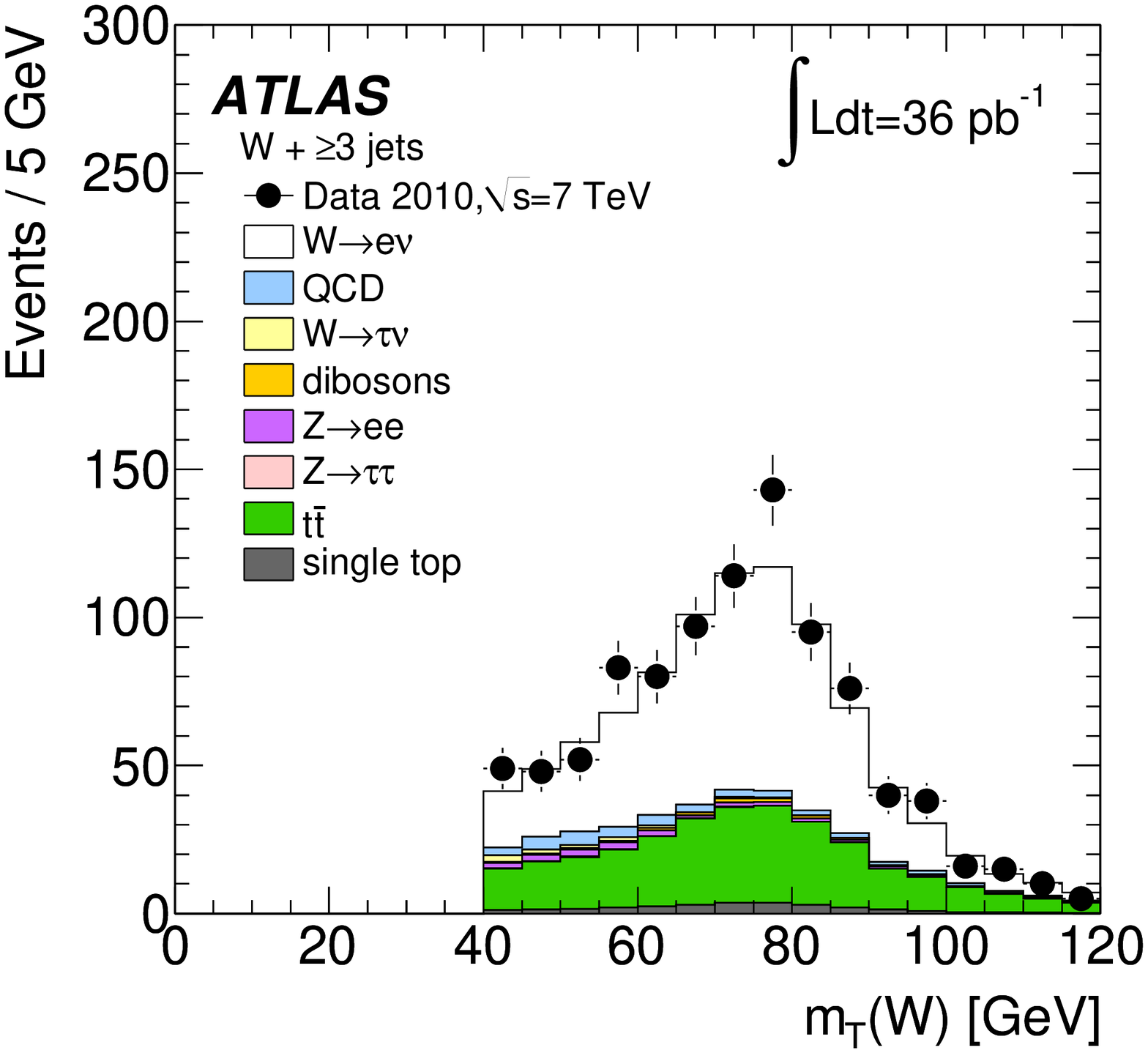}
\includegraphics[width=0.32\linewidth]{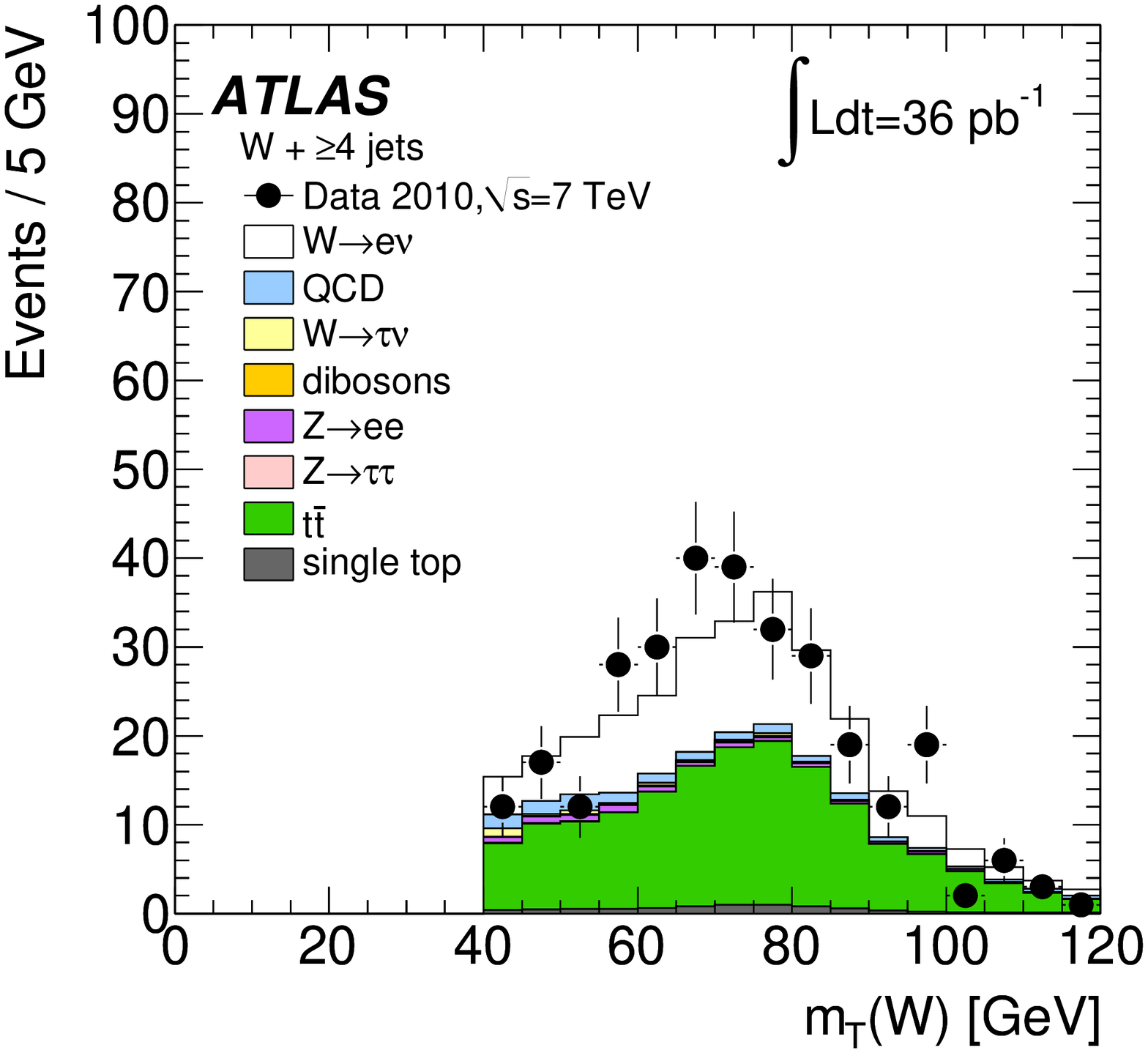}
\includegraphics[width=0.32\linewidth]{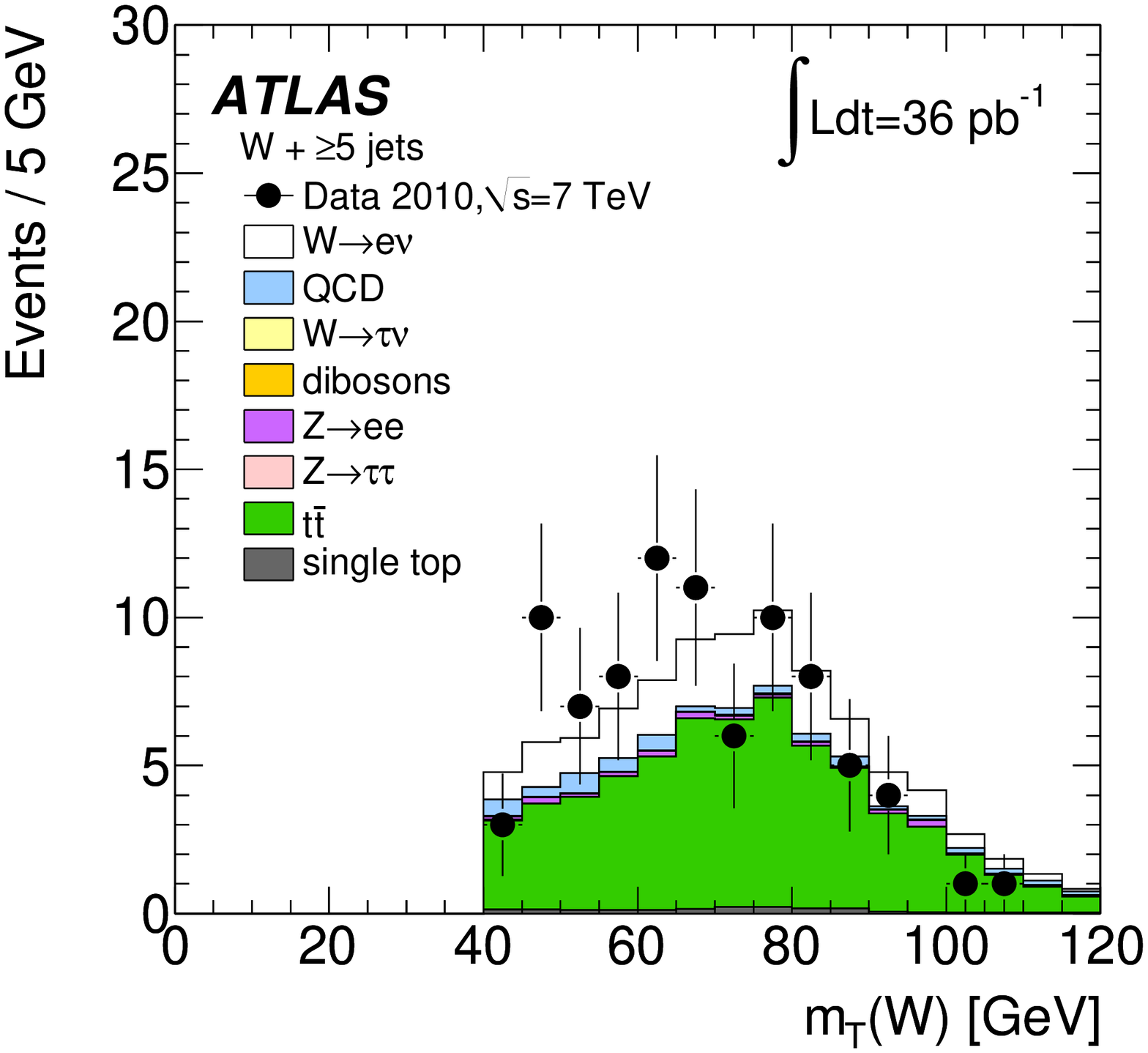}
\caption{Transverse mass distributions $m_{\rm T}(W)$ for selected
\Wen\ events in bins of inclusive jet multiplicity. MC predictions for
the signal and leptonic backgrounds are normalized to luminosity using
(N)NLO cross sections and the multijet background is estimated from
data (method I).}
\label{fig:Etmiss_WMt}
\end{figure*}

\par Electrons were required to pass the standard ``tight'' electron
selection criteria~\cite{incWv16,ElePerf} with \ET~$>$~20~\GeV\ and
$|\eta| < 2.47$. Electrons in the transition region between the barrel
and endcap calorimeter ($1.37<|\eta|<1.52$) were rejected.

\par To suppress multijet events containing non-isolated electrons
such as those from semileptonic decays of hadrons containing charm and
bottom quarks, a calorimeter-based isolation requirement was applied.
The transverse energy within a cone of radius $R=0.2$ around the
electron, corrected for contributions from the electron, was required
to be less than 4~\GeV. This isolation requirement is more than 96\%
efficient over all jet multiplicities for prompt electrons originating
from decays of $W$ bosons and reduces the non-isolated electron
background by a factor of two.

\par To remove backgrounds from \Zee\ decays, events were also
rejected if there was a second electron passing the ``medium''
electron selection criteria~\cite{incWv16,ElePerf} and the same
kinematic selections and isolation requirements as above.

\subsubsection{Electron channel background estimates}
\label{sec:ele_backgr}

\par The principal backgrounds in the electron channel arise from
multijet QCD events, other leptonic decays of gauge bosons and, at
higher jet multiplicities, \ttbar\ production. The background from
gauge bosons includes \Wtau\ where the $\tau$ lepton decays to an
electron and \Zee\ where one electron is not identified and hadronic
energy in the event is mismeasured. Leptonic \ttbar\ decays ($\ttbar
\rightarrow b\overline{b} q q' e \nu$), single-top events and diboson
($WW$, $WZ$, $ZZ$) processes were also evaluated. The number of
leptonic background events surviving the above selection requirements
was estimated with simulated event samples that were introduced
earlier in Section~\ref{sec:mc}.  Specifically, {\sc Pythia} was used
for \Wtau\ and \Ztau\ and {\sc Alpgen} for the other vector boson
samples. The simulated leptonic background samples were normalized to
the integrated luminosity of the data using the predicted cross
sections shown in Table~\ref{tab:MCbkg}. The \ttbar\ background is
discussed in more detail later in Section~\ref{sec:det_lvl_compar}.

\par The multijet background in the electron channel has two
components, one where a light flavor jet passes the electron selection
and additional energy mismeasurement results in large \met, and the
other where a bottom or charm hadron decays to an electron. The number
of multijet background events was estimated by fitting, for each
exclusive jet multiplicity, the \met\ distribution in the data
(without the \met\ selection requirement) to a sum of two templates:
one for the multijet background and another which included signal and
the leptonic backgrounds. The fits determined the relative 
normalizations of the two templates for each exclusive jet 
multiplicity. The shapes for the second template were obtained from 
simulation and their relative normalization was  fixed to the ratio of 
their predicted cross sections.

\par The template for the multijet background was obtained from the
data because the mechanisms by which a jet fakes an electron are
difficult to simulate reliably. The template was derived by loosening
some of the electron identification requirements. Two approaches were
taken so their results could be compared.

\par In the first, the requirements on shower shape in the calorimeter
were relaxed. The ``loose'' electron identification criteria of
Ref.~\cite{incWv16,ElePerf} were applied to the shower shapes. The
track-cluster matching requirements applied in the standard ``tight''
electron selection were still applied but the remaining ``tight''
requirements with respect to the ``medium'' requirements were required
to fail~\cite{incWv16,ElePerf}; the selection favors electron
candidates from conversions or from charged hadrons overlapping
electromagnetic showers.

\par In the second method, the requirement that a track matched the
energy deposition in the calorimeter was relaxed and loose photon
identification requirements were used instead of those of an electron.

\par To suppress any residual signal contribution, the isolation
requirement was also reversed in both methods. A large simulated dijet
sample was used to verify that these requirements do not bias the
\met\ shape of the background templates.

\par The results of the two methods were compared for each jet
multiplicity and agreed within their statistical uncertainties. For
the zero-jet bin they agreed to better than 17\% with respect to the
total number of candidate background events. Residual differences are
included in the estimates of systematic uncertainty described
below. The range of \met\ used to fit the templates was also varied to
estimate systematic effects. The first method was used to calculate
the central values of the multijet backgrounds for the various jet
multiplicities.

\par The comparisons of the template fits to the \met\ distributions
are shown in Fig.~\ref{fig:fitresultsI} for the first type of multijet
template. Figure~\ref{fig:Etmiss_WMt} shows the final $m_{\rm T}(W)$
distributions in the various bins of inclusive jet multiplicity.

\begin{figure*}[!htb]
\centering
\includegraphics[width=0.32\linewidth]{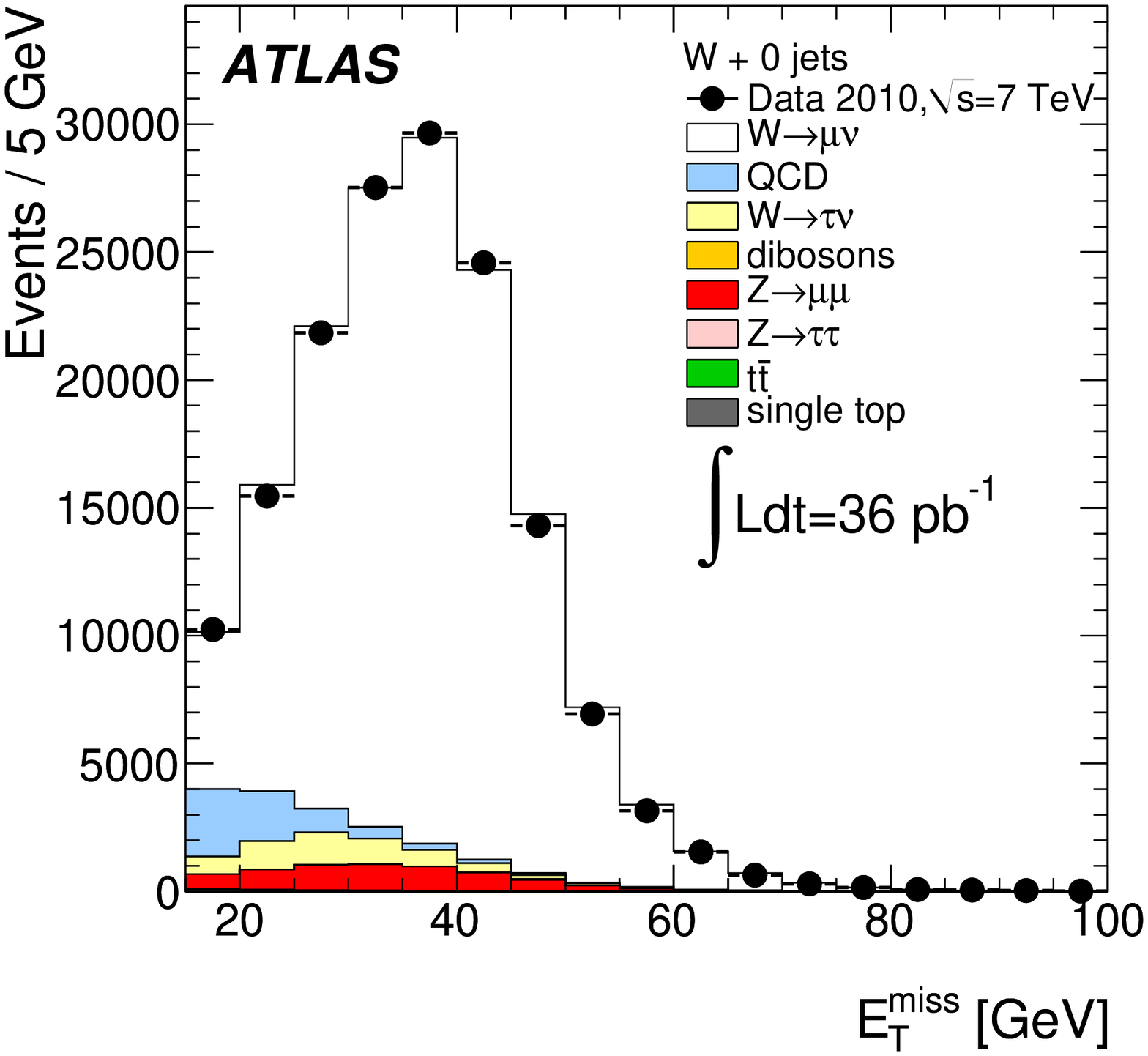}
\includegraphics[width=0.32\linewidth]{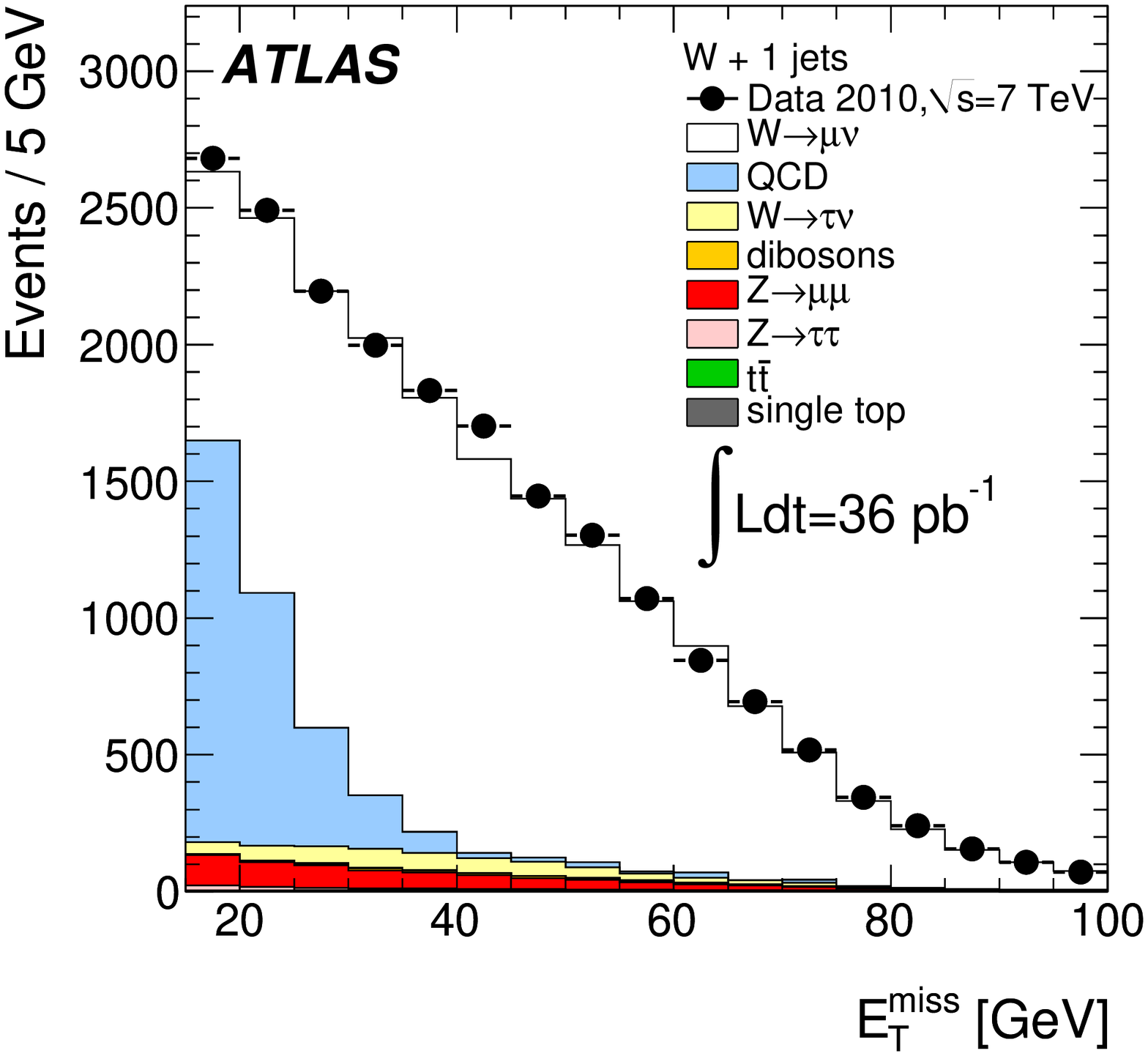}
\includegraphics[width=0.32\linewidth]{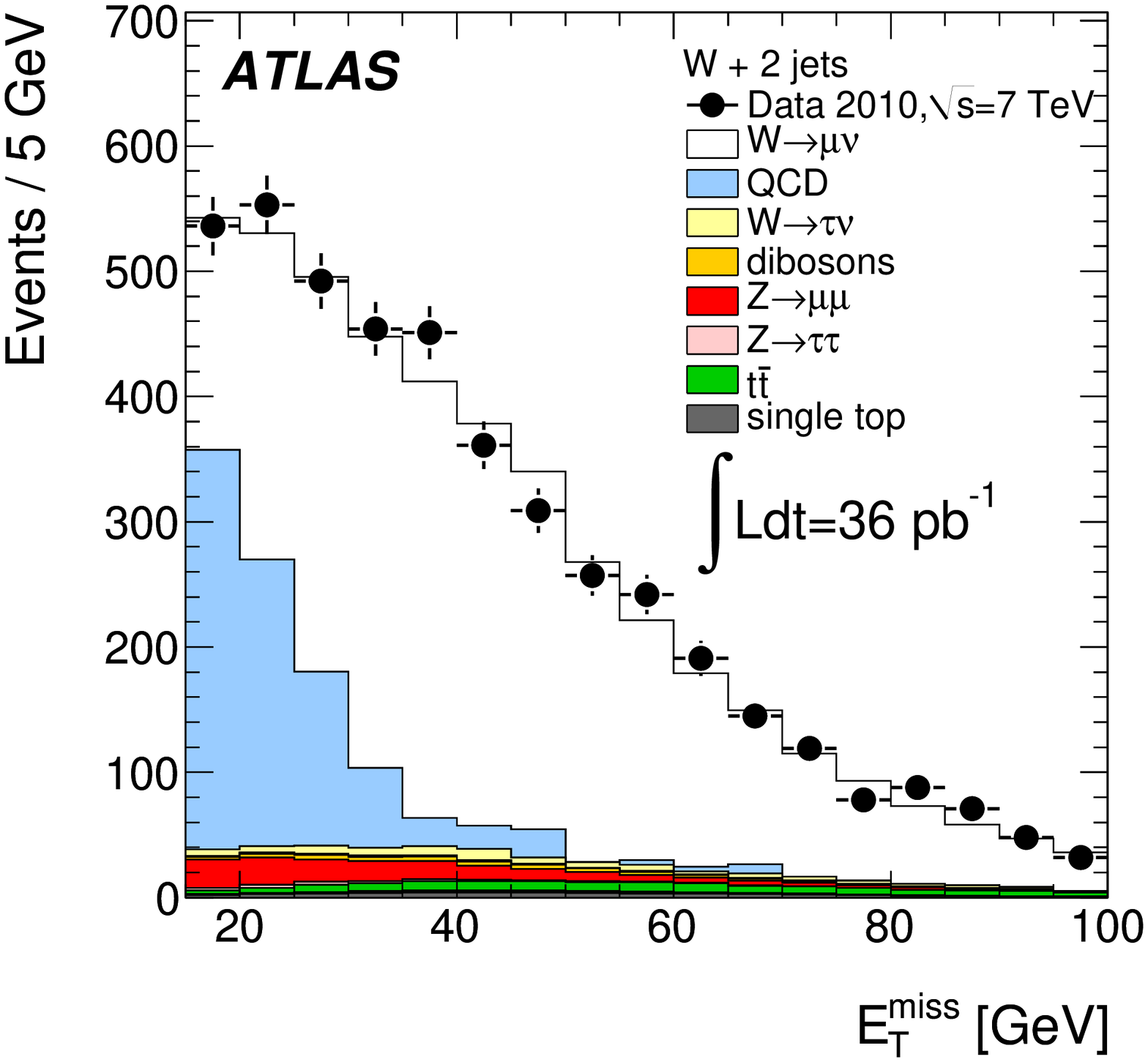}
\includegraphics[width=0.32\linewidth]{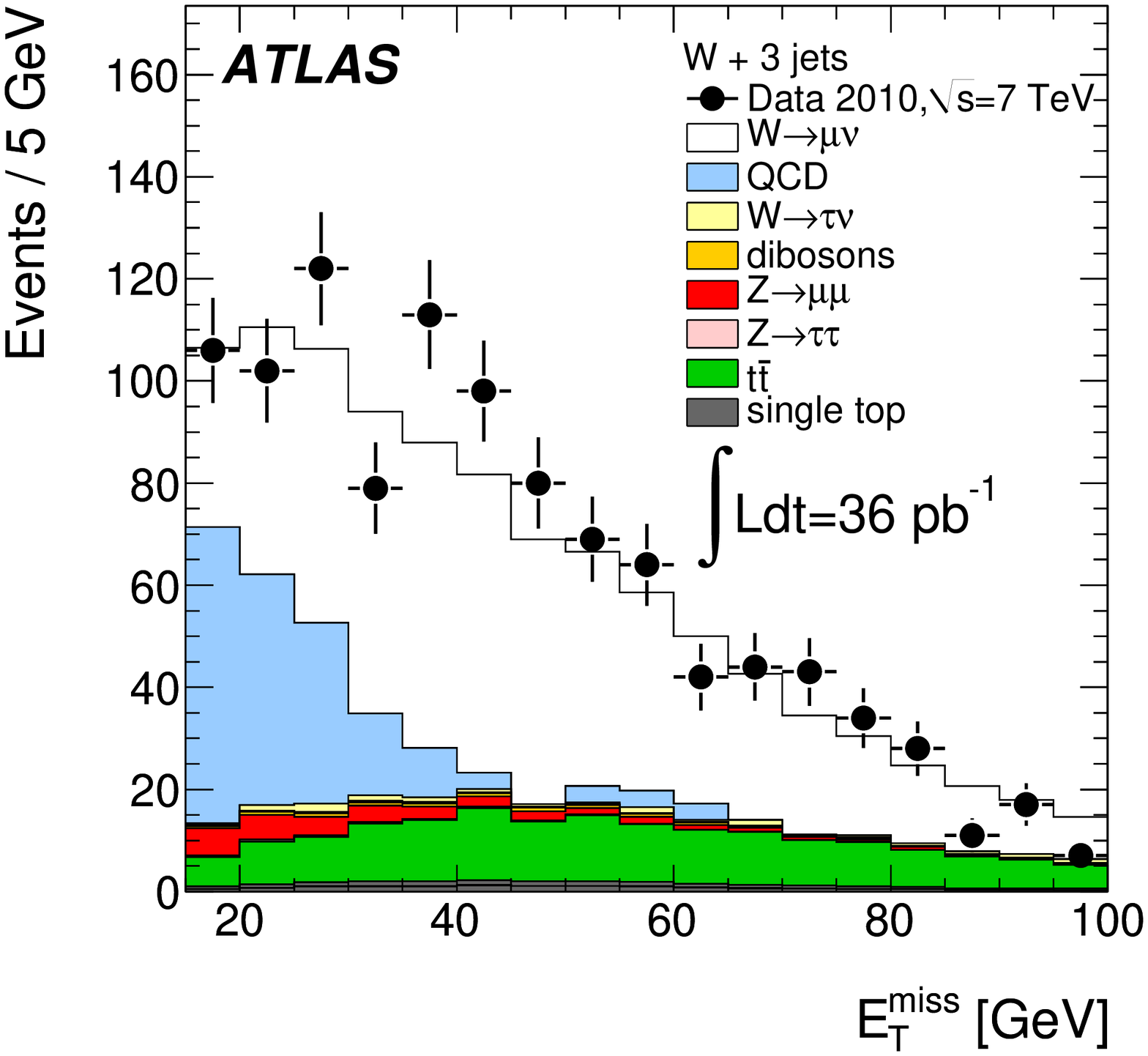}
\includegraphics[width=0.32\linewidth]{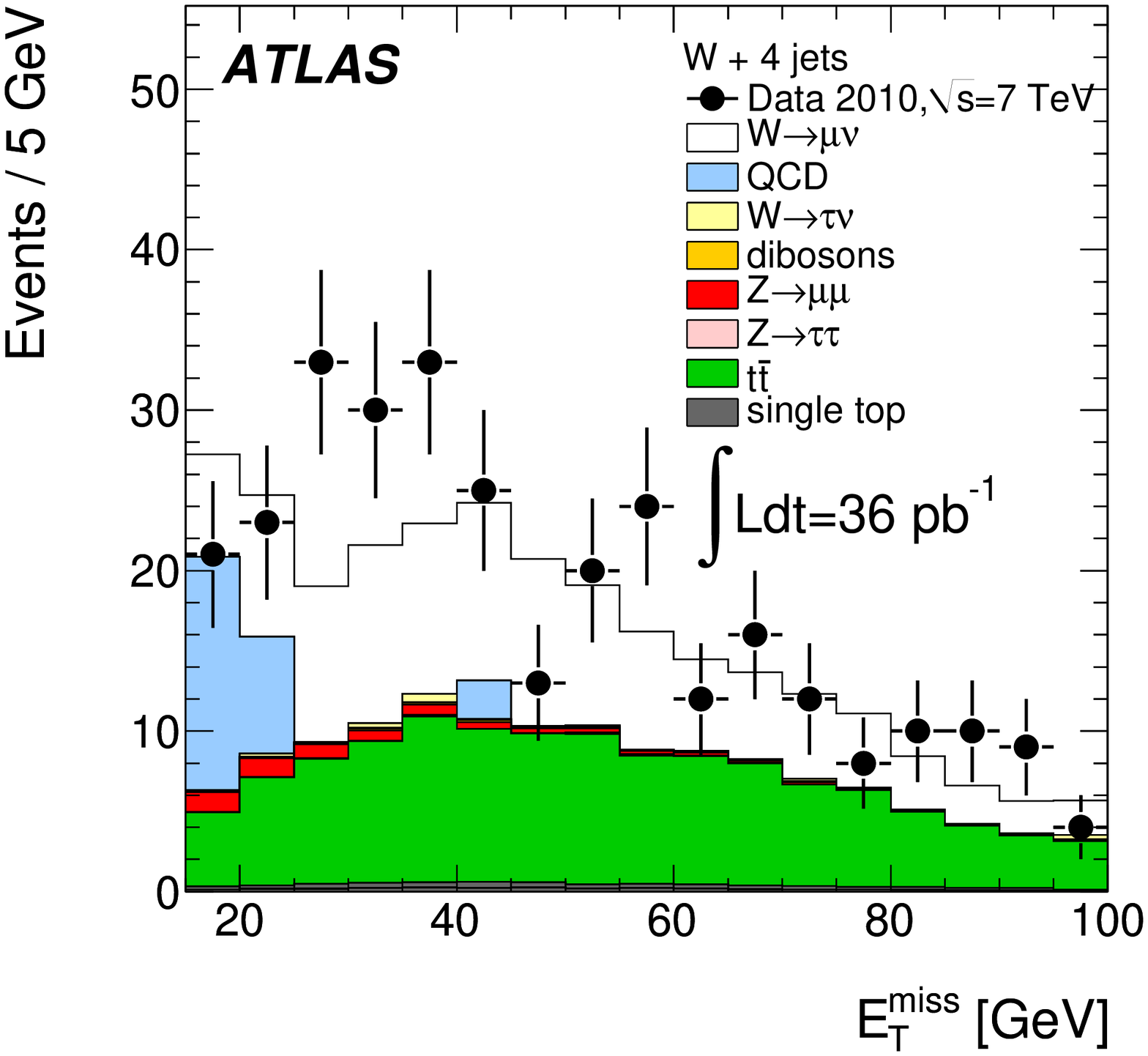}
\includegraphics[width=0.32\linewidth]{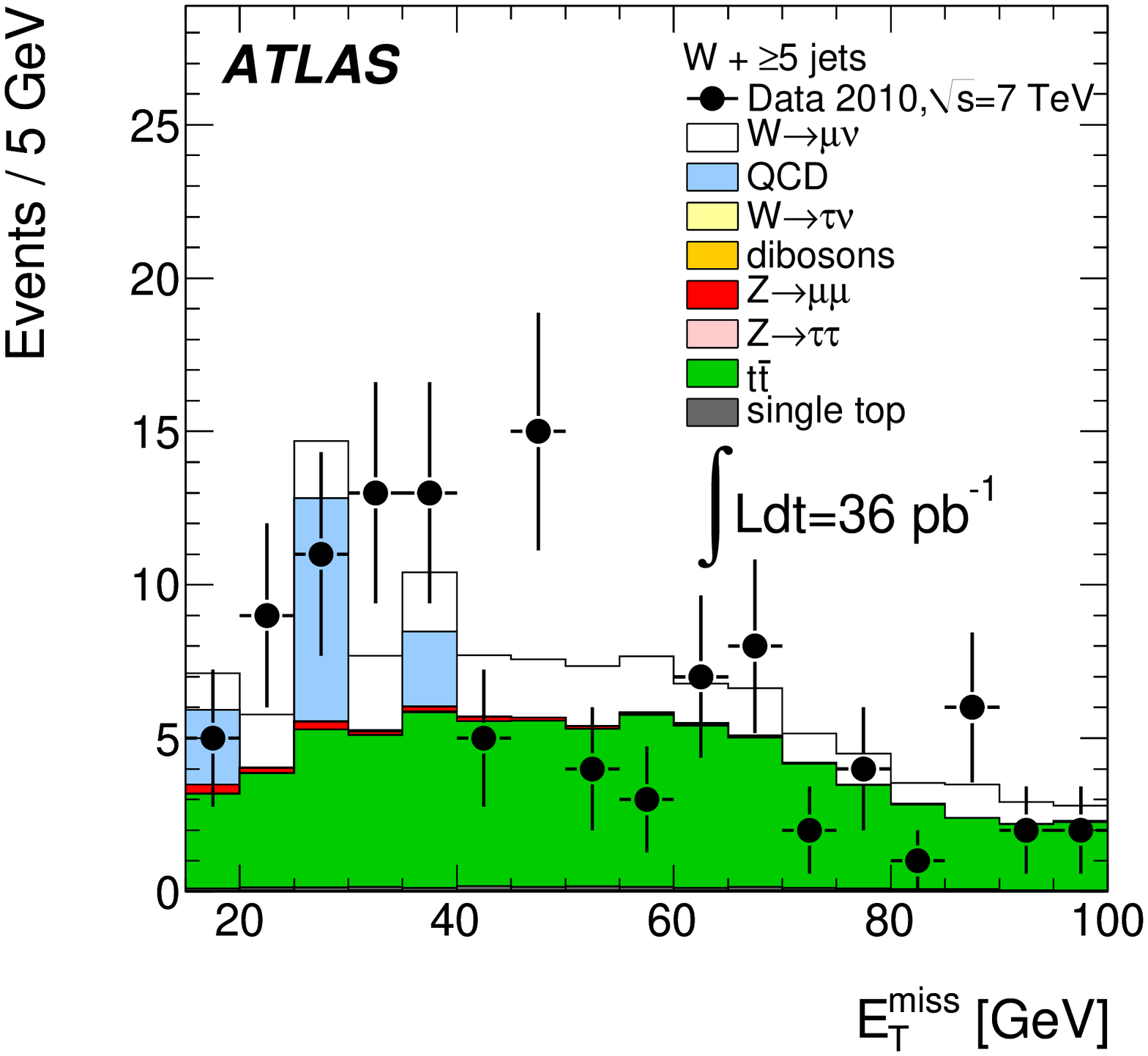} 
\caption{Result of the \met\ template fits used to obtain an estimate
of the multijet background for \Wmn\ events with relaxed kinematic
requirements, $m_{\rm T}(W)$~$>$~25~\GeV\ and
\met~$>$~15~\GeV. Results are shown in bins of exclusive jet
multiplicity. In this case the multijet template was obtained with a
reversed requirement on the significance of muon's impact
parameter. The data with $\ge$ 5 jets are not used for measurements
because of the low event count and a poor signal-to-background ratio.}
\label{fig:muo_met_fit}     
\end{figure*}

\begin{figure*}[!htb]
\centering
\includegraphics[width=0.32\linewidth]{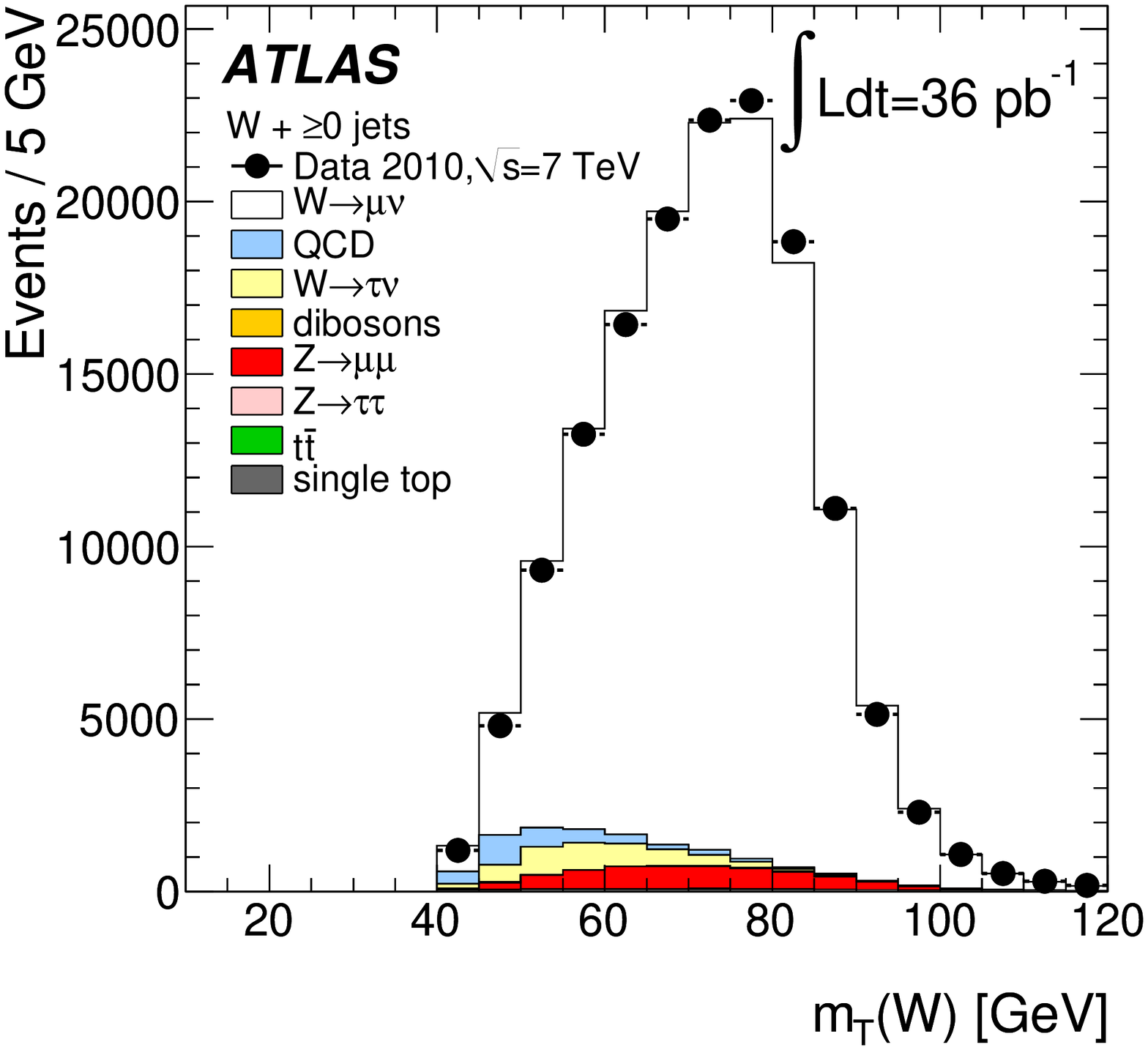}
\includegraphics[width=0.32\linewidth]{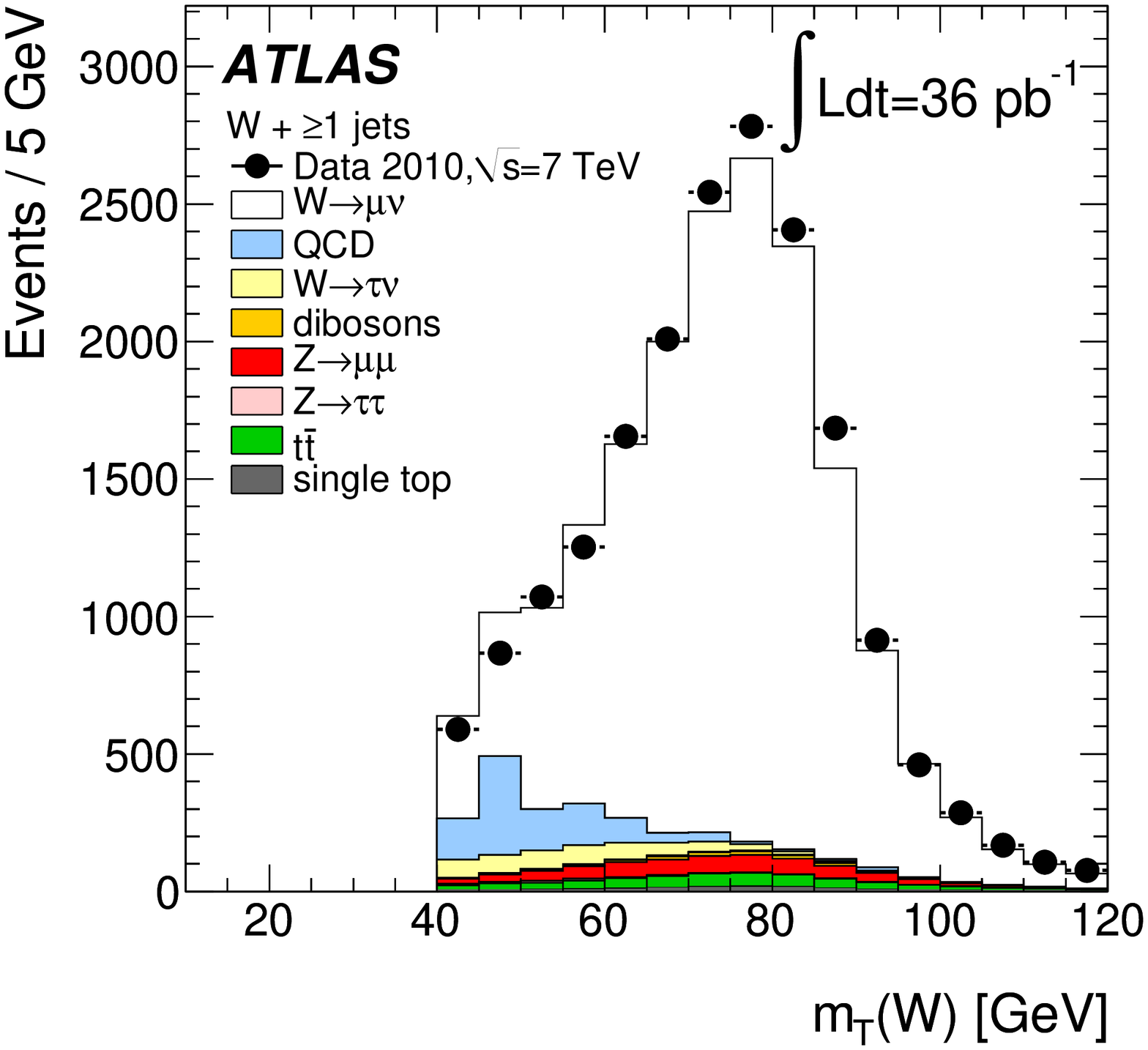}
\includegraphics[width=0.32\linewidth]{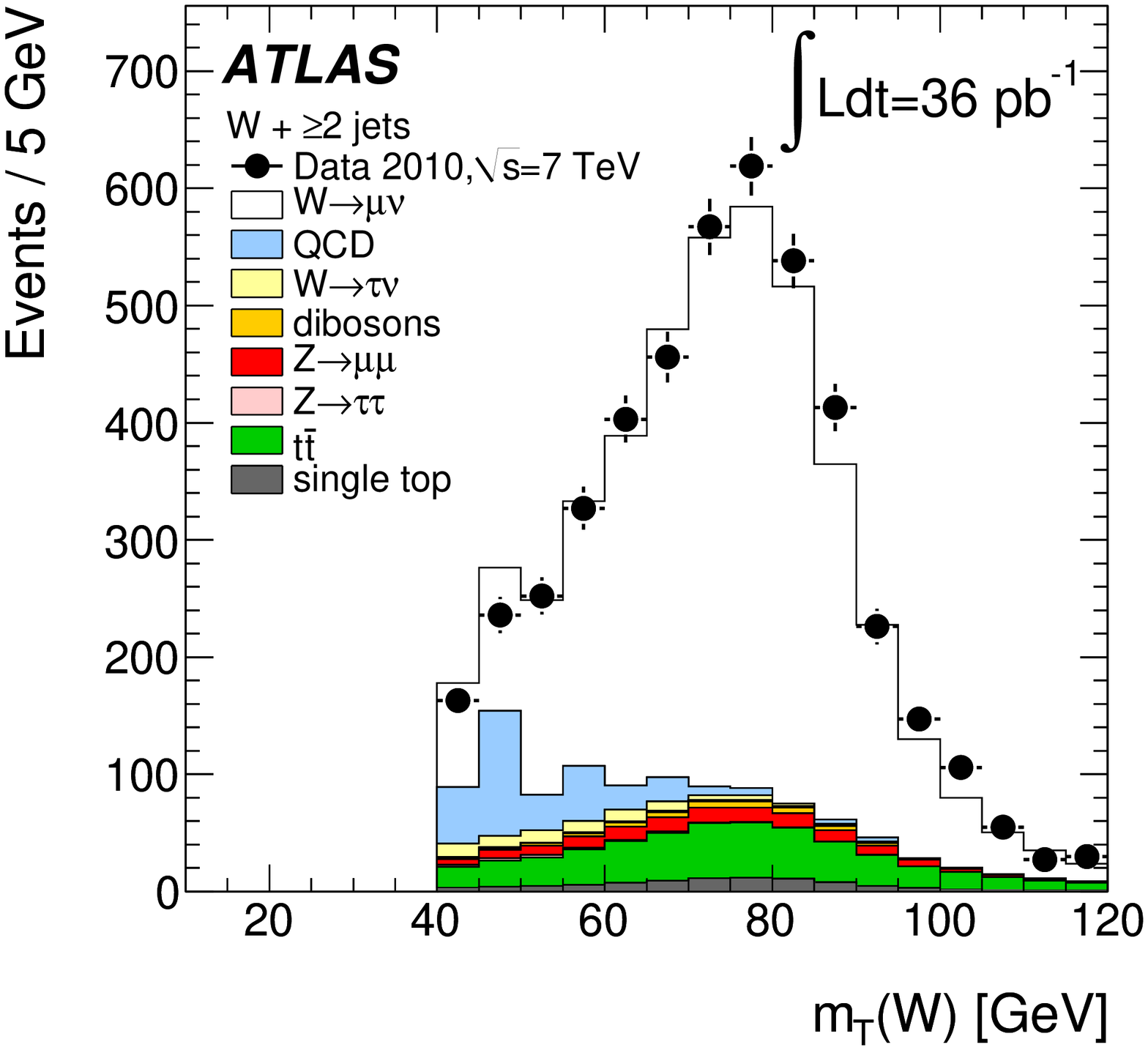}
\includegraphics[width=0.32\linewidth]{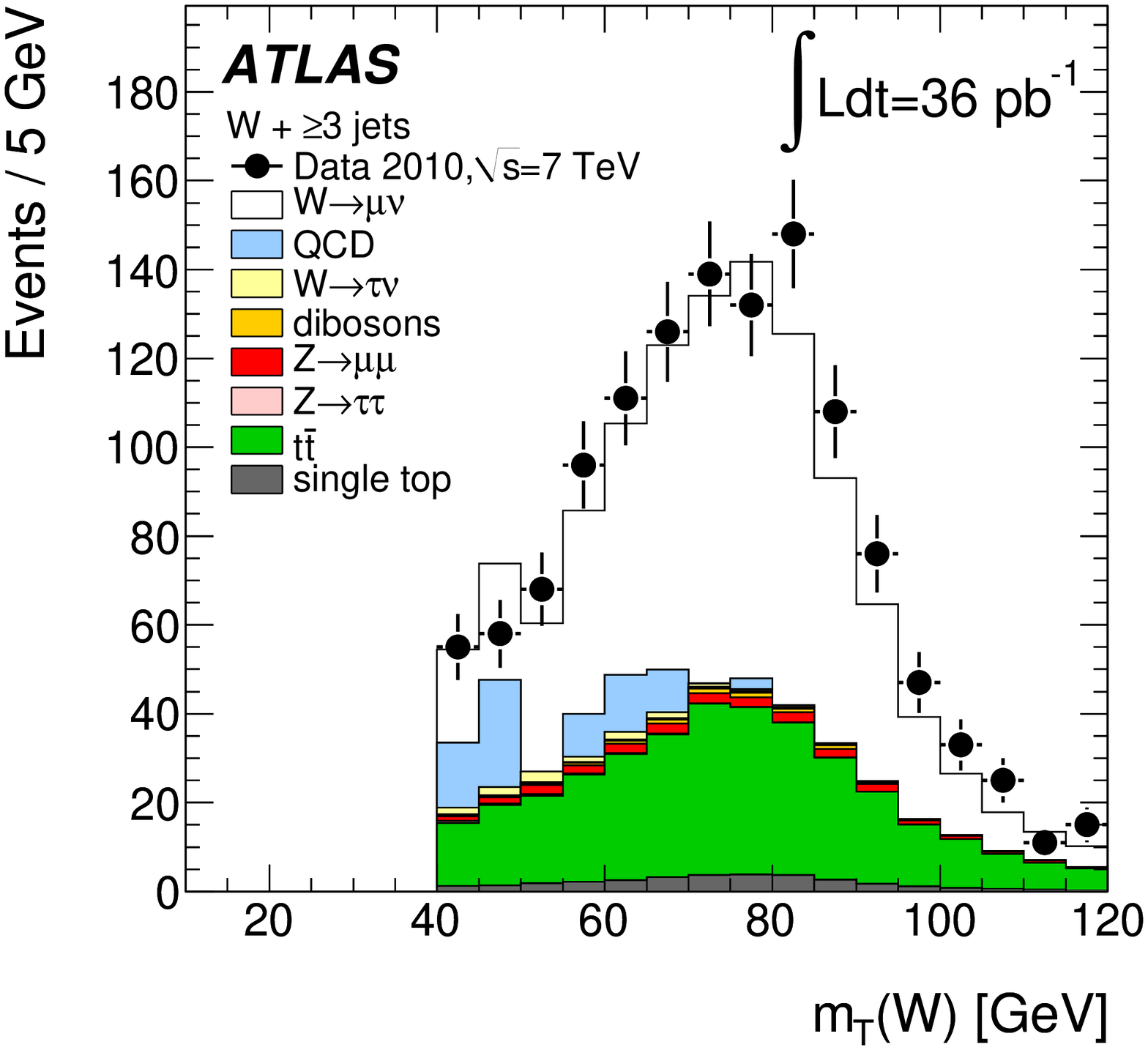}
\includegraphics[width=0.32\linewidth]{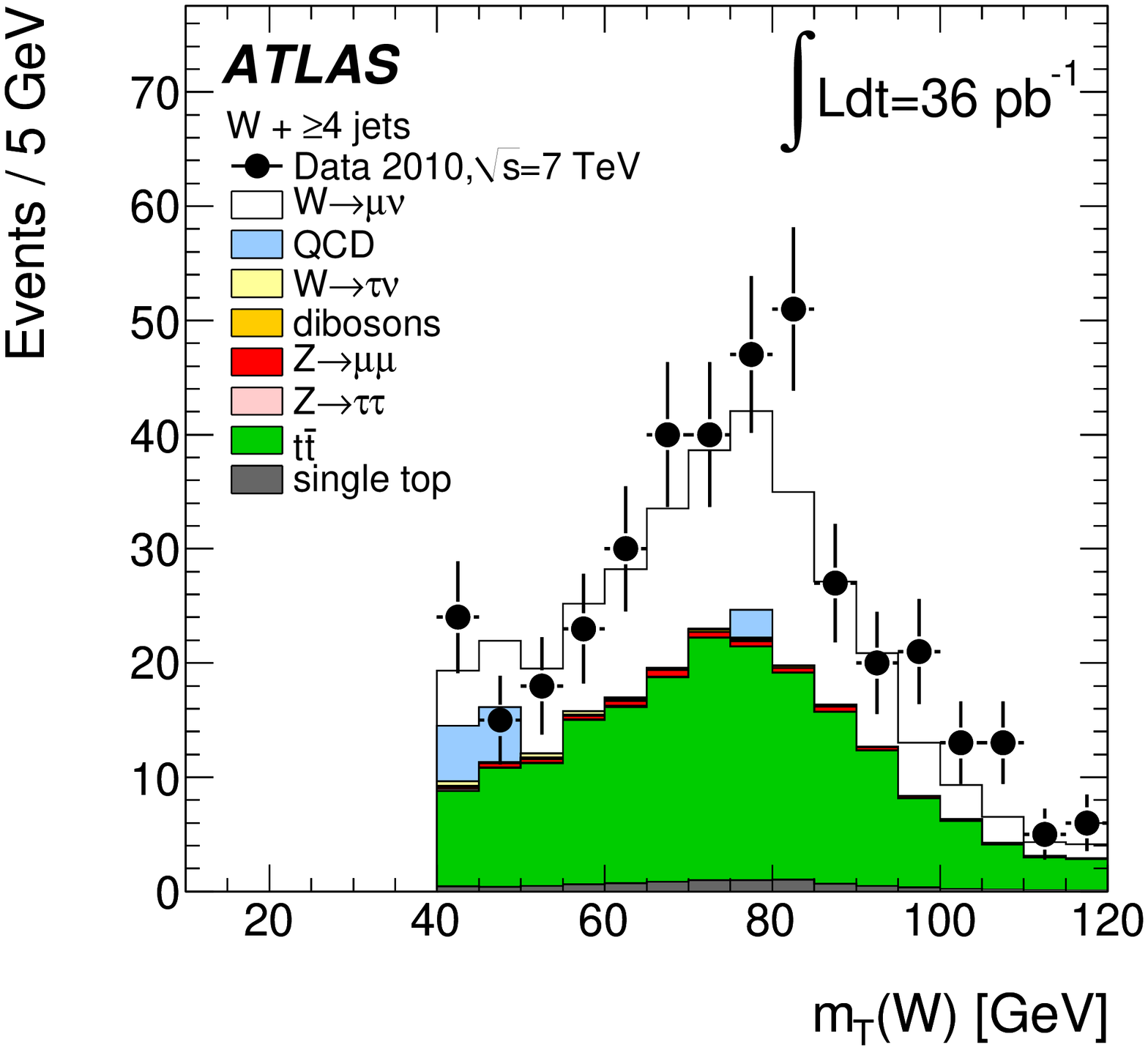}
\includegraphics[width=0.32\linewidth]{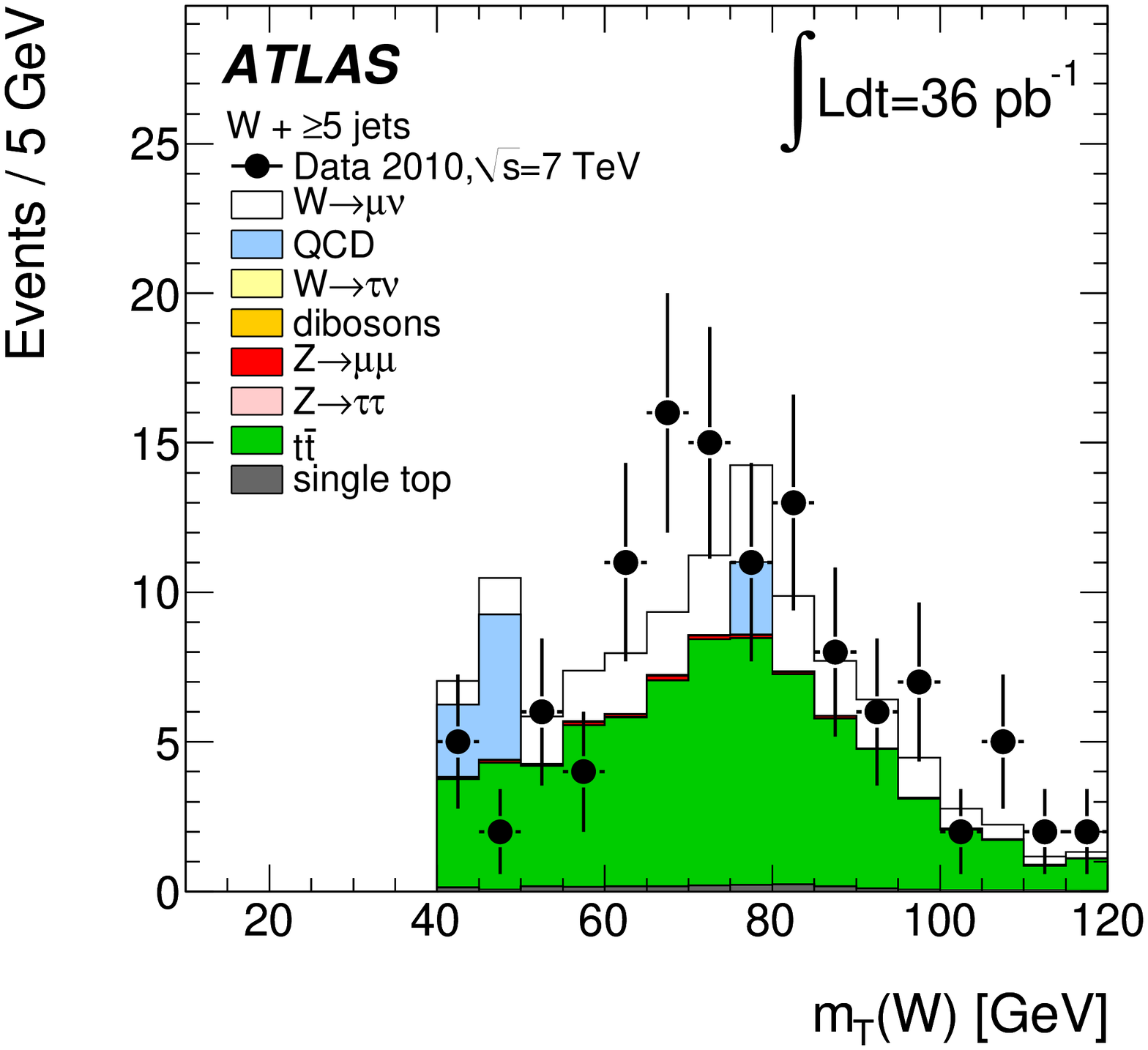}
\caption{Comparison of transverse mass distributions $m_{\rm T}(W)$
for \Wmn\ events. Results are shown in bins of inclusive jet
multiplicity for events passing the normal selection requirements. MC
predictions for the \Wmn\ signal and leptonic backgrounds are
normalized to luminosity using (N)NLO cross sections and the multijet
background is estimated from data.}
\label{fig:muo_mt_full}
\end{figure*}

\subsubsection{Electron channel systematic uncertainties}
\label{sec:ele_syst}

\par The systematic uncertainties for the electron channel are
summarized in Table~\ref{tbl:elec-bkg-sys}. The calculation of
uncertainty on the number of multijet background events was introduced
in Section~\ref{sec:ele_backgr}.

\par The electron trigger efficiency was measured using \Zee\ events
triggered by an object other than the electron under study
(tag-and-probe method). A scale factor of $99.5\pm0.5\%$ relative to
the value predicted by the Monte Carlo simulation was determined. The
same event samples were used to determine the electron reconstruction
and identification efficiencies relative to the Monte Carlo
prediction. The reconstruction efficiencies were consistent with the
Monte Carlo values within a systematic uncertainty of $1.5\%$.
Data-driven corrections to the simulated identification efficiencies
were characterized by a two-dimensional matrix in $\eta$ and \et. The
\Zee\ events were also used to test the electron identification
efficiency for any dependence on accompanying jet activity and none
was found.

\par The measured electron energy scale and resolution were also
studied with \Zee\ events. In the data, electron energies were
adjusted with an $\eta$-dependent correction with typical values of
about 2\%~\cite{ElePerf}. The electron energy resolution was similarly
tested and adjusted in simulated events. The residual systematic
uncertainties are shown in Table~\ref{tbl:elec-bkg-sys}.

\begin{table}[htb]
\centering
\caption{Summary of relative systematic uncertainties associated with
the electron channel.}
\begin{tabular}{|l|c|} 
\hline\hline
  Quantity & Uncertainty \\ \hline
  Trigger efficiency 		& 	$\sim0.5\%$			\\
  Electron reconstruction 	& 	$\sim1.5\%$	\\
  Electron identification 	& 	$2-8\%$\footnote{$\eta-\pt$\ dependent} \\
  Electron energy scale 	& 	$0.3-1.6\%$\footnotemark[1] \\
  Electron energy resolution 	&  	$<0.6\%$  of the energy  \\
  Multijet QCD background       &	$17-100$\%\footnote{increased with jet multiplicity}; difference between the\\ 
                                &	two methods, see Section~\ref{sec:ele_backgr}\\ 
\hline\hline
\end{tabular}
\label{tbl:elec-bkg-sys} 
\end{table}

\subsection{\Wmn\ + jets final state}
\label{sec:muo_offline}

\par The muons were required to be reconstructed in both the ID and MS
subsystems and to have \pT~$>$~20~\GeV\ and $|\eta|<2.4$. The ID track
requirements were those of Ref.~\cite{incWv16}. An ID-based muon
isolation was applied which required a relative isolation of
$\Sigma\pT^{\rm{ID}}/\pT^{\mu}<0.1$, using a cone size of $\Delta
R<0.2$, where $\Sigma\pT^{\rm{ID}}$ included all ID tracks in the cone
except the muon track. To help ensure that the muon is prompt it was
required that the transverse impact parameter of the track $d_0$ and
its uncertainty $\sigma(d_0)$ satisfied $|d_0/\sigma(d_0)|<3$. Also
the longitudinal impact parameter $\Delta z$ was required to satisfy
$|\Delta z|<10$~mm to reduce contributions from in-time pile-up and
cosmic ray muons. These impact parameters were measured with respect
to the primary vertex.  Events were rejected if there was a second
muon passing the same kinematic selections and isolation requirements
as above. These muon selection criteria are similar to those applied
in Ref.~\cite{WJets}.

\subsubsection{Muon channel background estimates}
\label{sec:muo_backgr}

\par For the muon channel, the main backgrounds arise from
semileptonic decays of heavy flavor hadrons in multijet events, other
leptonic decays of heavy gauge bosons, and \ttbar\ production. The
backgrounds from gauge bosons include \Wtau\ where the tau decays to a
muon, \Zmm\ where one muon is not identified, \Ztau, and diboson
production. For low jet multiplicities the largest backgrounds are
\Wtau\ and \Zmm, while for higher multiplicities \ttbar\ production
dominates ($\ttbar\rightarrow b\overline{b} q q'\mu\nu$). Similarly to
the electron channel, the number of leptonic background events
surviving the selection criteria was estimated with simulated event
samples described in Section~\ref{sec:mc}. {\sc Pythia} was used only
for inclusive production of \Wtau\ and \Ztau\ and {\sc Alpgen} for the
other vector boson samples. The simulated leptonic background samples
were normalized to the integrated luminosity of the data using the
predicted NNLO, NLO+NNLL or NLO cross sections. Discussion of the
\ttbar\ background follows in Section~\ref{sec:det_lvl_compar}.

\par The multijet QCD background in the muon channel is dominated by
leptonic decays of bottom or charm hadrons in jets where the hadron
decay involves a muon and neutrino. The number of background events
was estimated by fitting, for each exclusive jet multiplicity, the
\met\ distribution in the data (with relaxed selection requirements on
\met\ and $m_{\rm T}(W)$: \met~$>$~15~\GeV\ and $m_{\rm
T}(W)$~$>$~35~\GeV) to a sum of two templates: one for the multijet
background and another which included signal and the leptonic
backgrounds. The fit determined the relative normalization of the two
templates. The shapes for the second template were obtained from
simulation and their relative normalization was fixed to the predicted
cross sections. The full kinematic selection, \met~$>$~25~\GeV\ and
$m_{\rm T}(W)$~$>$~40~\GeV, was imposed on the multijet background
samples to convert their normalization coefficients from the relaxed
to full selection.

\par The template for the multijet background was obtained from data
by applying all the standard muon selection requirements, except that
the requirement on the significance of the transverse impact parameter
was reversed to $|d_0/\sigma(d_0)|>3$. In addition, the impact
parameter was required to be within $0.1<|d_0|<0.4$~mm. The lower cut
on the impact parameter reduces signal \Wmn\ events leaking into the
background sample. The upper cut on $|d_0|$ was placed to minimize
bias from multijet events where an isolated muon is accompanied by a
nearby energetic jet; the isolated muons from decays of heavy hadrons
tend to have large impact parameter. The background events with a muon
and an energetic jet do not survive the standard muon selection due to
the stringent requirement on the impact parameter, in conjunction with
the isolation cut.

\par The comparisons of the template fits to the \met\ distributions
are presented in Fig.~\ref{fig:muo_met_fit} for \Wmn\ events with the
relaxed selection requirements on \met\ and $m_{\rm T}(W)$.
Figure~\ref{fig:muo_mt_full} shows the final $m_{\rm T}(W)$
distributions in the various bins of inclusive jet multiplicity for
events passing the normal selection requirements.

\par Another set of templates for the multijet background was obtained
using a simulated dijet sample from {\sc Pythia} where the event
record was required to contain at least one muon with \pt~$>$~8~\GeV.
The second set of templates was fitted to data in the same manner as
the first in order to estimate a systematic uncertainty in the number
of multijet background events. The uncertainty increased with the jet
multiplicity from 15\% for the inclusive $W$-boson sample up to 76\%
for events with a $W$ boson and four or more jets.

\subsubsection{Muon channel systematic uncertainties}
\label{sec:muo_syst}

\begin{table}[htb]
\centering
\caption{Summary of relative systematic uncertainties associated with
the muon channel.}
\begin{tabular}{|l|c|} 
\hline\hline Quantity & Uncertainty \\ \hline
  Trigger efficiency 		& 0.6--0.7\%\footnote{$\eta-\pt$~dependent}			\\
  Muon reconstruction 	        & $\sim$1.1\%\footnote{$\eta-\phi$~dependent}	\\
  and identification 	        & \\
  Muon \pt\ scale 	        & $\sim$0.4\%\footnotemark[1] \\ 					
  Muon \pt\ resolution 	        & $<6$\%\footnote{$\eta-\pt$~dependent relative to the measured resolution}\\
  Multijet QCD background	& $15-76$\%\footnote{varies with jet multiplicity}; difference between the \\
                                & two templates, see Section~\ref{sec:muo_backgr}\\ 
\hline\hline
\end{tabular}
\label{tbl:muon-bkg-sys} 
\end{table} 

\par The muon trigger efficiencies were measured using a \Zmm\ sample
triggered by a muon candidate other than the muon under
study~\cite{incWv16}. Scale factors close to unity, relative to the
value predicted by the Monte Carlo simulation, were obtained for the
muon triggers. The scale factors were calculated as a function of muon
\eta\ and \pt. The same sample of events was used to determine the
muon reconstruction and identification efficiencies as a
two-dimensional matrix in $\eta$ and
$\phi$~\cite{ATLAS-CONF-2011-008,ATLAS-CONF-2011-063}. The measured
efficiencies were used to correct the simulated samples. The average
efficiency correction is consistent with unity within a systematic
uncertainty of $1.1\%$.

\par The measured momentum scale and resolution for the muons were
studied with \Zmm\ events~\cite{ATLAS-CONF-2011-046}. The muon
transverse momentum and its resolution were calibrated as a function
of $\eta$ and \pt. The systematic uncertainties for the muon channel
are summarized in Table~\ref{tbl:muon-bkg-sys}.

\subsection{Detector-level comparisons between final states of \Wen\ + jets and \Wmn\ + jets}
\label{sec:det_lvl_compar}

\begin{figure}[htb]
\centering
    \includegraphics[width=0.868\linewidth]{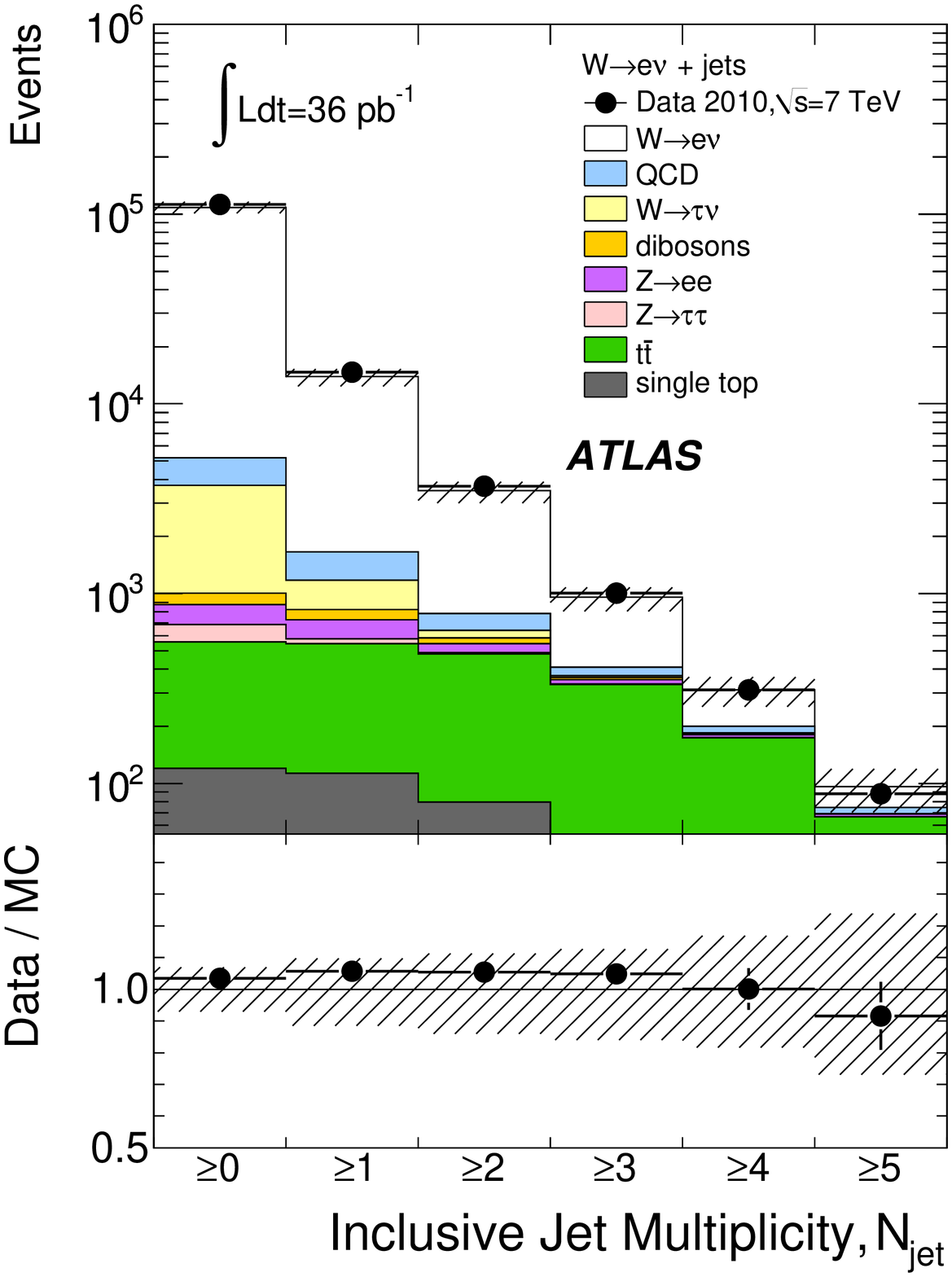}
    \includegraphics[width=0.868\linewidth]{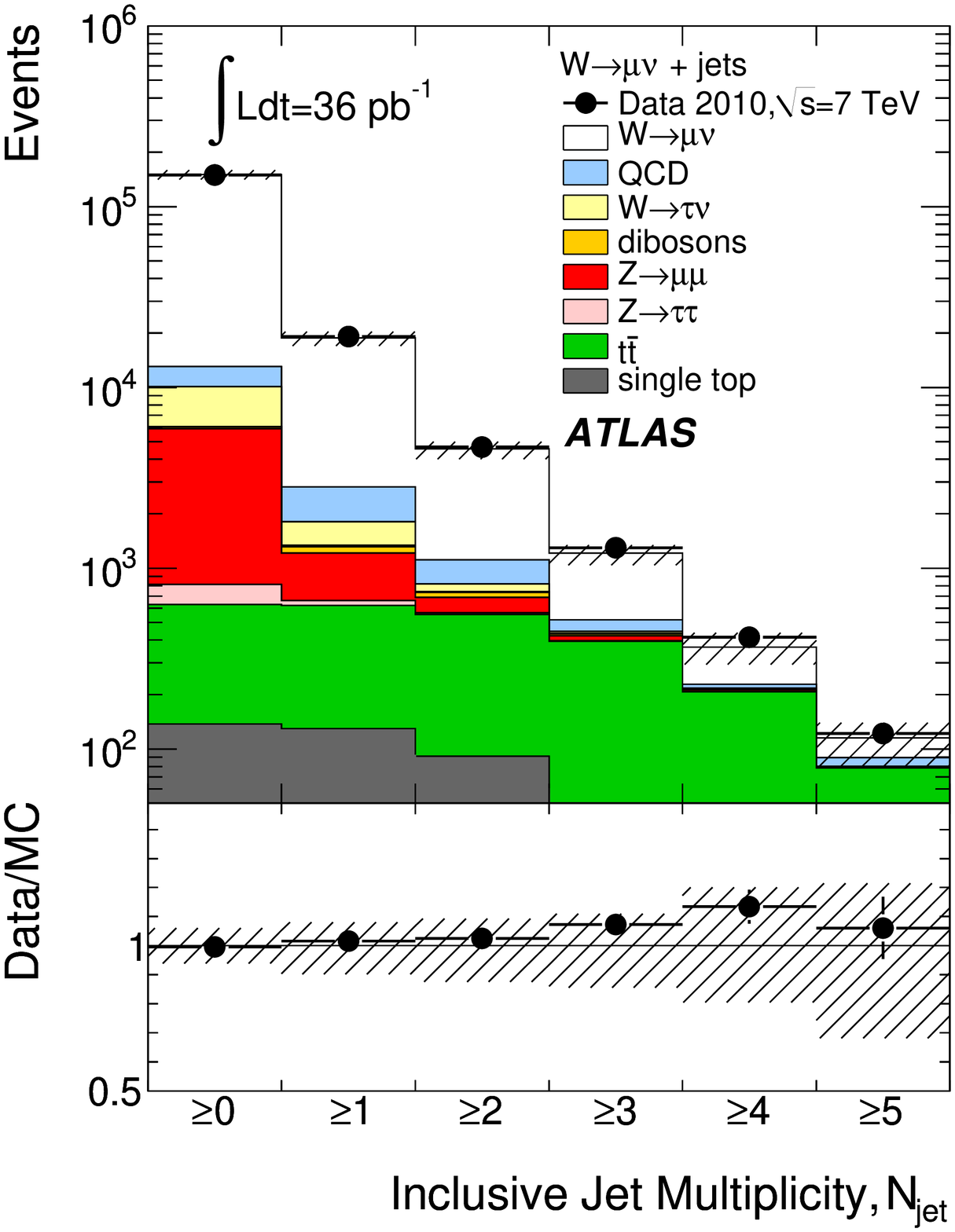}
  \caption{The uncorrected inclusive jet multiplicity
  distribution. The following remarks apply to this and subsequent
  figures. Top: electron channel. Bottom: muon channel. The signal and
  leptonic backgrounds are shown using simulations, whereas the
  multijet background uses the method described in the text. The
  signal and leptonic backgrounds are normalized to the predicted
  cross sections. The black-hashed regions illustrate the experimental
  uncertainties on the predicted distributions.}
  \label{fig:det-njet}
\end{figure}

\begin{figure}[htb]
\centering
  \includegraphics[width=0.868\linewidth]{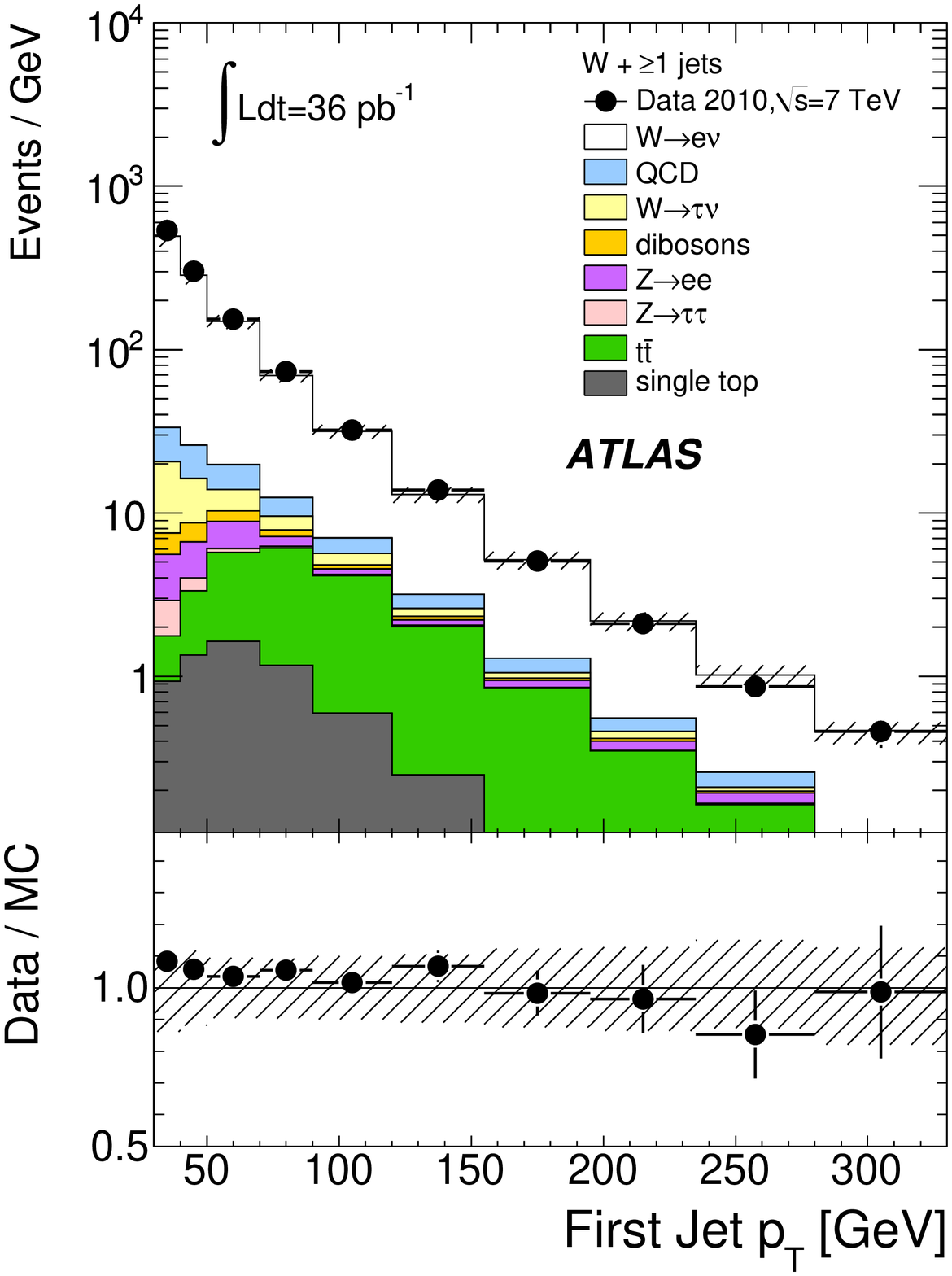}
  \includegraphics[width=0.868\linewidth]{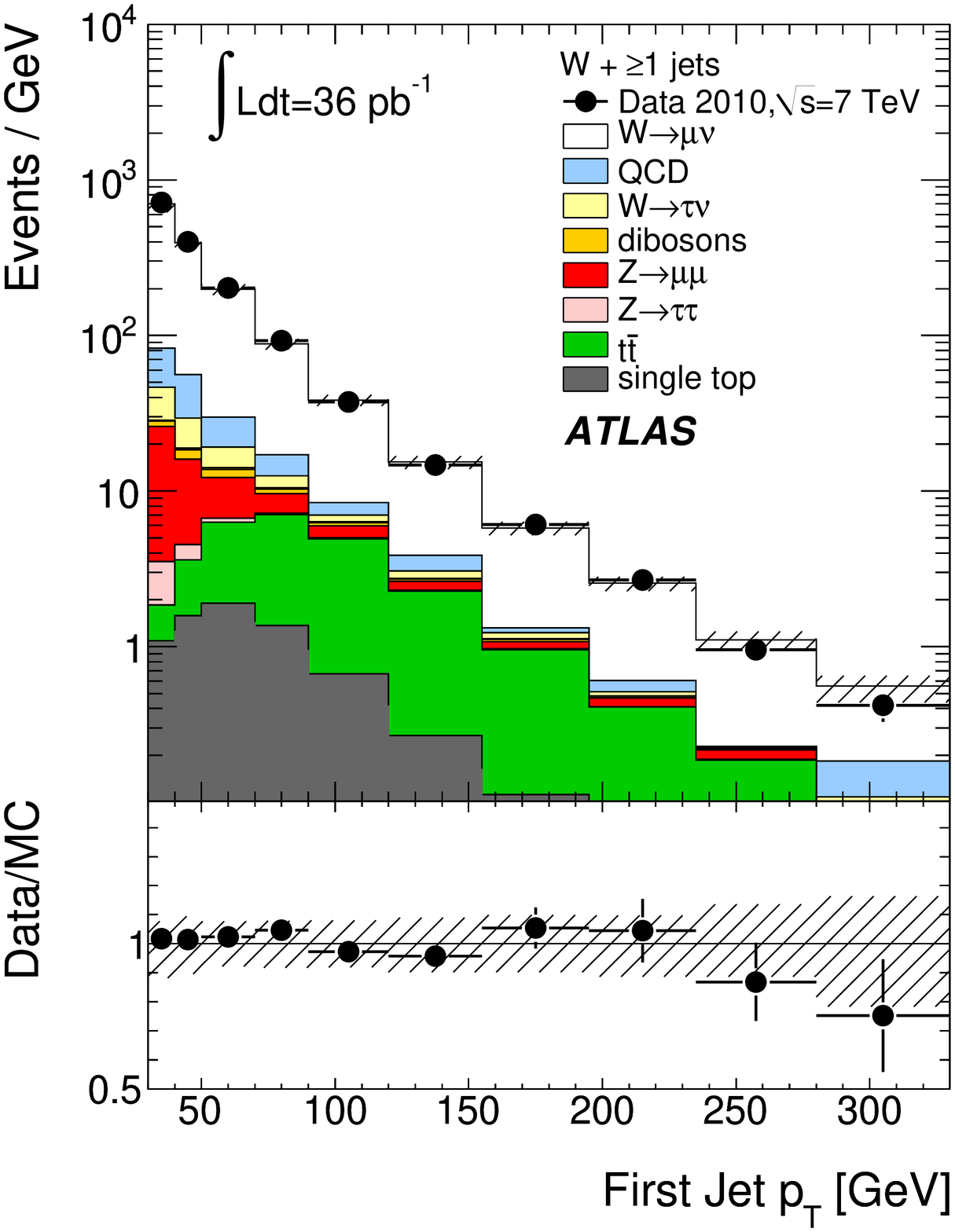}
  \caption{The uncorrected distribution in \pt\ of the jet with the
  highest \pt, in events with one or more jets.\\
\\
\\
\\
\\
\\
\\}
  \label{fig:det-pt1}
\end{figure}

\begin{figure}[htb]
\centering
  \includegraphics[width=0.868\linewidth]{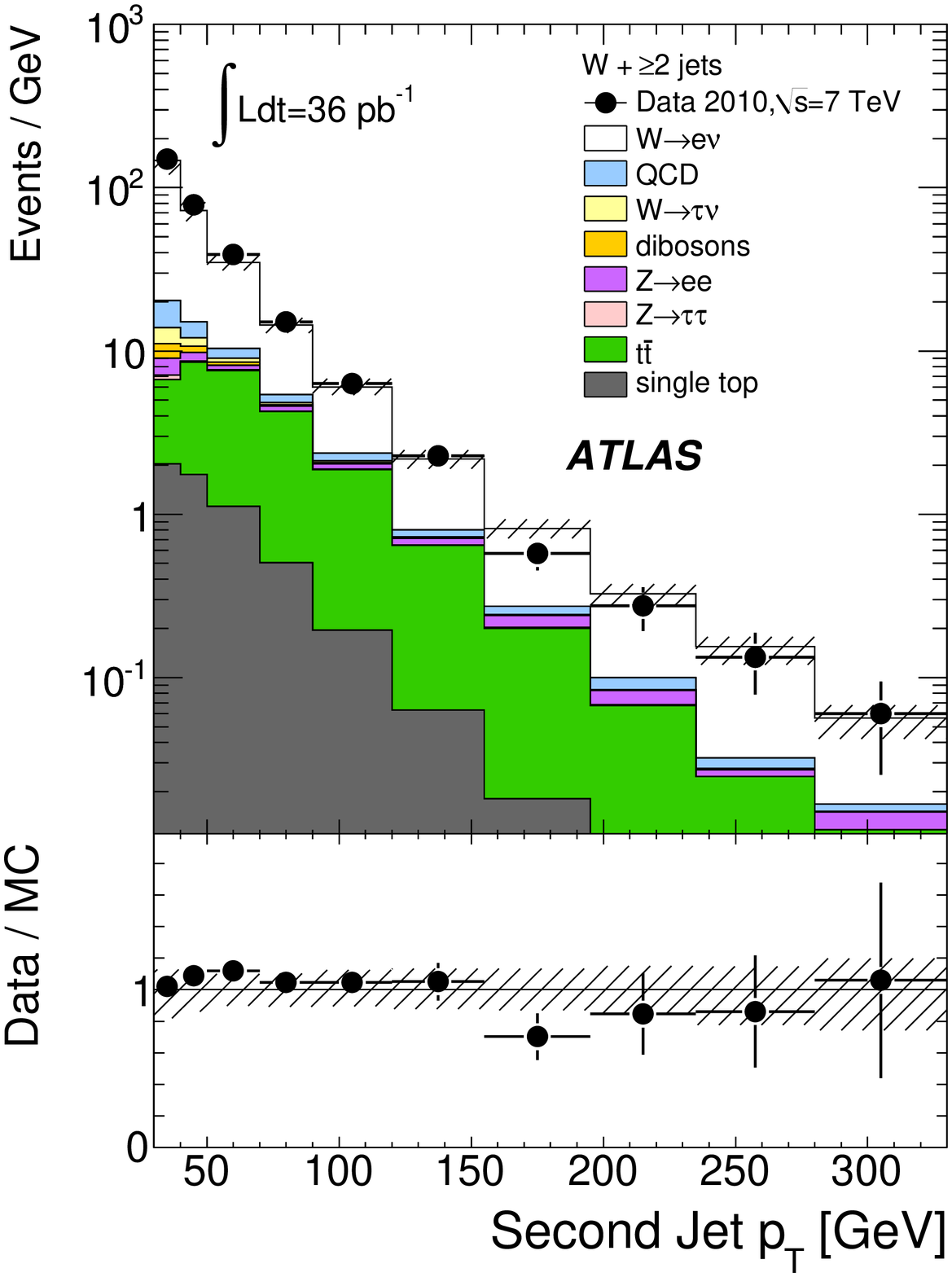} 
  \includegraphics[width=0.868\linewidth]{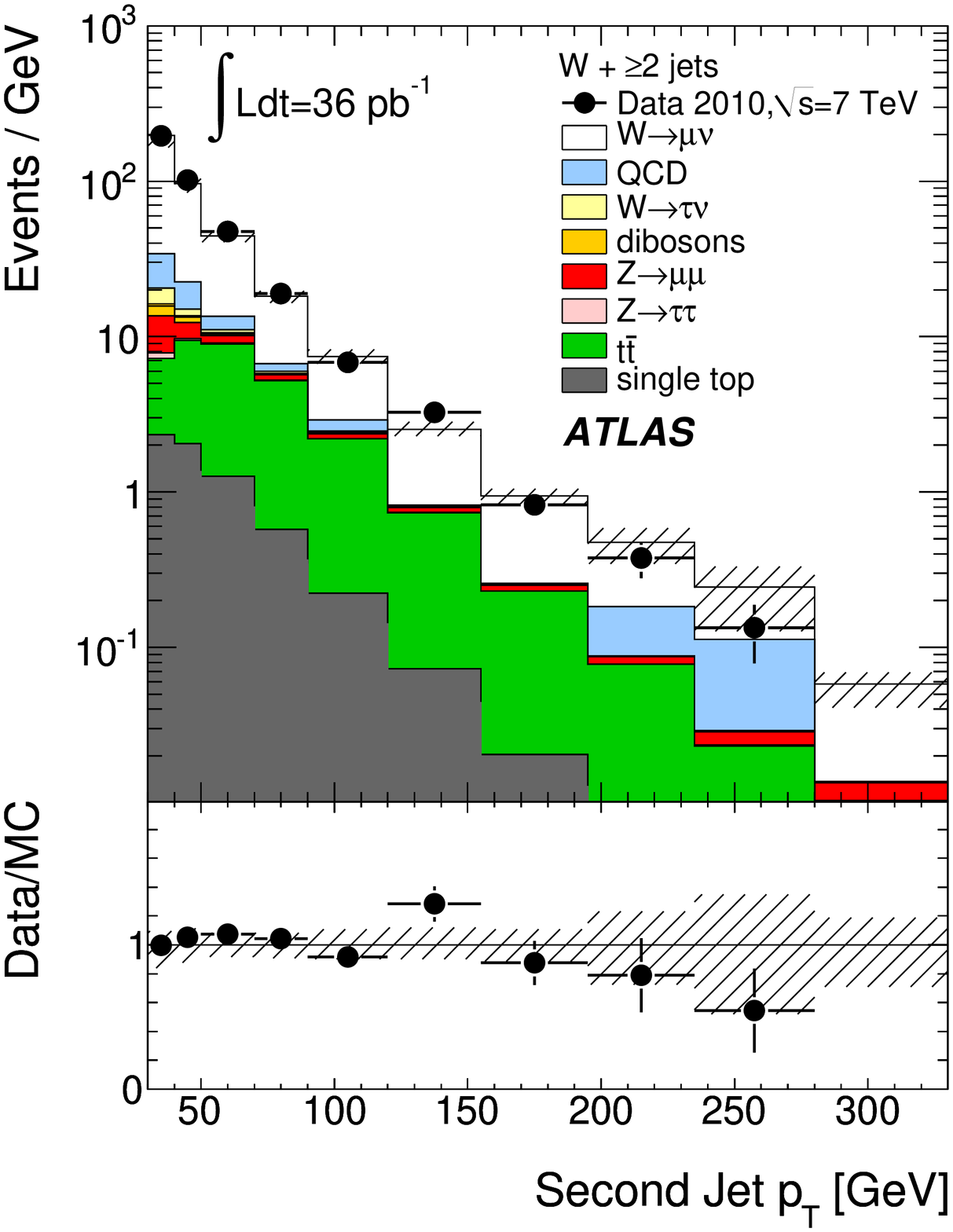}
  \caption{The uncorrected distribution in \pt\ of the jet with the
  second highest \pt, in events with two or more jets.\\
\\
\\
\\
\\
\\
\\}
  \label{fig:det-pt2}
\end{figure}

\begin{figure}[htb]
\centering
  \includegraphics[width=0.868\linewidth]{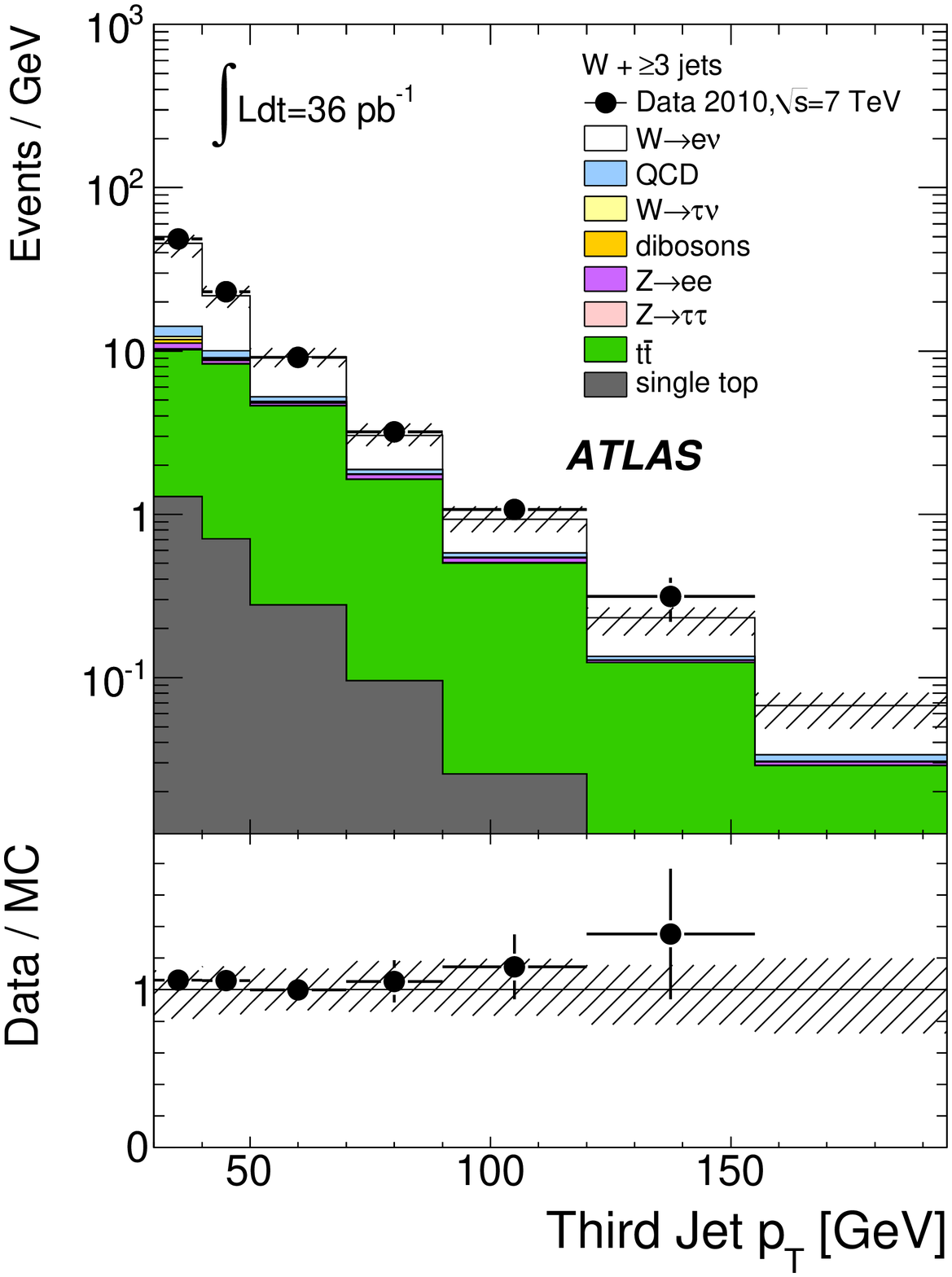}
  \includegraphics[width=0.868\linewidth]{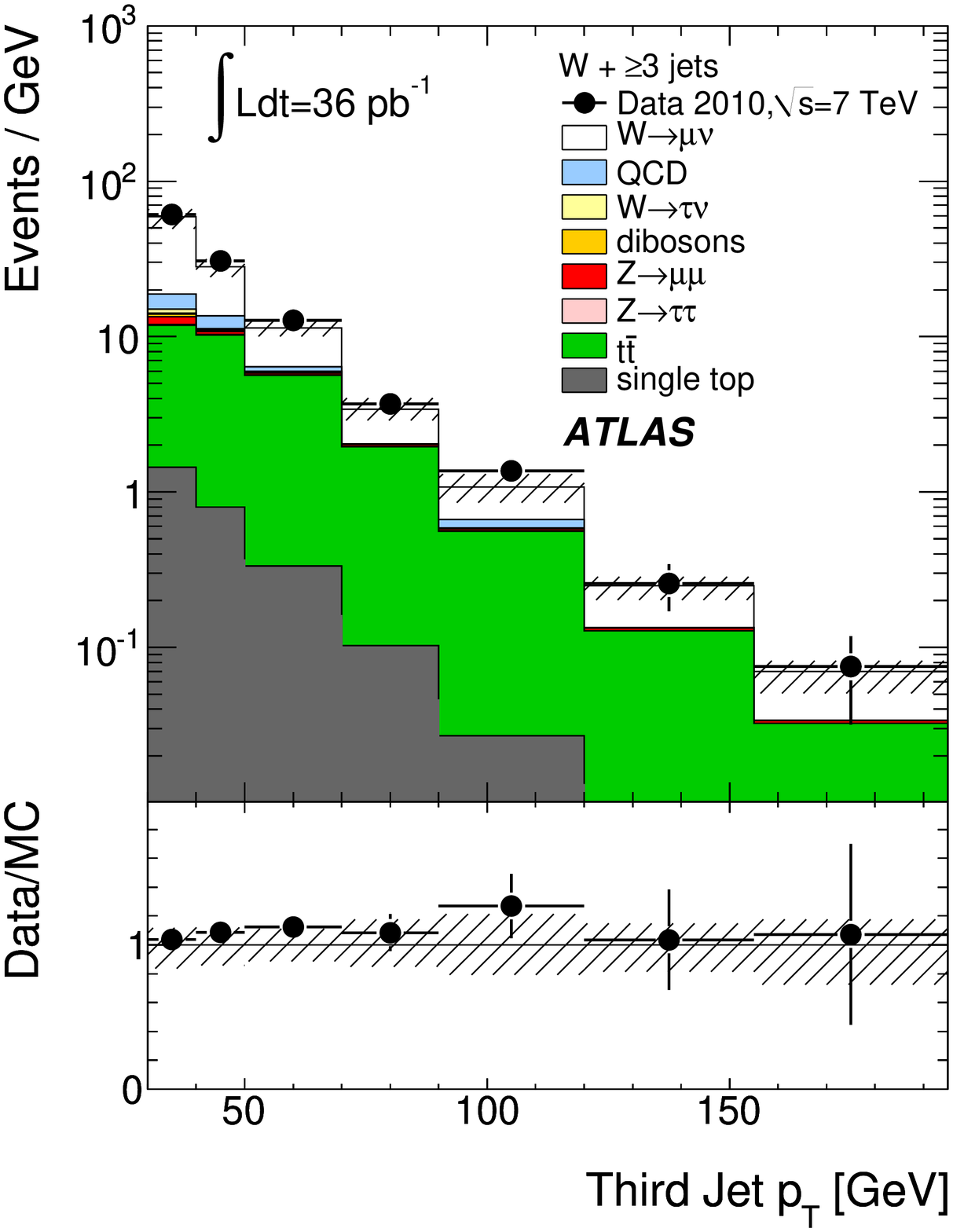}
  \caption{The uncorrected distribution in \pt\ of the jet with the
  third highest \pt\, in events with three or more jets.\\
\\
\\
\\
\\
\\
\\}
  \label{fig:det-pt3}
\end{figure}

\begin{figure}[htb]
\centering
  \includegraphics[width=0.868\linewidth]{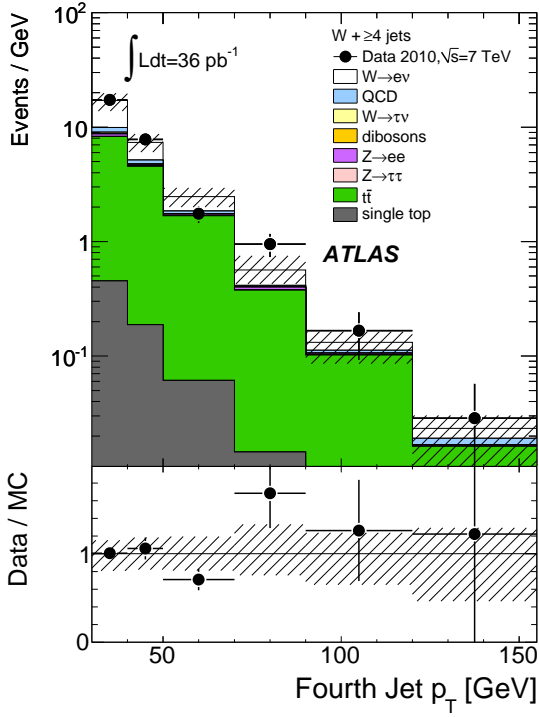} 
  \includegraphics[width=0.868\linewidth]{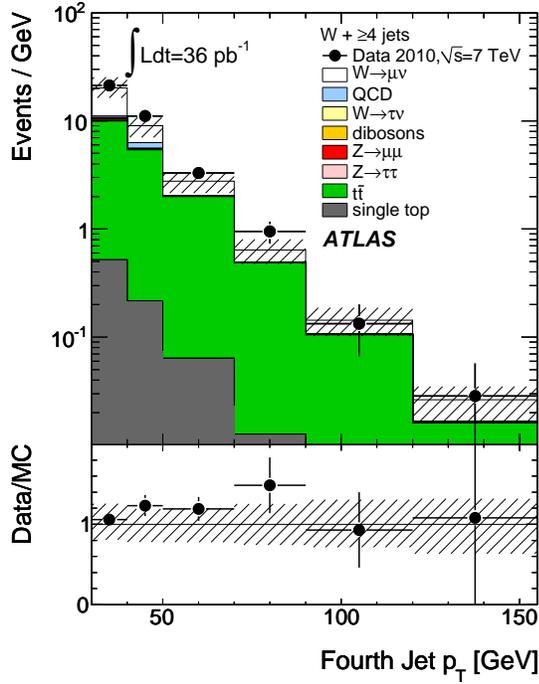}
  \caption{The uncorrected distribution in \pt\ of the jet with the
  fourth highest \pt, in events with four or more jets.\\
\\
\\
\\
\\
\\
\\}
  \label{fig:det-pt4}
\end{figure}

\begin{figure}[htb]
\centering
  \includegraphics[width=0.868\linewidth]{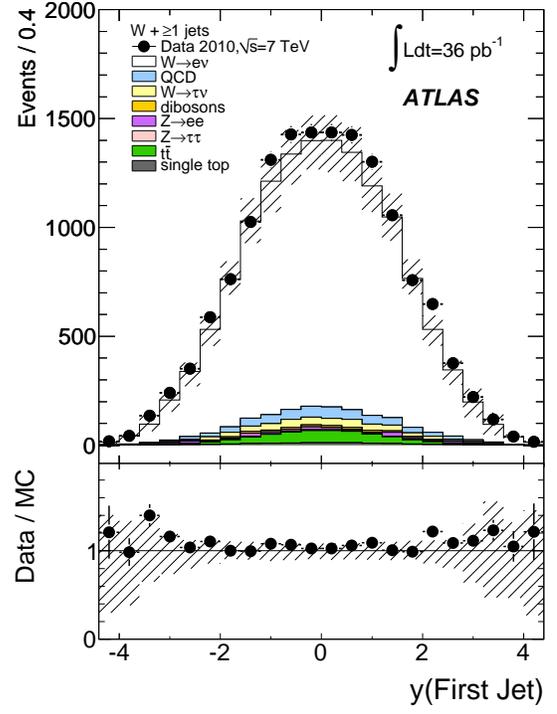}
  \includegraphics[width=0.868\linewidth]{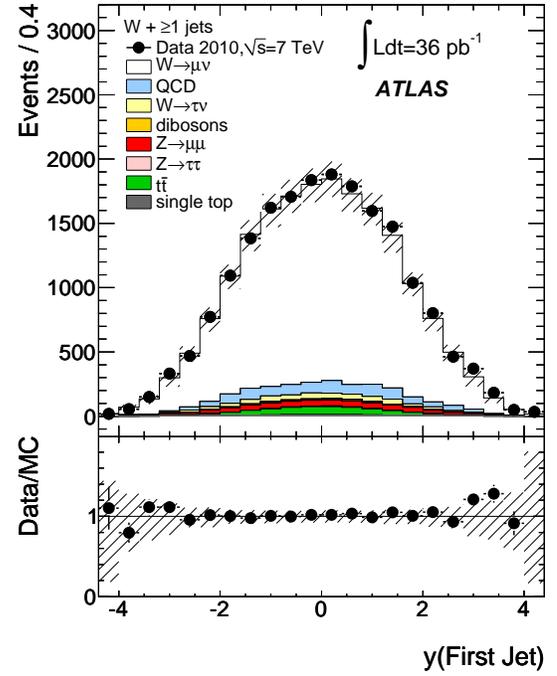}
  \caption{The uncorrected distribution in rapidity of the leading
  jet, $y({\rm first\:jet})$, in events with one or more jets.\\
\\
\\
\\
\\
\\
\\}
  \label{fig:det-y1}
\end{figure}

\begin{figure}[htb]
\centering
  \includegraphics[width=0.868\linewidth]{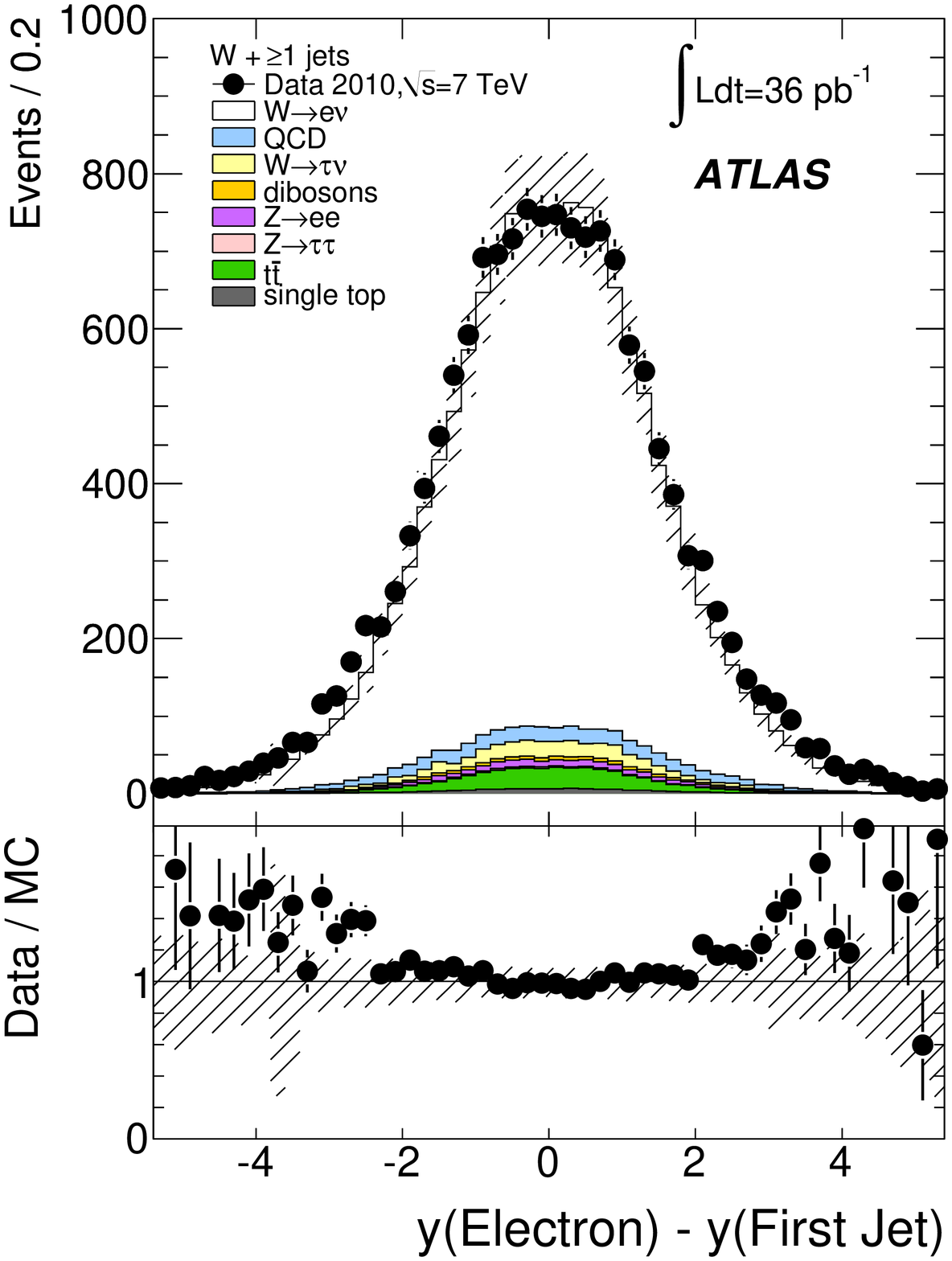}
  \includegraphics[width=0.868\linewidth]{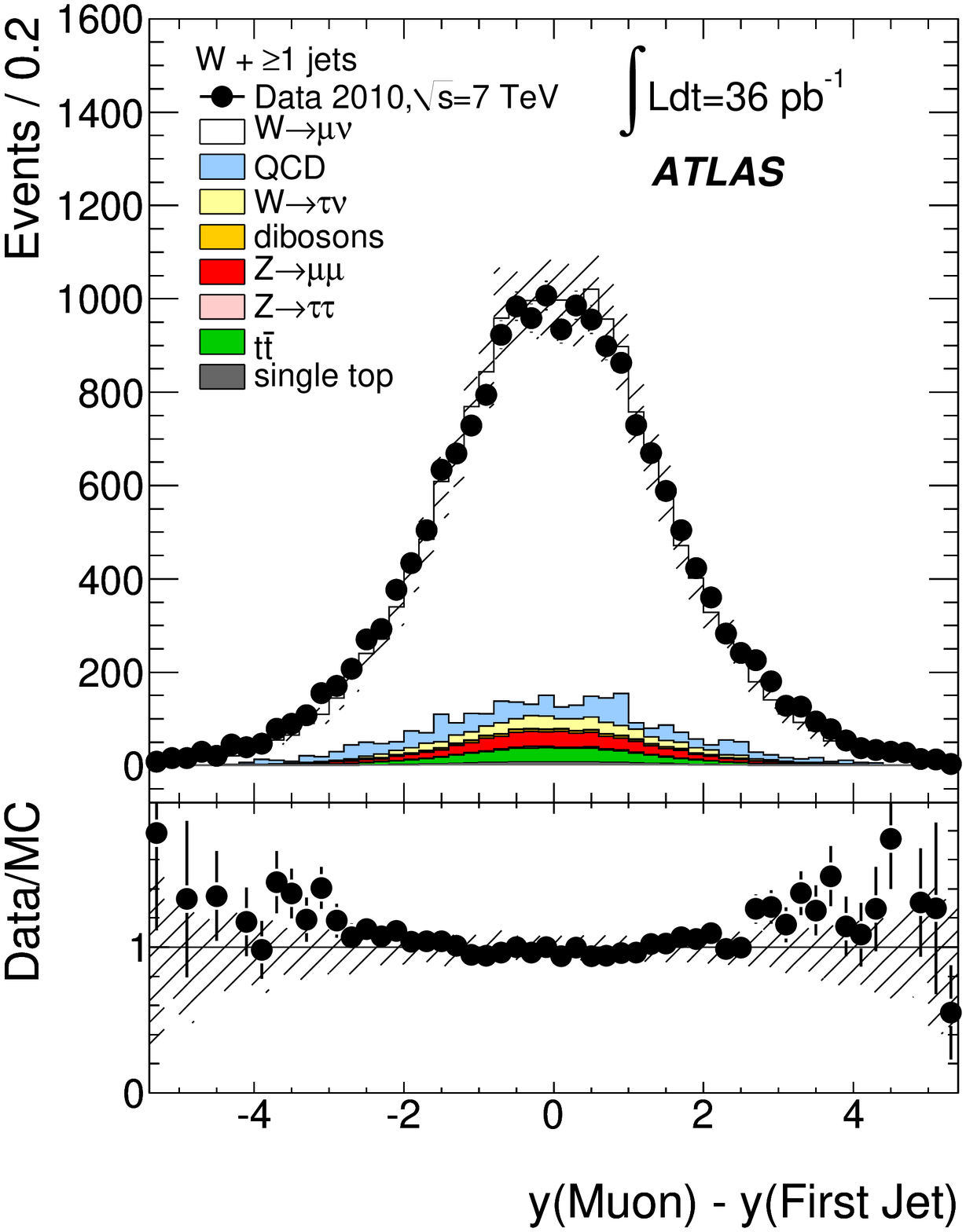}
  \caption{The uncorrected distribution in $y(\ell)\:-\:y({\rm
  first\:jet})$, rapidity difference between the lepton and the
  leading jet, for events with one or more jets.\\
\\
\\
\\
\\
\\}
  \label{fig:det-dylj}
\end{figure}

\begin{figure}[htb]
\centering
  \includegraphics[width=0.868\linewidth]{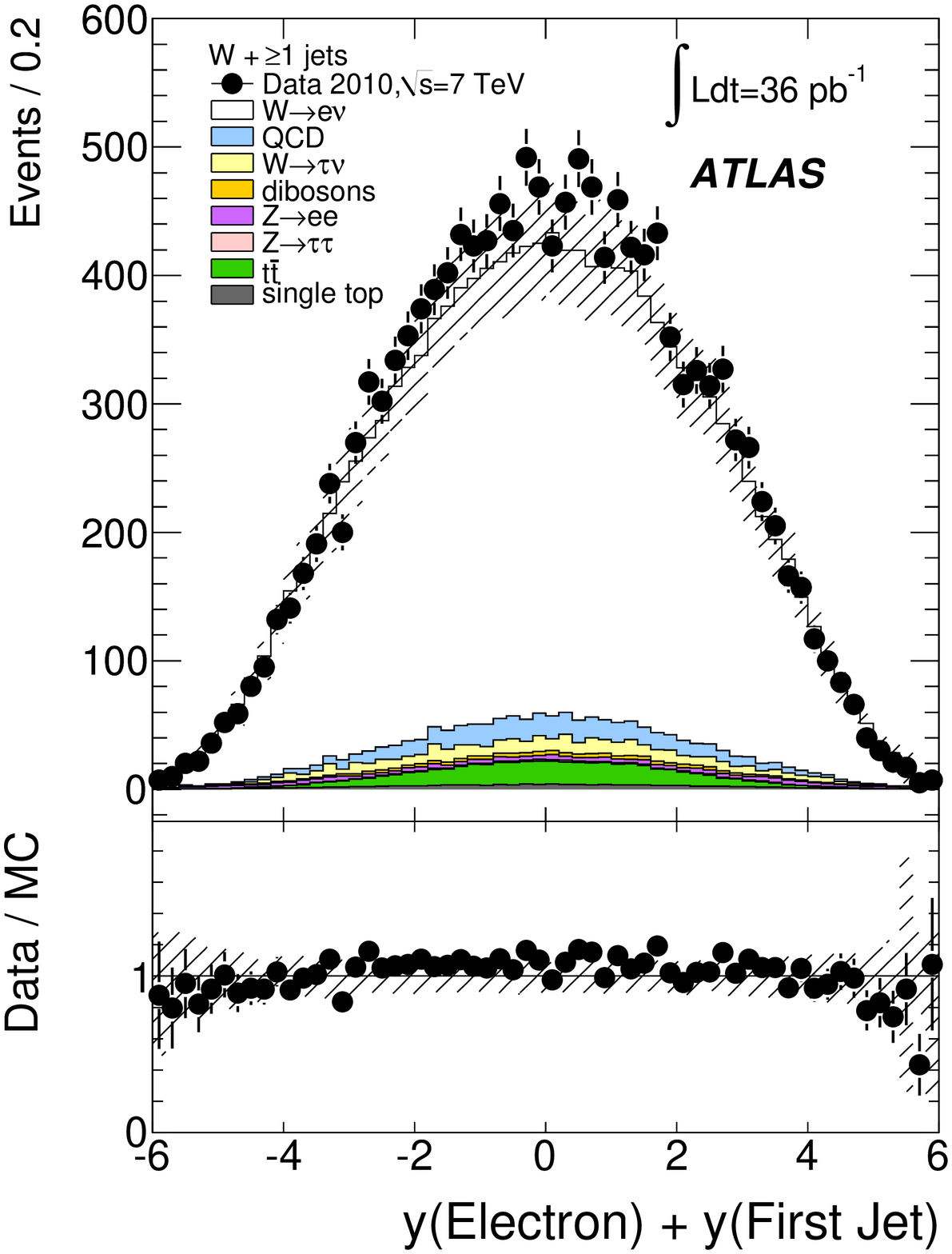}
  \includegraphics[width=0.868\linewidth]{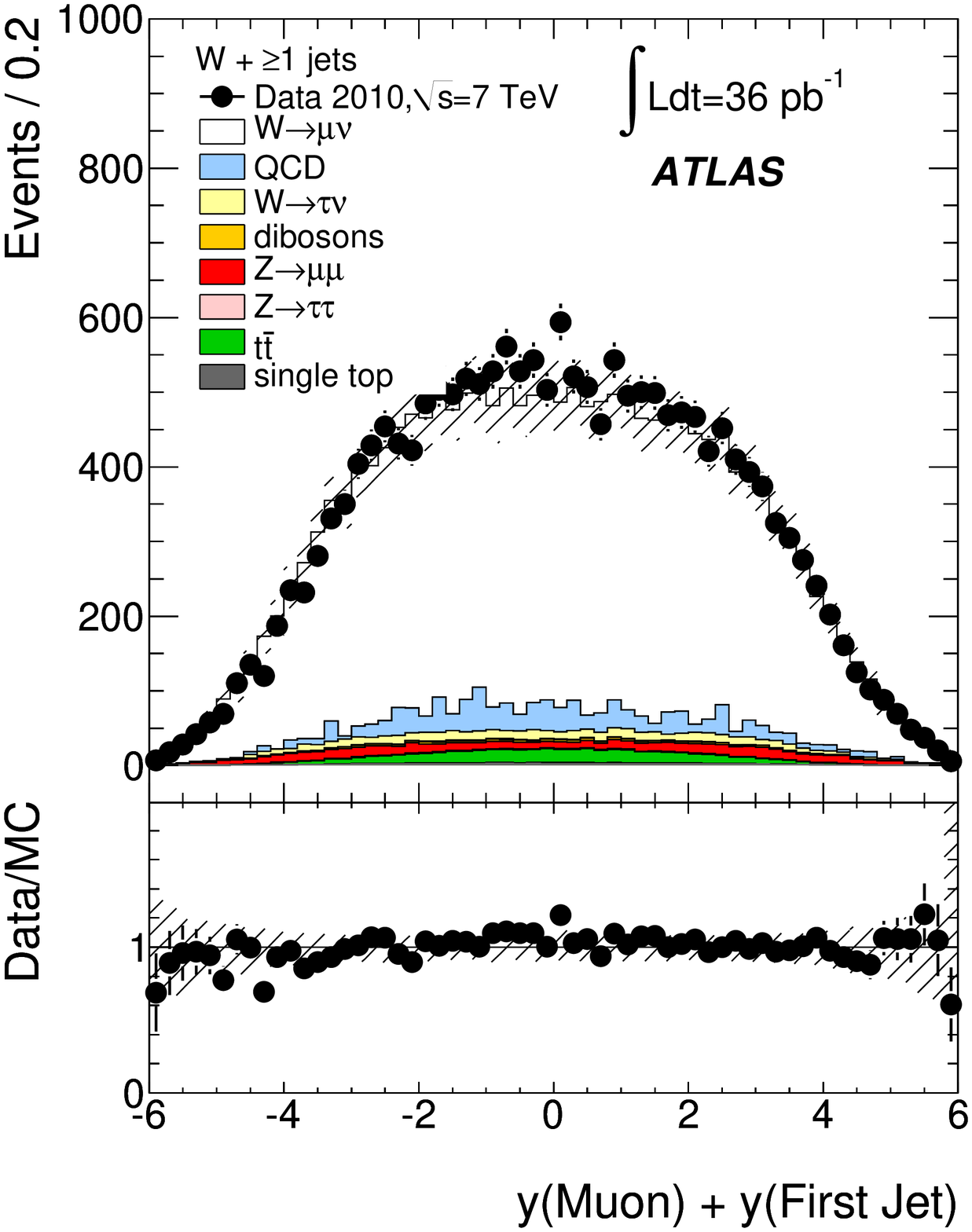}
  \caption{The uncorrected distribution in $y(\ell)\:+\:y({\rm
  first\:jet})$, sum of rapidities of the lepton and the leading jet,
  for events with one or more jets.\\
\\
\\
\\
\\
\\}
  \label{fig:det-sylj}
\end{figure}

\begin{figure}[htb]
\centering
  \includegraphics[width=0.868\linewidth]{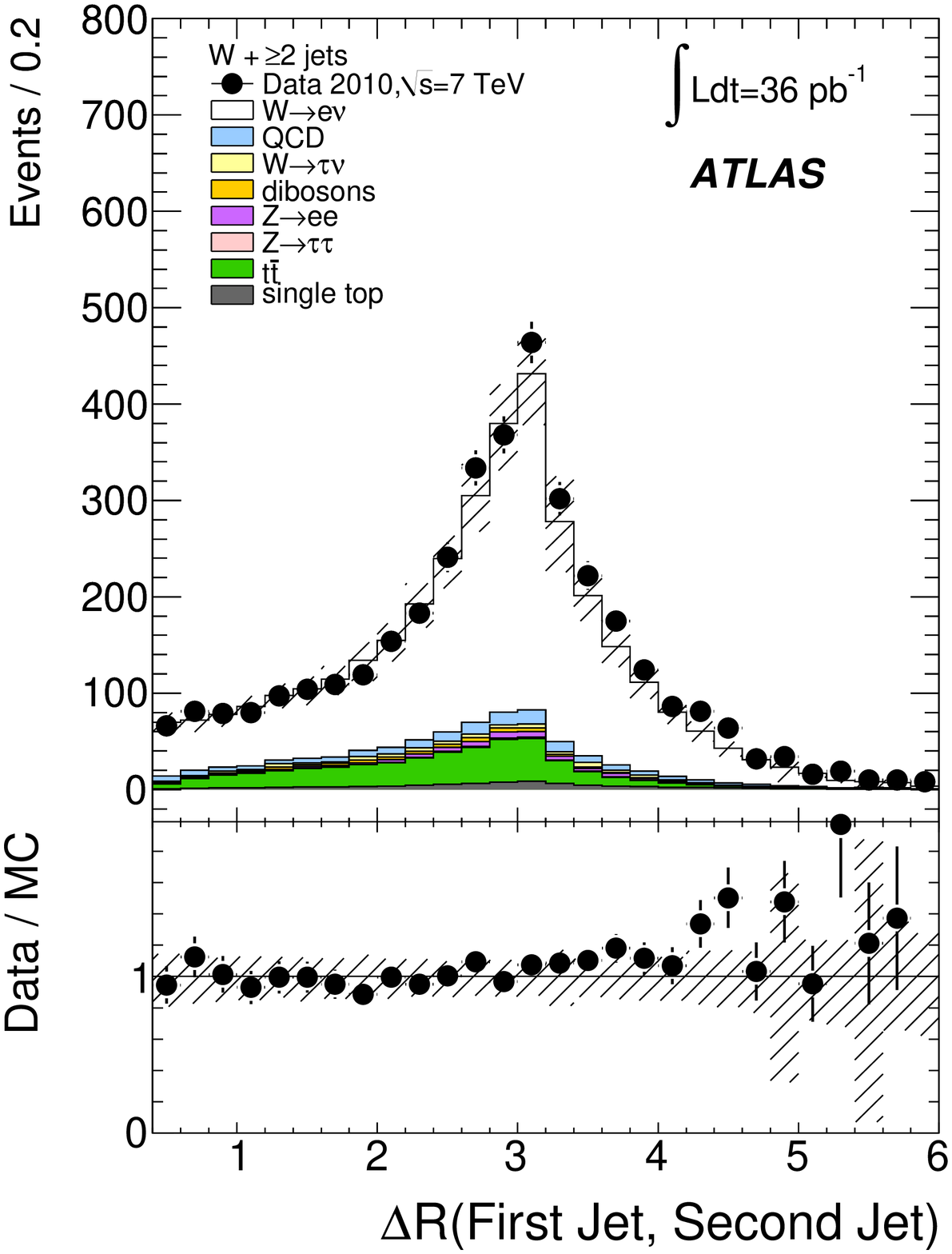}
  \includegraphics[width=0.868\linewidth]{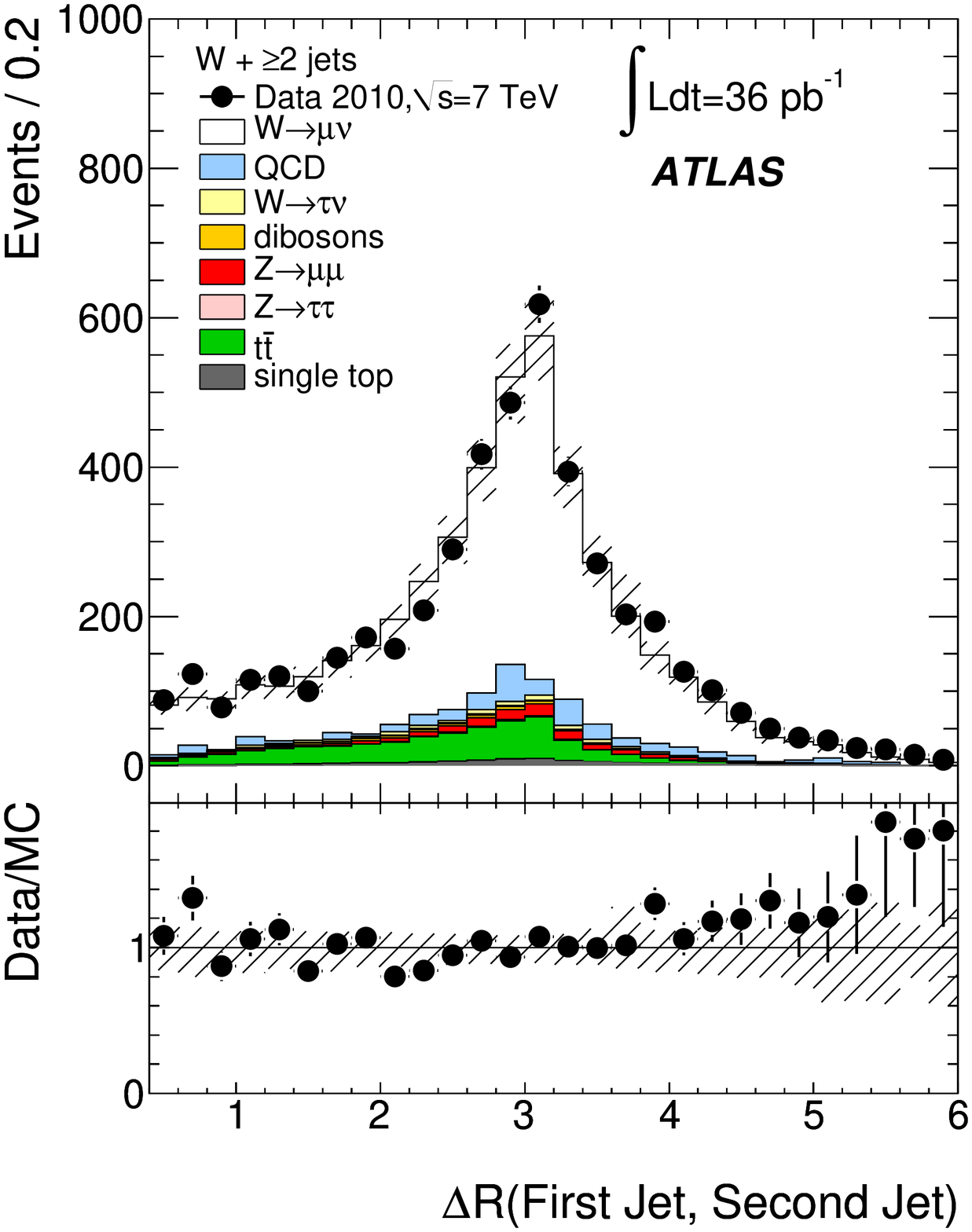}
  \caption{The uncorrected distribution as a function of $\Delta
  R({\rm first\:jet,\:second\:jet})$, distance between the first two
  jets, for events with two or more jets.\\
\\
\\
\\
\\
\\}
  \label{fig:det-dRjj}
\end{figure}

\begin{figure}[htb]
\centering
  \includegraphics[width=0.868\linewidth]{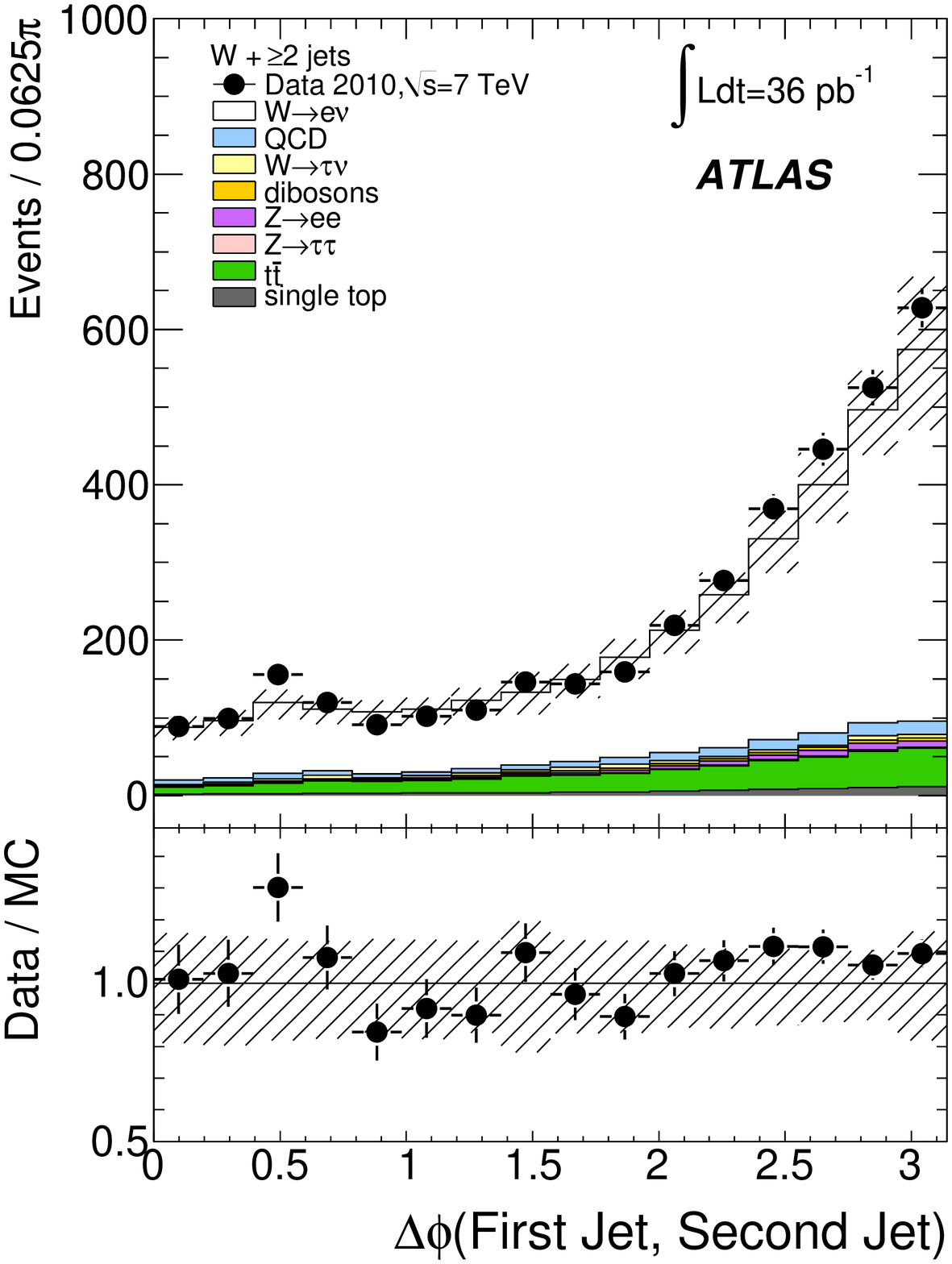}
  \includegraphics[width=0.868\linewidth]{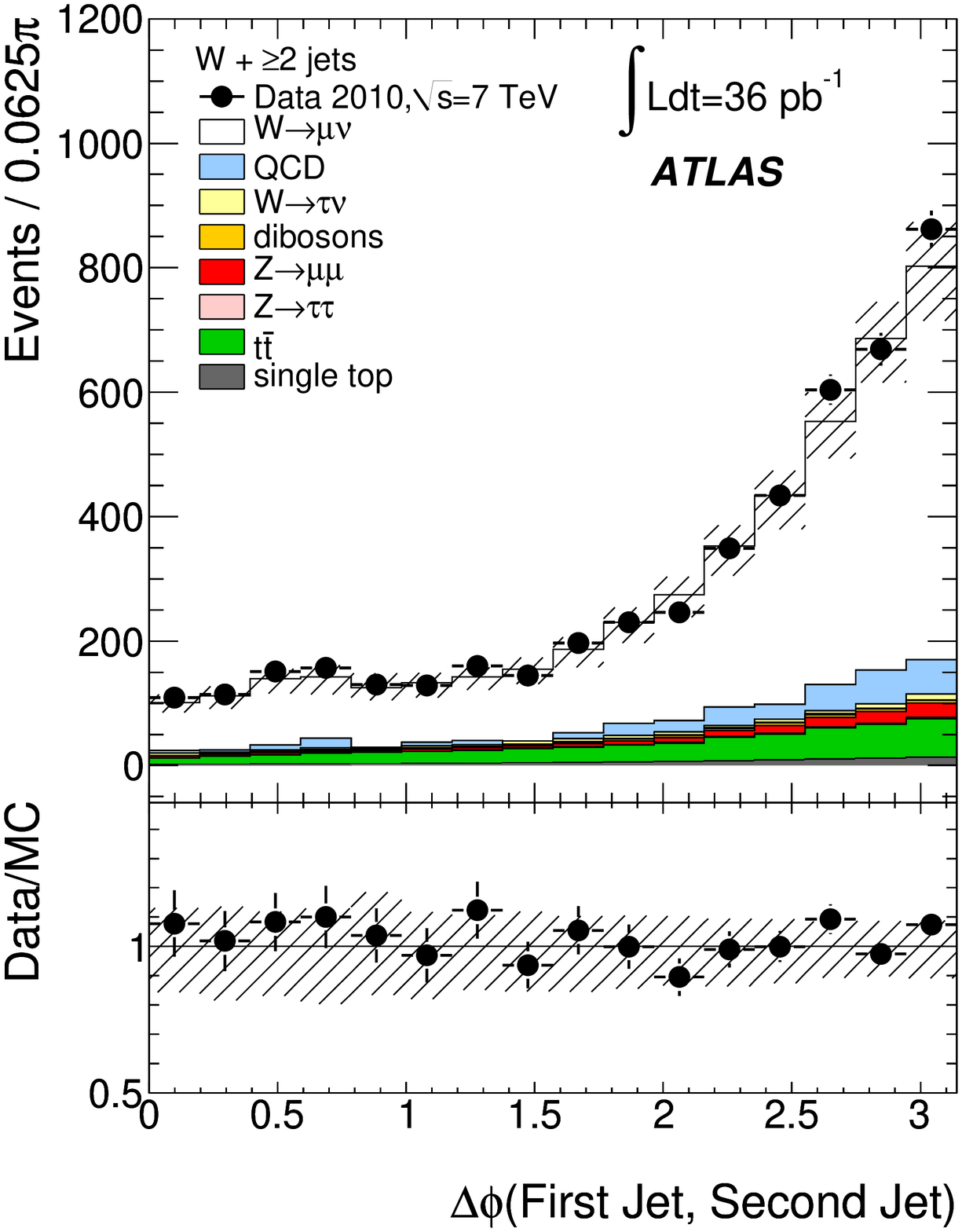}
  \caption{The uncorrected distribution as a function of
  $\Delta\phi({\rm first\:jet,\:second\:jet})$, azimuthal separation
  between the first two jets, for events with two or more jets.\\
\\
\\
\\
\\
\\}
  \label{fig:det-dphijj}
\end{figure}

\begin{figure}[htb]
\centering
  \includegraphics[width=0.868\linewidth]{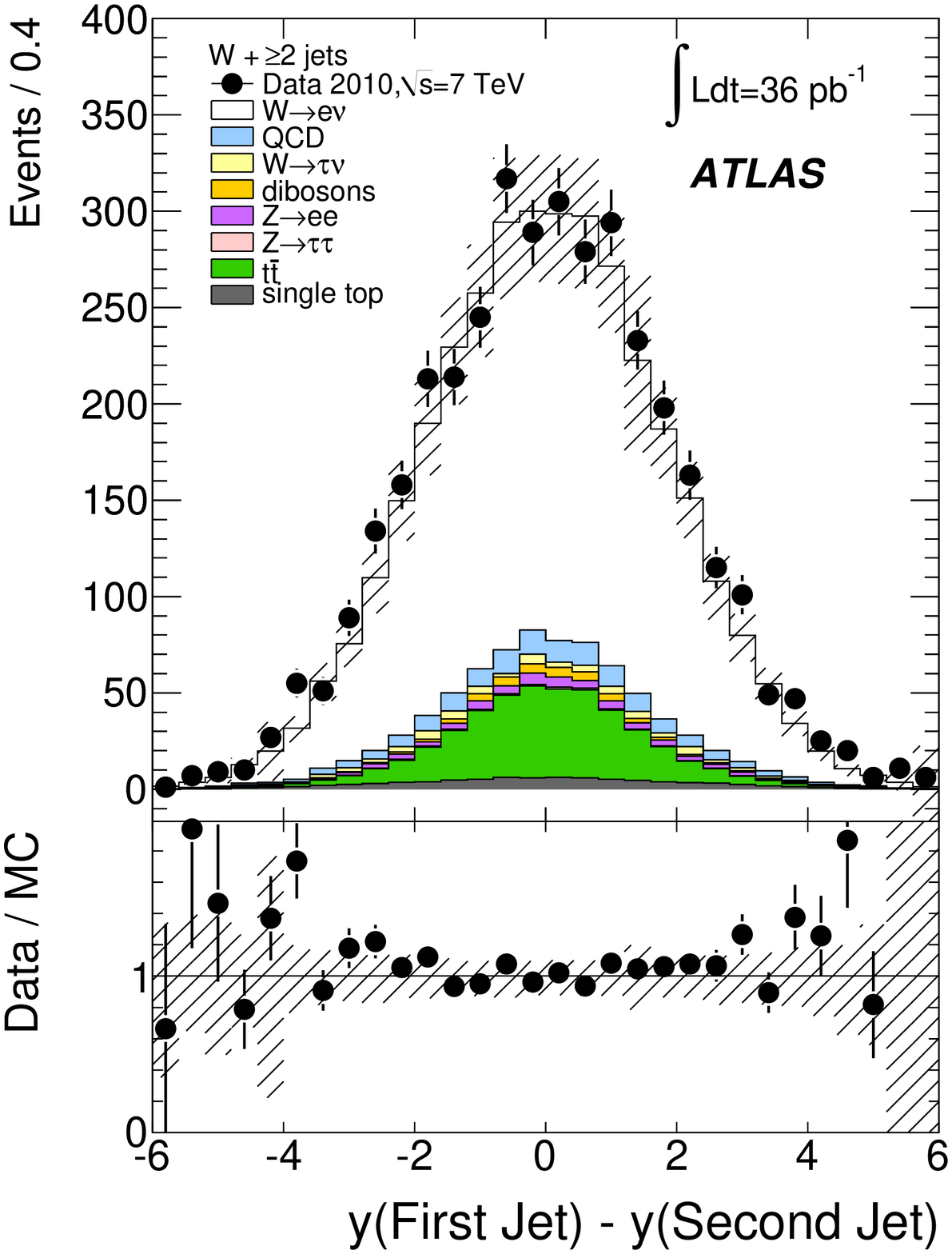}
  \includegraphics[width=0.868\linewidth]{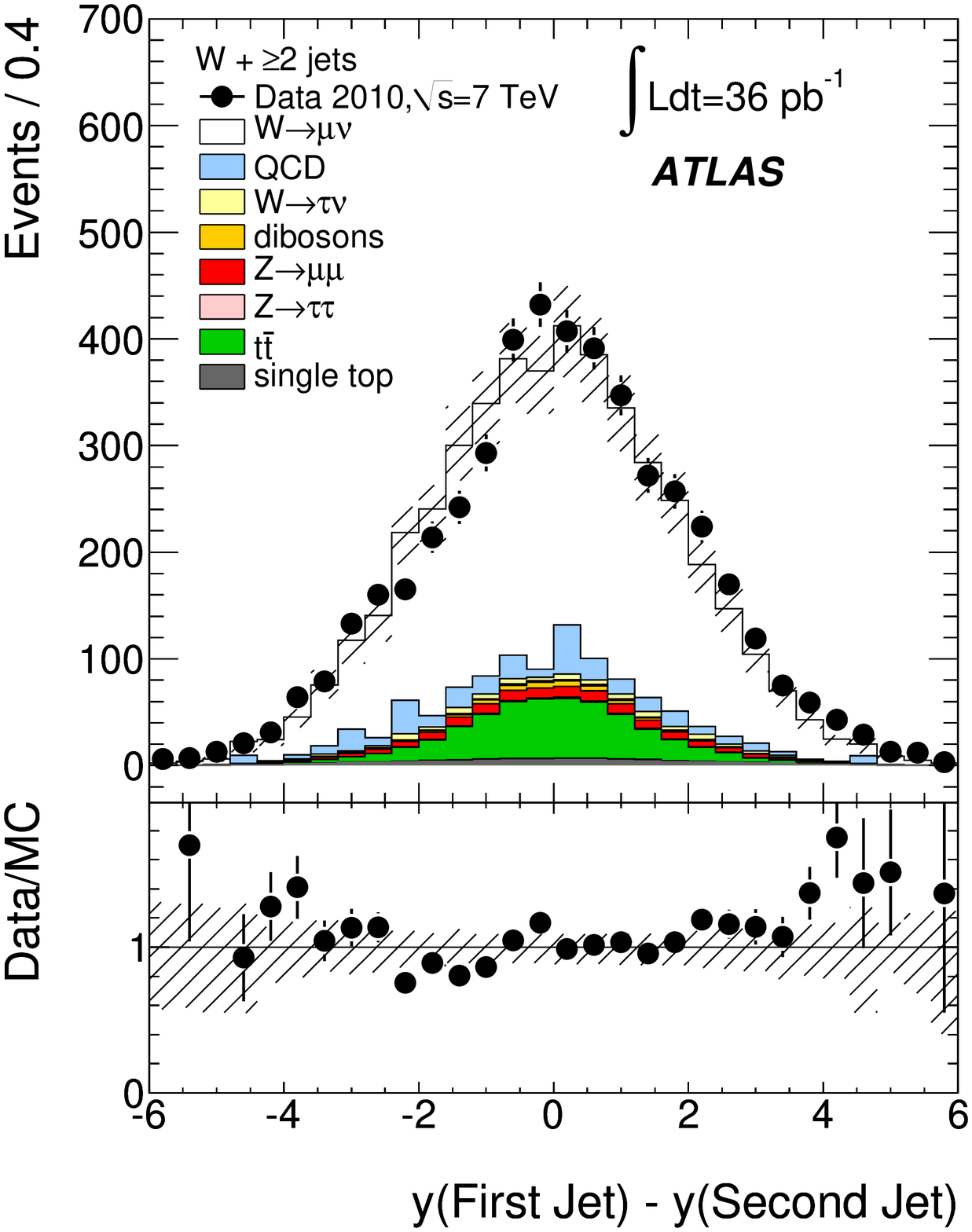}
  \caption{The uncorrected distribution as a function of $y({\rm
  first\:jet})\:-\:y({\rm second\:jet})$, rapidity separation between
  the first two jets, for events with two or more jets.}
  \label{fig:det-dyjj}
\end{figure}

\par Observed and expected distributions for several variables have
been compared for the electron and muon channels. The observed
distributions are shown with statistical uncertainties. The expected
distributions are presented with experimental uncertainties that
include those described later in Section~\ref{sec:syst_uncert} in
addition to the uncertainties specific to the two channels from
Sections~\ref{sec:ele_syst} and~\ref{sec:muo_syst}. Distributions of
the inclusive jet multiplicity are shown in Fig.~\ref{fig:det-njet}.
Figures~\ref{fig:det-pt1}, \ref{fig:det-pt2}, \ref{fig:det-pt3},
and~\ref{fig:det-pt4} show distributions in \pt\ of the first four
(highest \pt) jets. The rapidity of the first jet is shown in
Fig.~\ref{fig:det-y1}. The difference and sum of the rapidities of the
lepton and the first jet are shown in Figs.~\ref{fig:det-dylj}
and~\ref{fig:det-sylj}, respectively. Variables dependent on the
azimuthal and rapidity separations between the first two jets are
featured in Figs.~\ref{fig:det-dRjj}, \ref{fig:det-dphijj},
and~\ref{fig:det-dyjj}. Overall, a good agreement is seen between
measured and predicted distributions. Minor discrepancies appear for
jet pairs with large rapidity separation in Figs.~\ref{fig:det-dRjj}
and~\ref{fig:det-dyjj}. Fig.~\ref{fig:det-dylj} illustrates
discrepancies for events with the first jet separated in rapidity from
the lepton. Predictions in Figs.~\ref{fig:det-y1},~\ref{fig:det-dylj},
and~\ref{fig:det-sylj} are found to be sensitive to the choice of PDF.

\par Top quark pair production is a substantial background to $W$+jets
in events with four or more jets as can be seen in
Fig.~\ref{fig:det-njet}. The predicted \ttbar\ cross section of
165$^{+11}_{-16}$~pb~\cite{ttbar_th} is fully consistent with the
measured value of
171$\pm$20(stat.)$\pm$14(syst.)$^{+8}_{-6}$(lum.)~pb, obtained with
the same 2010 data sample~\cite{ttbar2010}. Here the predicted one was
used to obtain the cross section results. 

\par Several kinematic distributions were used to check the
normalization of the \ttbar\ component in the channels with a $W$
boson plus four or more jets. These included the rapidity of the
charged lepton and the mass of the $W$-jet system. The normalizations
obtained were consistent with the expected value but had a statistical
uncertainty too large to usefully constrain the \ttbar\ cross section.

\subsection{Unfolding of efficiency and resolution effects}
\label{sec:unfold}

\par The yield of signal events was corrected back to the particle
level separately for the two lepton channels, taking into account
detector acceptance and reconstruction efficiency. The correction was
made using an iterative Bayesian method of
unfolding~\cite{BayesUnfold}. Bin sizes in each histogram were chosen
to be a few times larger than the resolution of the corresponding
variable. Migration matrices were computed using the {\sc Alpgen}
$W$+jets event generator plus full detector simulation, restricting
the events to the common phase space:
\begin{itemize}
\item $\pt^\ell>20$ \GeV\ ($\ell$ = electron or muon),
\item $|\eta^\ell|<2.5$,
\item $\pt^\nu>25$ \GeV,
\item $m_{\rm T}(W)>40$ \GeV,
\item $\pt^{\rm jet}>30$ \GeV,
\item $|y^{\rm jet}|<4.4$ and $\Delta R(\ell,\rm jet)>0.5$.
\end{itemize}
The common phase space requirements were applied to generated objects
before the detector simulation. In this analysis, particle level jets
were constructed in simulated events by applying the anti-$k_{\rm{t}}$
jet finder to all final state particles with a lifetime longer than
10~ps, whether produced directly in the $pp$ collision or from the
decay of particles with shorter lifetimes. Neutrinos, electrons, and
muons from decays of the massive $W$ bosons were not used for the jet
finding. Final state QED radiation differs for electrons and muons,
and its effects were corrected in the combined cross
sections. Fiducial cross sections for each channel were defined using
final-state leptons for which collinear radiation in a cone of $R=0.1$
is added to the lepton four-momentum~\cite{LesHouches2009}. This
accounts for the most significant effects of collinear QED
radiation. A residual correction for large-angle radiation outside
this cone is then applied to bring both electrons and muons to the
Born level for the combined cross sections. These correction factors
range from 0.985 to 0.995 and are similar for both electrons and
muons.

\par Instead of inverting the migration matrix, the unfolded
distributions were determined using Bayes' theorem to recalculate the
particle level distributions from the detector level distributions.
The unfolded values were calculated using different numbers of
iterations for different bins of a distribution. The standard Bayesian
approach treats all bins using the same number of iterations. Fewer
iterations were performed for bins with few events than for bins with
large numbers of events to avoid large statistical fluctuations in the
tails of the distributions. The number of iterations was limited for a
bin once the statistical uncertainty becomes substantially larger than
the change due to the last application of the unfolding
matrix~\footnote{The iterative procedure was stopped when the
statistical uncertainty was five times larger than the last change in
the unfolded value.}. Tests with simulated data showed that the
iterative Bayesian method was sufficient to recover particle-level
distributions. The dominant detector to particle level corrections in
the electron channel come from electron reconstruction efficiency
($\approx$30\% correction). In the muon channel, the dominant
corrections come from trigger and reconstruction efficiency
(corrections of $\approx 10-20$\% and $\approx 10$\% respectively).
The statistical uncertainty on the unfolding was estimated using toy
simulations. The systematic uncertainties on the unfolding included
the uncertainty on the migration matrix which was estimated by using
the alternative {\sc Sherpa} simulation for $W$+jets production (see
Table~\ref{tab:MC}).

\begin{table*}[htb] 
\centering
\caption{Summary of systematic uncertainties on the cross sections.
The uncertainties are shown for $N_{\rm{jet}} \ge 1$ and $N_{\rm{jet}}
\ge 4$. The sign convention for the JES and lepton energy scale
uncertainties is such that a positive change in the energy scale
results in an increase in the jet or lepton energy observed in the
data.}
\begin{tabular}{l|c|c|c} \hline\hline
\multicolumn{4}{c}{\Wen\ channel} \\\hline
         &       & \multicolumn{2}{c}{Cross Section Uncertainty (\%)}\\
  Effect & Range &  $N_{\rm{jet}}\ge1$ &  $N_{\rm{jet}}\ge4$ \\\hline
  Jet and cluster energy scales& 2.5--14\% (dependent on jet
         $\eta$ and \pT) & $+9.0,-6.6$ & $+37,-35$\\
  Jet energy resolution & $\sim$10\% on each jet (dependent on jet $\eta$ and \pT)  & $\pm$1.6 & $\pm$6\\
  Electron trigger & $\pm0.5$\% & $+0.6,-0.5$ & $\pm$1\\
  Electron reconstruction & $\pm1.5$\% & $+1.7,-1.6$ & $\pm$4\\
  Electron identification & $\pm$2--8\% (dependent on electron \eta\ and \pT) & $+4.3,-4.0$ & $+10,-9$\\
  Electron energy scale & $\pm$0.3--1.6\% (dependent on \eta\ and \pt)  & $\pm$0.6 & $+1,-3$\\
  Electron energy resolution & $< 0.6\%$ of the energy & $\pm$0.0 & $<$1\\
  Pile--up removal requirement & $\sim1.5$\% in lowest jet \pT~bin & $\pm$1.1 & $\pm$3\\
  Multijet QCD background shape & from template variation & $\pm$0.7 & $\pm$11\\
  Unfolding & {\sc Alpgen} vs. {\sc Sherpa} & $\pm$1.5 & $\pm$6\\
  Luminosity & $\pm3.4$\% &  $+3.8,-3.6$ & $+9,-8$\\
  NNLO cross section for $W/Z$ 	& $\pm$5\% & $\pm$0.2 & $<$1\\
  NLO cross section for \ttbar\	& 	$^{+7}_{-10}$\%	& $\pm$0.3 & $\pm$10\\
  Simulated \ttbar\ shape & from samples with more or less ISR&$\pm$0.1 & $+12,-21$\\
  \hline\hline
  \multicolumn{4}{c}{    } \\
  \multicolumn{4}{c}{    } \\
  \hline\hline
  \multicolumn{4}{c}{\Wmn\ channel} \\\hline
         &       & \multicolumn{2}{c}{Cross Section  Uncertainty (\%)}\\
  Effect & Range &  $N_{\rm{jet}} \ge 1$  &  $N_{\rm{jet}} \ge 4$\\\hline
  Jet and cluster energy scales & 2.5--14\% (dependent on jet $\eta$ and \pT) & $+8.2,-6.2$ & $+33,-26$ \\
  Jet energy resolution & 10\% on each jet (dependent on jet $\eta$ and \pT) & $\pm$1.5 & $\pm$5\\
  Muon trigger & $\pm0.7$\% ($\pm0.6$\%) in barrel (endcap)&$\pm$0.6 &$\pm$1\\
  Muon reconstruction & $\pm1.1$\% &  $\pm$1.1 & $\pm$2\\
  and identification & & & \\
  Muon momentum scale & $\pm0.4$\%  & $+0.2,-0.3$  & $<$1\\
  Muon momentum resolution & $\pm6$\%  & $\pm$0.1  & $<$1\\
  Pile--up removal requirement & $\sim1.5$\% in lowest jet \pT~bin & $\pm$1.0 & $\pm$3\\
  Multijet QCD background shape & from template variation & $+0.8$ & $-20$\\
  Unfolding & {\sc Alpgen} vs. {\sc Sherpa} & $\pm$0.2 & $<$1\\
  Luminosity & $\pm3.4$\% & $+3.7,-3.5$ & $\pm$7\\
  NNLO cross section for $W/Z$ 	& $\pm$5\% & $\pm$0.4 & $<$1\\
  NLO cross section for \ttbar\	& $^{+7}_{-10}$\%	& $+0.4,-0.3$ & $+10,-7$\\
  Simulated \ttbar\ shape &  from samples with more or less ISR &$<$0.1 & $+13,-15$\\
\hline\hline
\end{tabular}
\label{tab:effsys_summary}
\end{table*}

\subsection{Overall Systematic Uncertainties}
\label{sec:syst_uncert}

\par In addition to the systematic uncertainties specific to the
electron and muon channels documented earlier in
Sections~\ref{sec:ele_syst} and~\ref{sec:muo_syst}, respectively,
there are a number of common sources of uncertainty. As a brief
reminder, the uncertainty on the identification efficiency for
electrons results in $^{+4.0}_{-4.3}$\% variation of the
$N_{\rm{jet}}\ge1$ cross section, giving the largest variation among
the electron-specific uncertainties. Similarly, the uncertainty on
reconstruction and identification efficiency of muons corresponds to a
variation of $\pm$1.1\% in the $N_{\rm{jet}}\ge1$ cross section and
represents the single largest muon-specific uncertainty.

\par The dominant source of systematic uncertainty in the cross
section measurement for both electron and muon channels is the
uncertainty in the jet energy scale~\cite{JES}. For $N_{\rm{jet}} \ge
4$, uncertainties on the predicted \ttbar\ cross section and \ttbar\
shape also become significant and can be as high as 10\% and 21\%,
respectively. The luminosity uncertainty enters primarily through the
signal normalization but also has a small effect on the estimation of
the leptonic backgrounds.

\par Uncertainties in the jet energy scale (JES) and jet energy
resolution (JER) were determined from data and simulation~\cite{JES}.
The JER uncertainty was 10\% of the jet energy
resolution~\cite{JES}. The JES uncertainty varies as a function of jet
\pT\ and $\eta$, and ranges from $\sim 2.5$\% at 60~\GeV\ in the
central region to $\sim 14$\% below 30~\GeV\ in the forward regions;
the uncertainty increases monotonically with the absolute value of jet
pseudorapidity. The uncertainty on the correction of the JES for
pile-up $pp$ interactions is less than 1.5\% per additional
interaction for jets with \pt~$>$~50~\GeV. To take into account the
differences in calorimeter response to quark- and gluon-initiated
jets, the uncertainty on the fraction of gluon-initiated jets, the
flavor composition~\cite{JES}, was estimated by comparing the
fractions in {\sc Sherpa} and {\sc Alpgen} simulations for $W$+jets
production. For jets accompanied by a second jet within $\Delta
R<0.7$, an additional uncertainty is added to the JES uncertainty; the
additional uncertainty is less than 2.8\%. To estimate the impact of
the JES uncertainty, jet energies in the simulated events were
coherently shifted by the JES uncertainty, and the \met\ vector was
recomputed. In addition, simulated energy clusters in the calorimeters
not associated with a jet or electron, such as those coming from the
underlying event and pile-up interactions, were scaled using a \pT\
and $|\eta|$ dependent uncertainty~\cite{incWv16}, ranging from $\pm
5.5$\% for central clusters at \pT~$\simeq 500$~\MeV\ to $\pm 3$\% at
high \pT. Similarly the simulated jet energies were smeared by the JER
uncertainty and the \met\ vector was recomputed. The full analysis was
repeated with these variations, and the cross sections were
recomputed; the change in the cross section was taken as the
systematic uncertainty. The uncertainty on the measured cross sections
caused by the uncertainties on the JES and cluster energy scale
increases with jet multiplicity from 9\% for $N_{\rm{jet}}\ge1$ to
37\% for $N_{\rm{jet}}\ge4$. The impact of the JES uncertainty is
amplified for events with high jet multiplicities due to the large
subtraction of \ttbar\ events, corresponding to $\sim$54\% of these
events. The simulated jet multiplicity of the top background is
sensitive to the JES. The magnification is somewhat smaller when jets
are selected with $\pt^{\rm jet}$~$>$~20~\GeV\ instead of 30~\GeV; the
JES-related uncertainty on the $N_{\rm{jet}}\ge4$ cross section is up
to 29\%.

\par The uncertainty due to jets originating from pile-up interactions
and the influence of the JVF selection requirement includes the
efficiency of the requirement and how well the rate of pile-up jets is
modeled in the simulation. As a conservative estimate, the percentage
of jets in the data removed by the JVF requirement is applied as the
uncertainty. This results in a 1.5\% uncertainty for jets with
\pT~$<$~40~\GeV\ with a resulting uncertainty on the cross section of
1\% for $N_{\rm{jet}} \ge1$.

\par Other uncertainties which were considered include the jet
reconstruction efficiency and biases in the procedure for correcting
for detector effects (by comparing correction factors obtained with
{\sc Alpgen} to those obtained with {\sc Sherpa}). Their effect on the
cross section was found to be smaller than the uncertainties described
before. All of these systematic uncertainties were also applied to the
estimates of the multijet and leptonic backgrounds in both electron
and muon channels. In addition, for the leptonic backgrounds the
uncertainty in the NNLO cross sections was taken to be 5\% for $W/Z$
production as in Ref.~\cite{incWv16}. The \ttbar\ cross section
uncertainty was taken to be $^{+7}_{-10}$\%~\cite{ttbar_th}. The
uncertainty on the shapes of the \ttbar\ distributions was estimated
using {\sc AcerMC} simulations where rates of ISR and FSR were altered
with respect to the default settings. Samples with altered ISR were
used to estimate the shape uncertainty since their impact on measured
cross sections was the largest among these samples. The procedure has
been used for ATLAS measurements involving top pair
production~\cite{ttbar2010}.

\par
\begin{figure}[htb]
\centering
\includegraphics[width=0.95\linewidth]{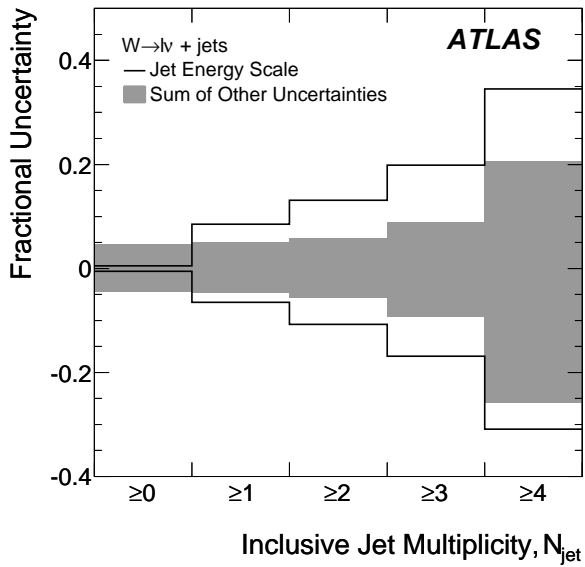}
\caption{Systematic uncertainties on the cross section as a function
of the inclusive jet multiplicity. The uncertainty due to the jet
energy scale is bounded by the two black lines. The quadratic sum of
the other systematic uncertainties is presented as the shaded
area. The uncertainties are for the sum of the electron and muon cross
sections.}
\label{fig:syst_Njet}
\end{figure}

 The systematic uncertainties in the cross section measurement are
summarized in Table~\ref{tab:effsys_summary} for $N_{\rm{jet}}\ge1$
and $N_{\rm{jet}}\ge4$; most of the uncertainties are approximately
independent of the jet multiplicity, except for the uncertainty due to
the jet energy scale and resolution, multijet background shape,
\ttbar\ production, and pile-up jet removal. The uncertainty due to
the jet energy scale dominates for events with at least one jet as
illustrated in Fig.~\ref{fig:syst_Njet}.

\par In the cross section ratio measurement, \mbox{$\sigma(W+\ge
N_{\rm{jet}})/\sigma(W+\ge N_{\rm{jet}}-1)$}, the uncertainty due to
the jet energy scale uncertainty remains the dominant effect,
amounting to approximately 5--20\% on the ratio. The luminosity
uncertainty does not completely cancel in the ratio because the
background estimates are affected by the luminosity uncertainty and
the background levels vary as a function of jet multiplicity.

\section{Next-to-leading-order QCD predictions}

\par The MCFM v5.8 \cite{MCFM} and {\sc
BlackHat-Sherpa}~\cite{BlackHat4jets} predictions were obtained with
the same jet algorithm and same kinematic selection requirements
applied to the data. In both cases, renormalization and factorization
scales were set to \HT/2, where \HT\ is the scalar sum of the \pT\ of
all the partons and of the lepton and neutrino from the $W$-decay. The
PDFs used for MCFM were CTEQ6L1~\cite{CTEQ6L1} and
CTEQ6.6M~\cite{CTEQ6.6M} for the LO and NLO calculations,
respectively. For {\sc BlackHat-Sherpa} CTEQ6.6M was used for both LO
and NLO calculations.

\par The systematic uncertainty in the MCFM and {\sc BlackHat-Sherpa}
cross section due to renormalization and factorization scales were
estimated by varying the scales by factors of two, up and down, in all
combinations. The ratio of one scale to the other was kept within the
range 0.5 to 2.0 to avoid the effects of large logarithms of the scale
ratios in some kinematic regions. The cross section ratio,
\mbox{$\sigma(W+\ge N_{\rm{jet}})/\sigma(W+\ge N_{\rm{jet}}-1)$}, was
recalculated for each variation of the scales and the resulting
uncertainty was determined using the recalculated values. Overall, the
asynchronous variations of scales resulted in bigger deviations from
the nominal values than the synchronous variations. The upper and
lower uncertainties were taken as the maximum deviations from the
nominal value.

\par Following the PDF4LHC recommendations~\cite{PDF4LHC}, PDF
uncertainties were computed by summing in quadrature the dependence on
each of the 22 eigenvectors characterizing the CTEQ6.6 PDF set; the
uncertainty in $\alpha_{\rm{s}}$ was also taken into account. The
uncertainties were scaled to a confidence level (C.L.) of 68\%. Two
alternative PDF sets, MSTW2008~\cite{MSTW2008}, with its set of 68\%
C.L. eigenvectors, and NNPDF2.0~\cite{NNPDF}, were also examined. The error
envelope of CTEQ6.6 was found to contain nearly all variations due to
the two alternative PDF sets.  The uncertainties due to the scale
variations were substantially larger than those due to PDFs.

\par As a cross-check, cross sections from {\sc BlackHat-Sherpa} and
{\sc MCFM} were compared for events with up to two jets, and found to
be nearly identical. Therefore, only distributions from {\sc
BlackHat-Sherpa} were compared to the measured cross sections.

\par Bin-by-bin corrections for non-pQCD effects, hadronization and
underlying event, were computed using simulated $W$+jets samples for
each predicted distribution for the NLO cross sections. The
corrections were taken to be the ratios of the distributions for
particle-level jets to the distributions for parton-level jets, where
the sample for parton-level jets was produced with the underlying
event turned off. To calculate the central values, samples from {\sc
Alpgen} v2.13 were showered with {\sc Herwig} v6.510 and {\sc Jimmy}
v4.31 set to the AUET2 tune~\cite{AUET2}. The systematic uncertainty
on the non-pQCD corrections was evaluated by comparing the central
values to corrections from samples where {\sc Alpgen} was showered
with {\sc Pythia} v6.4.21 set to the AMBT1~\cite{AMBT1} event
generator tune. The corrections and their uncertainties were applied
to all the NLO predictions presented in the paper.

\begin{figure}[htb]
  \centering
  \includegraphics[width=0.9\linewidth]{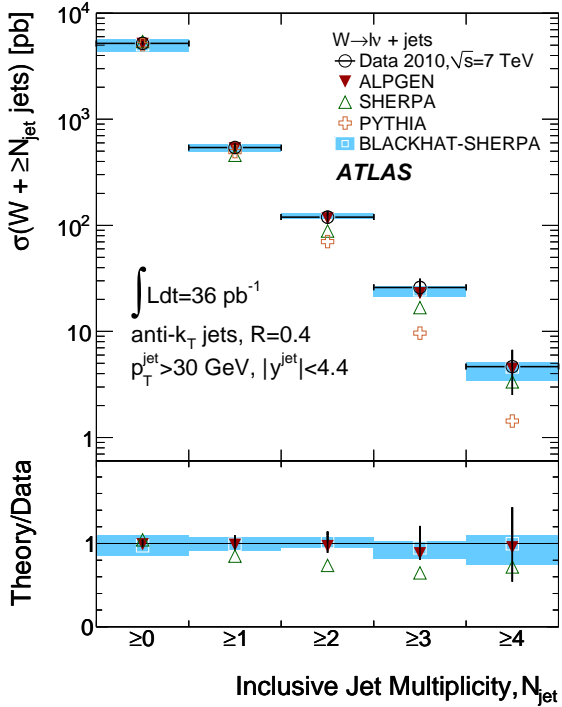}
  \caption{$W$+jets cross section results as a function of corrected
  jet multiplicity. The following remarks apply to this and subsequent
  figures unless specific comments are provided. The cross sections
  are quoted in the kinematic region described in
  Section~\ref{sec:unfold}. For the data, the statistical
  uncertainties are shown with a tick on the vertical bars, and the
  combined statistical and systematic uncertainties are shown with the
  full error bar. Also shown are predictions from {\sc Alpgen}, {\sc
  Sherpa}, {\sc Pythia} and {\sc BlackHat-Sherpa}, and the ratio of
  theoretical predictions to data ({\sc Pythia} is not shown in the
  ratio). The distributions from {\sc Sherpa}, {\sc Pythia}, and {\sc
  Alpgen} were normalized to the NNLO total $W$-boson production cross
  section.}
  \label{fig:result-xsec}
\end{figure}

\begin{figure}[htb]
  \centering
  \includegraphics[width=0.9\linewidth]{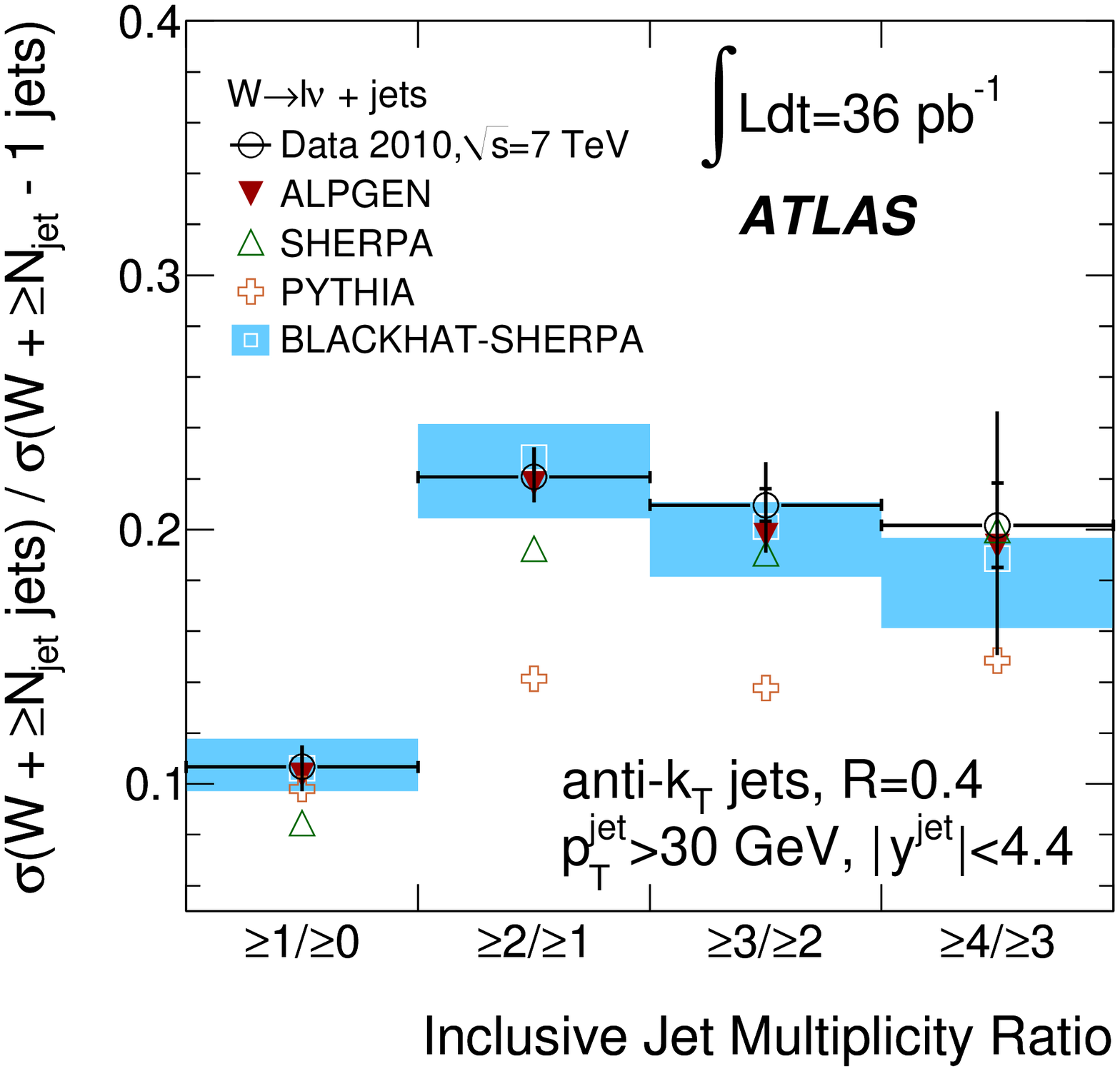}
  \caption{$W$+jets cross section ratio results as a function of
  corrected jet multiplicity.}
  \label{fig:result-xsecratio}
\end{figure}

\begin{figure}[htb]
  \centering
  \includegraphics[width=0.9\linewidth]{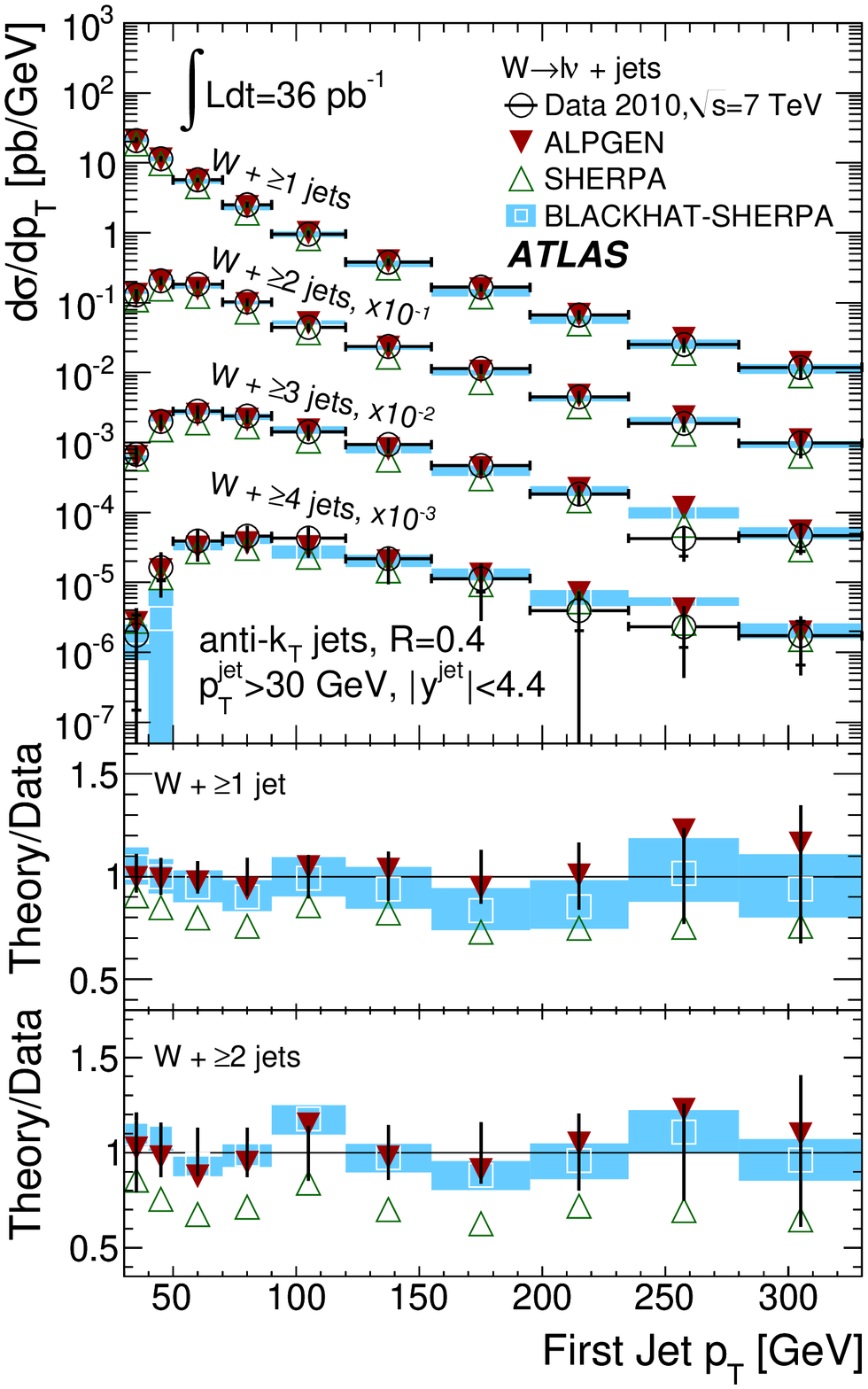}
  \caption{$W$+jets cross section as a function of the \pT\ of the
  first jet in the event. The \pT\ of the first jet is shown
  separately for events with $\ge1$~jet to $\ge4$~jet. The $\ge2$~jet,
  $\ge3$~jet, and $\ge4$~jet distributions have been scaled down by
  factors of 10, 100, and 1000 respectively. Shown are predictions
  from {\sc Alpgen}, {\sc Sherpa}, and {\sc BlackHat-Sherpa}, and the
  ratio of theoretical predictions to data for $\ge 1$~jet and $\ge
  2$~jet events.}
  \label{fig:result-xsec-jet1}
\end{figure}

\begin{figure}[htb]
  \centering
  \includegraphics[width=0.9\linewidth]{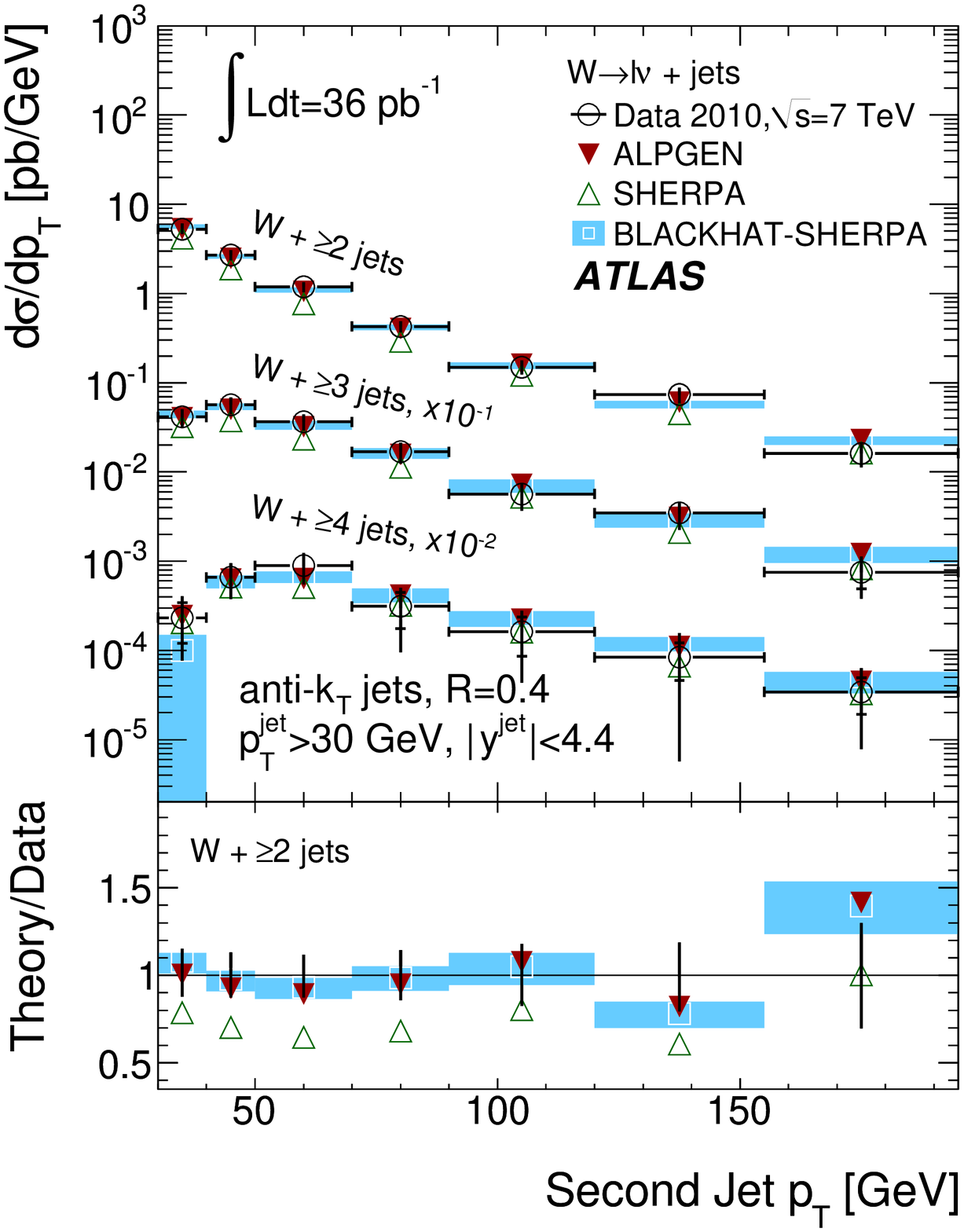}
  \caption{$W$+jets cross section as a function of the \pT\ of the
  second jet in the event. The \pT\ of the second jet is shown
  separately for events with $\ge 2$~jet to $\ge 4$~jet. The $\ge
  3$~jet and $\ge 4$~jet distributions have been scaled down by
  factors of 10 and 100 respectively. Shown are predictions from {\sc
  Alpgen}, {\sc Sherpa}, and {\sc BlackHat-Sherpa}, and the ratio of
  theoretical predictions to data for and $\ge 2$~jet events.}
  \label{fig:result-xsec-jet2}
\end{figure}

\begin{figure}[htb]
  \centering
  \includegraphics[width=0.9\linewidth]{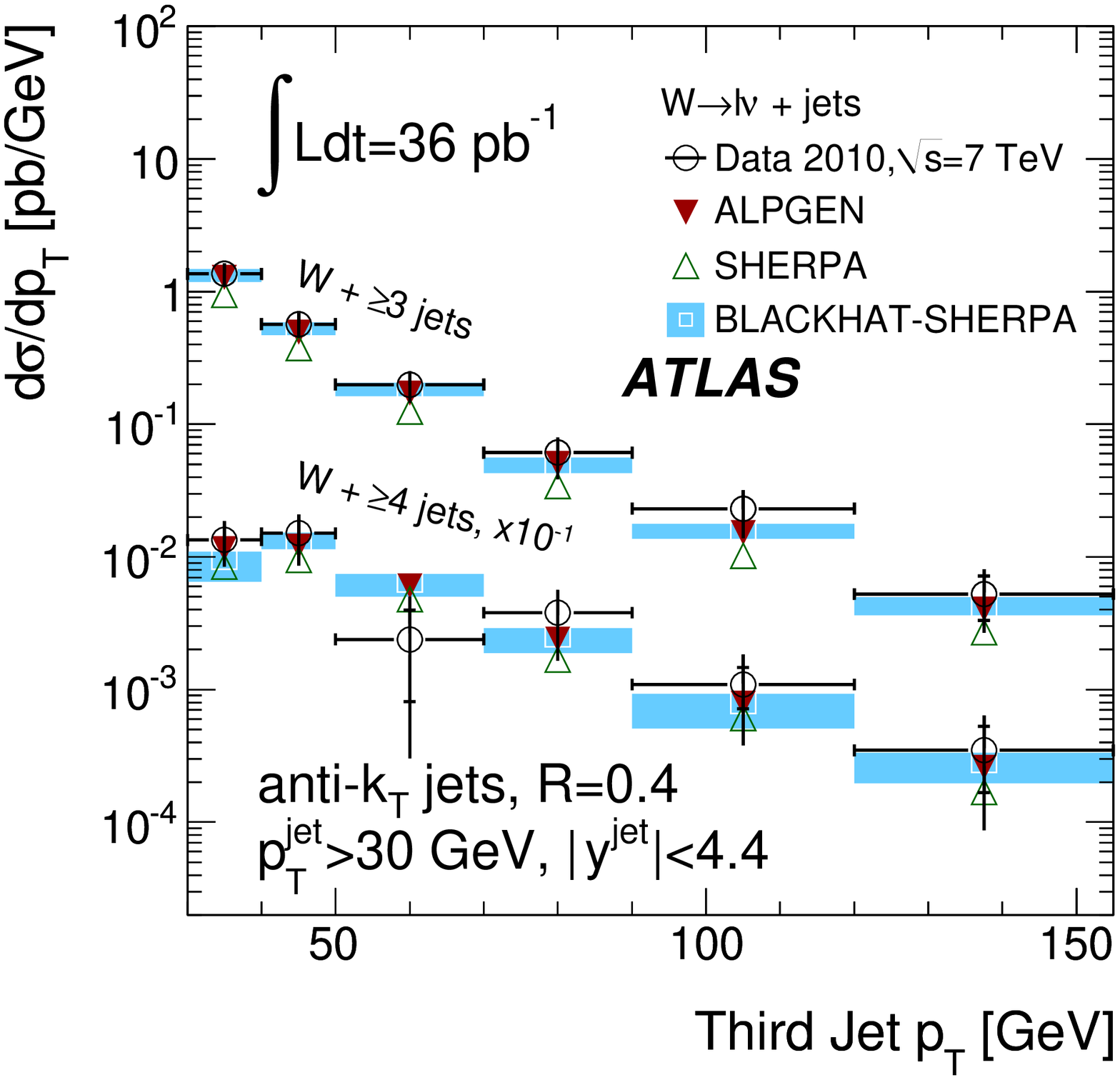}
  \caption{$W$+jets cross section as a function of the \pT\ of the
  third jet in the event. The \pT\ of the third jet is shown
  separately for events with $\ge3$~jet and $\ge4$~jet. The $\ge4$~jet
  distribution has been scaled down by a factor of 10. Shown are
  predictions from {\sc Alpgen}, {\sc Sherpa}, and {\sc
  BlackHat-Sherpa}.}
  \label{fig:result-xsec-jet3}
\end{figure}

\begin{figure}[htb]
  \centering
  \includegraphics[width=0.9\linewidth]{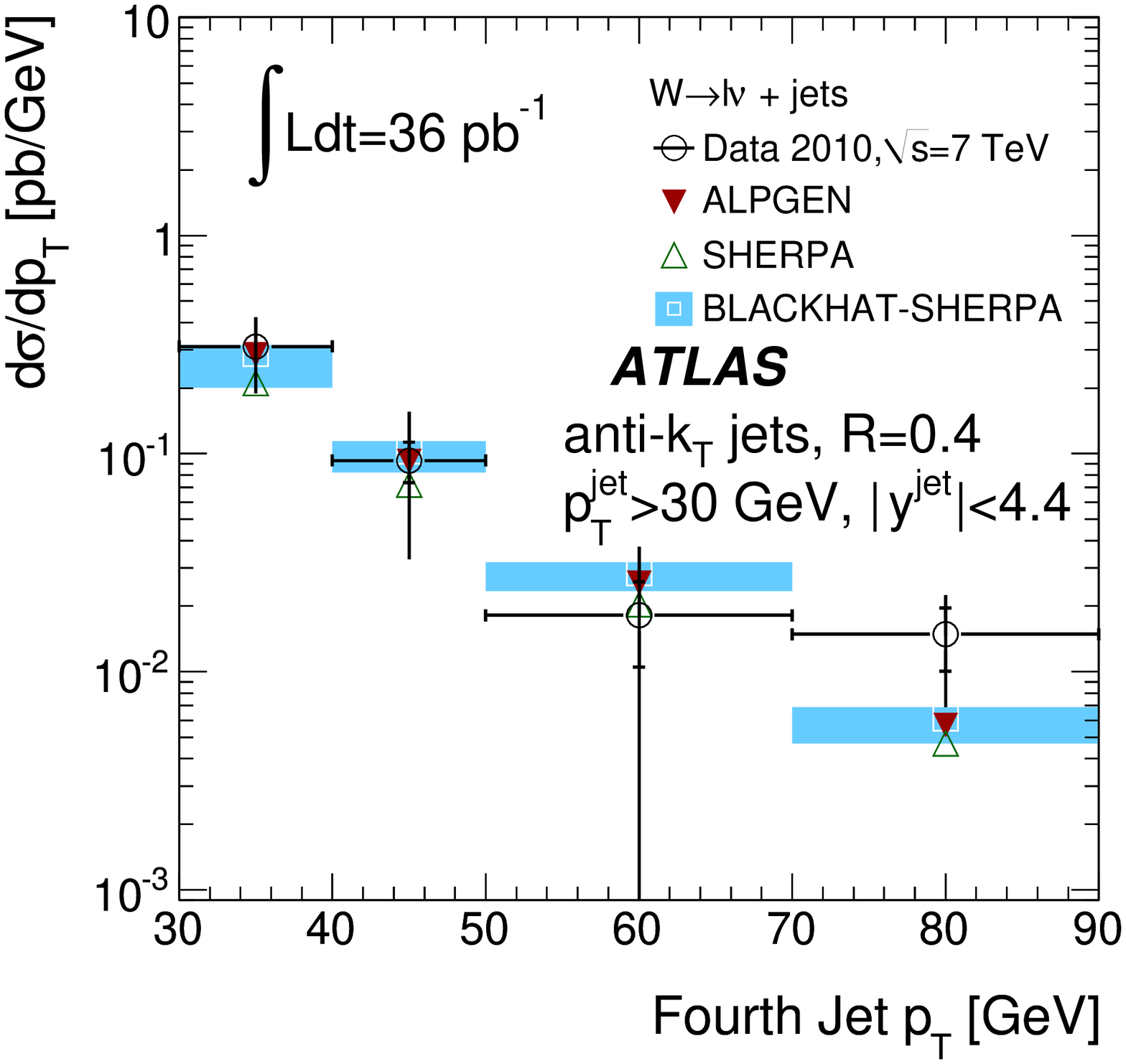}
  \caption{$W$+jets cross section as a function of the \pT\ of the
  fourth jet in the event. The distributions are for events with $\ge
  4$~jet. Shown are predictions from {\sc Alpgen}, {\sc Sherpa}, and
  {\sc BlackHat-Sherpa}.}
  \label{fig:result-xsec-jet4}
\end{figure}

\begin{figure}[htb]
  \centering
  \includegraphics[width=0.9\linewidth]{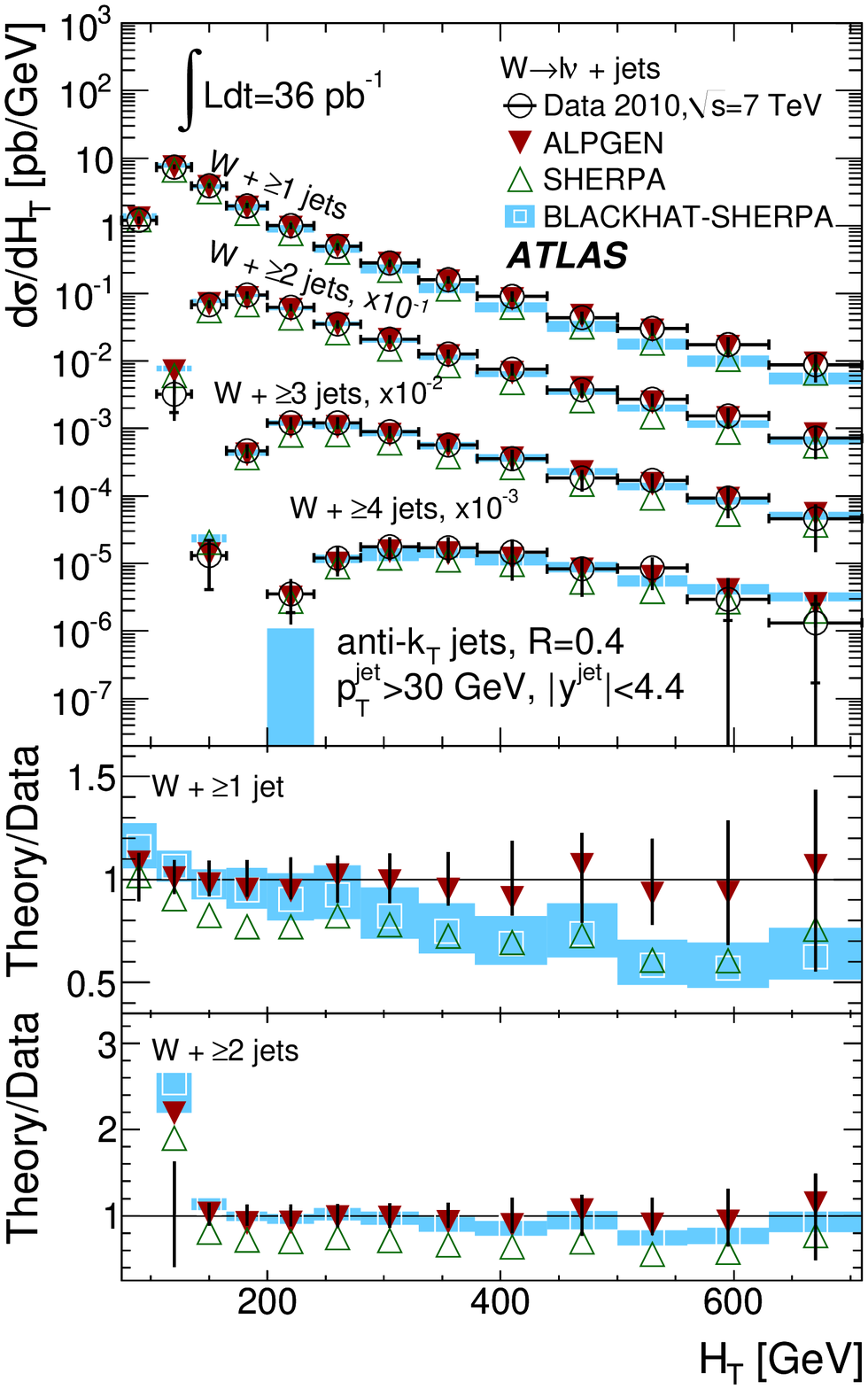}
  \caption{$W$+jets cross section as a function of \HT, shown
  separately for $\ge 1$~jets to $\ge 4$~jets. The $\ge 2$~jet, $\ge
  3$~jet, and $\ge 4$~jet distributions have been scaled down by
  factors of 10, 100, and 1000 respectively. Shown are predictions
  from {\sc Alpgen}, {\sc Sherpa}, and {\sc BlackHat-Sherpa}, and the
  ratio of theoretical predictions to data for $\ge 1$~jet and $\ge
  2$~jet events. The apparent discrepancy between the data and {\sc
  BlackHat-Sherpa} predictions is discussed in the text.}
  \label{fig:result-xsec-ht}
\end{figure}

\begin{figure}[htb]
  \centering
  \includegraphics[width=0.9\linewidth]{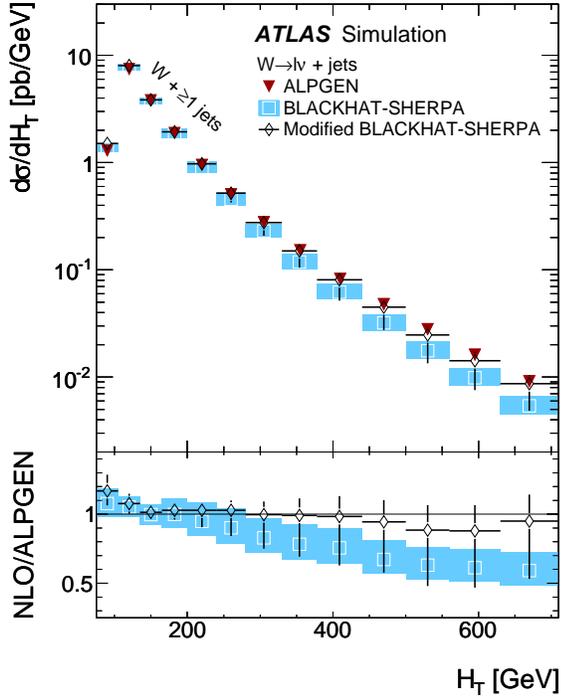}
  \caption{$W$+jets cross section as a function of \HT, shown for
  final states with $\ge 1$~jets. The cross sections are quoted in the
  kinematic region described in Section~\ref{sec:unfold}. Shown are
  predictions from {\sc Alpgen}, {\sc BlackHat-Sherpa}, and modified
  {\sc BlackHat-Sherpa} and the ratio of these NLO theoretical
  predictions to {\sc Alpgen}. The {\sc BlackHat-Sherpa} predictions
  were modified by introducing higher-order NLO terms with two, three,
  and four real emissions to the $N_{\rm jet}\geq1$ distribution. The
  distribution from {\sc Alpgen} was normalized to the NNLO total
  $W$-boson production cross section.}
  \label{fig:result-xsec-ht-compar}
\end{figure}

\begin{figure}[htb]
  \centering
  \includegraphics[width=0.9\linewidth]{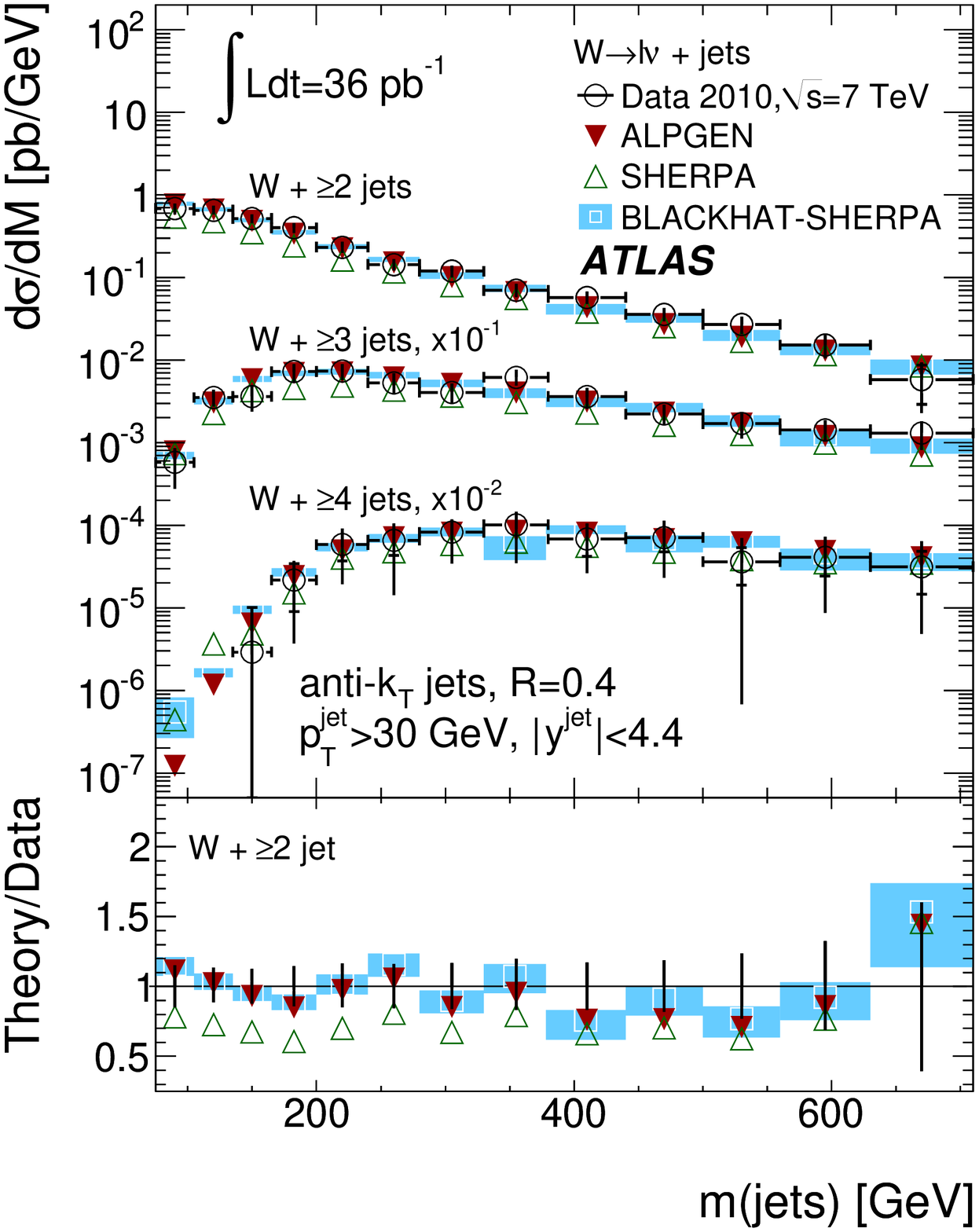}
  \caption{$W$+jets cross section as a function of $m({\rm jets})$,
  shown separately for $\ge 2$~jets to $\ge 4$~jets. The $\ge 3$~jet
  and $\ge 4$~jet distributions have been scaled down by factors of 10
  and 100, respectively. Shown are predictions from {\sc Alpgen}, {\sc
  Sherpa}, and {\sc BlackHat-Sherpa}, and the ratio of theoretical
  predictions to data for $\ge 2$~jet events.}
  \label{fig:result-xsec-mj}
\end{figure}

\begin{figure}[htb]
  \centering
  \includegraphics[width=0.9\linewidth]{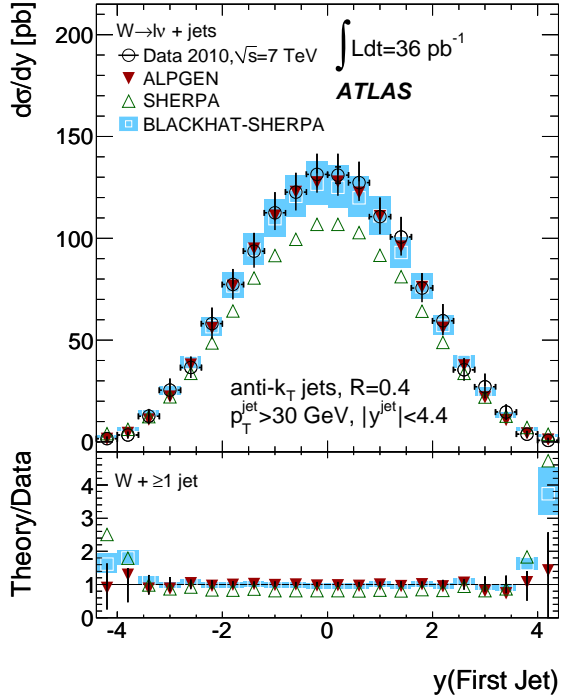}
  \caption{$W$+jets cross section as a function of $y({\rm
  first\:jet})$ for events with $\ge 1$~jets. Shown are predictions
  from {\sc Alpgen}, {\sc Sherpa}, and {\sc BlackHat-Sherpa}, and the
  ratio of theoretical predictions to data.  The apparent discrepancy
  between the data and {\sc BlackHat-Sherpa} predictions is described
  in the text.}
  \label{fig:result-xsec-yj}
\end{figure}

\begin{figure}[htb]
  \centering
  \includegraphics[width=0.9\linewidth]{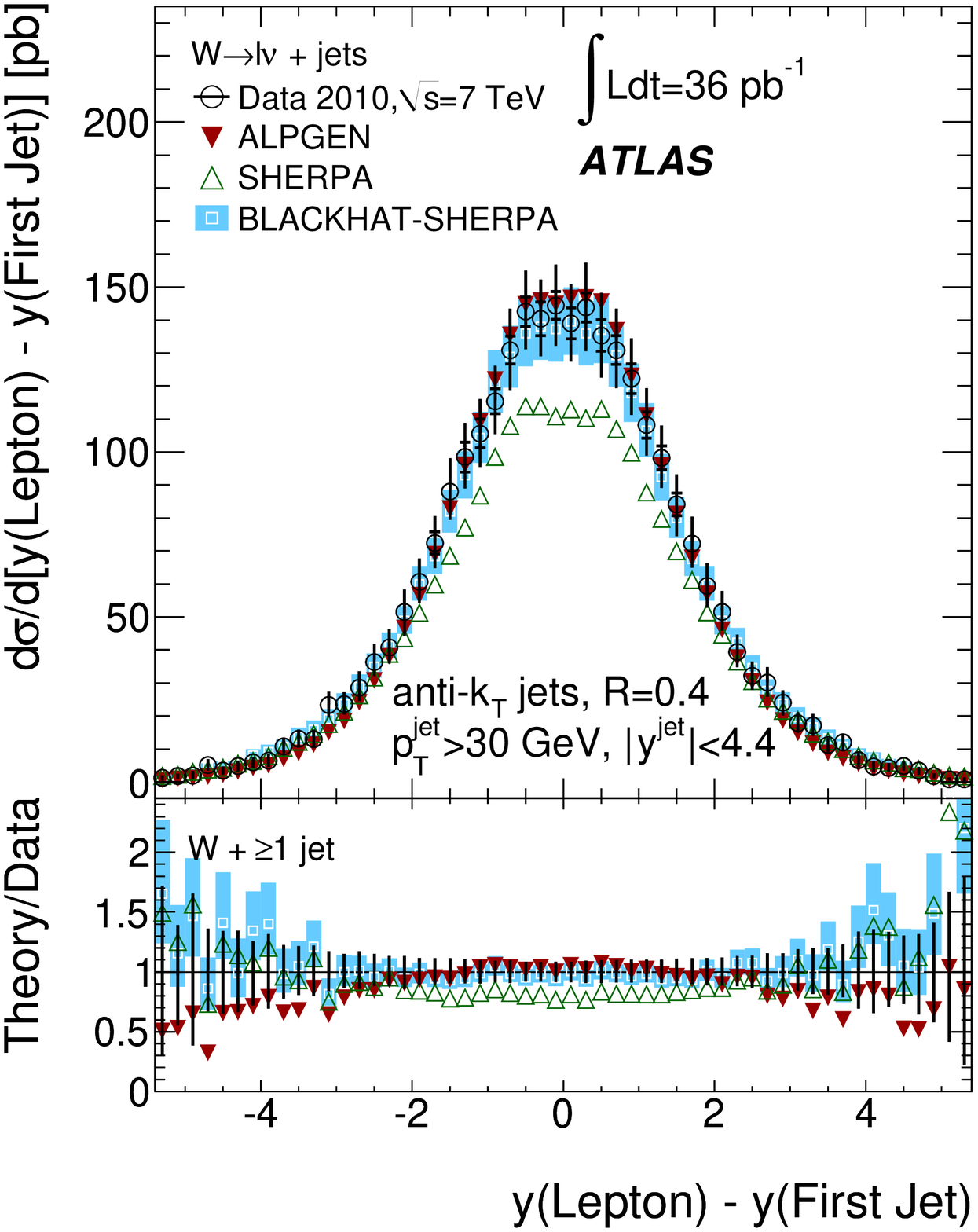}
  \caption{$W$+jets cross section as a function of $y(\ell)\:-\:y({\rm
  first\:jet})$ for events with $\ge 1$~jets. Shown are predictions
  from {\sc Alpgen}, {\sc Sherpa}, and {\sc BlackHat-Sherpa}, and the
  ratio of theoretical predictions to data.}
  \label{fig:result-xsec-dylj}
\end{figure}

\begin{figure}[htb]
  \centering
    \includegraphics[width=0.9\linewidth]{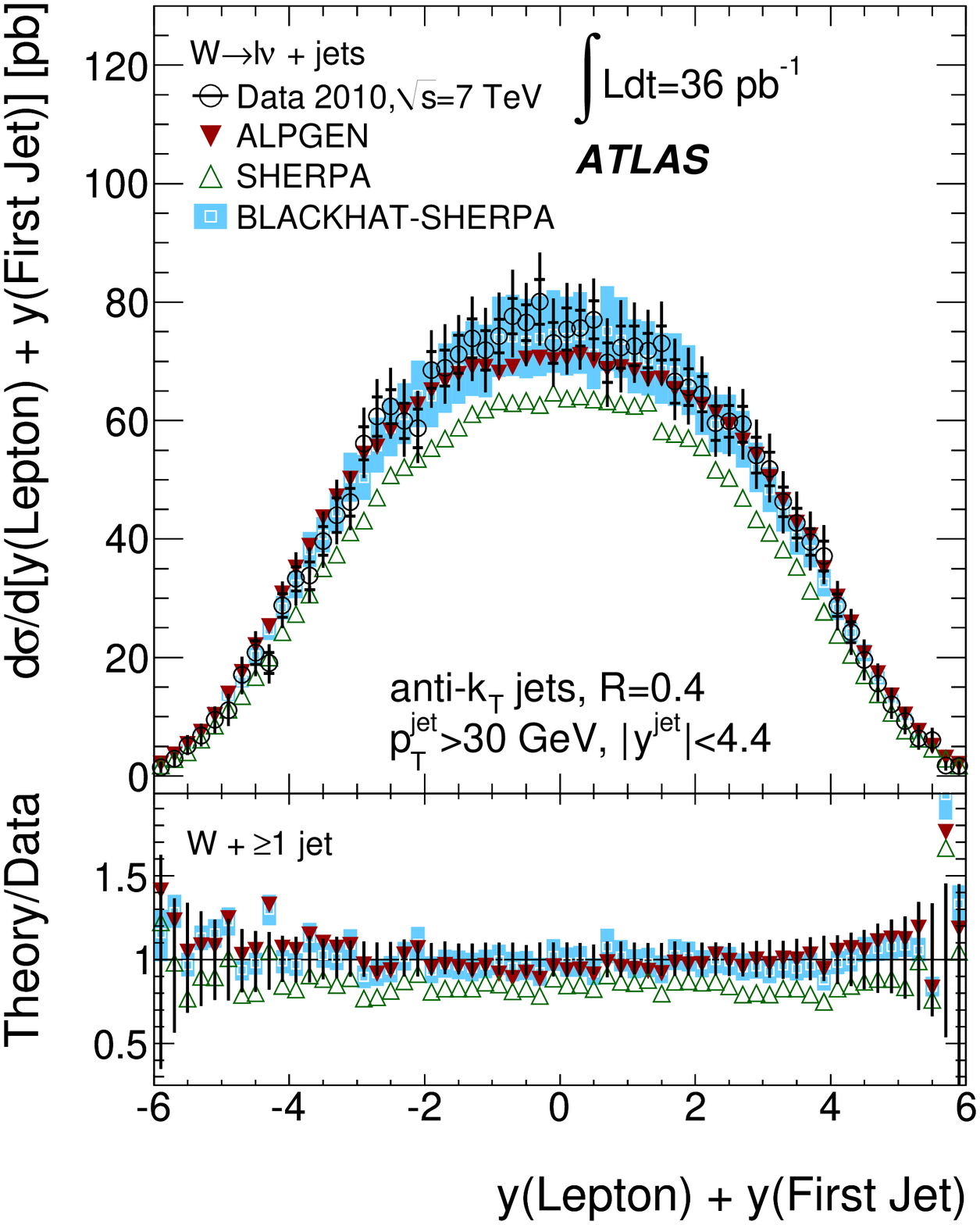}
  \caption{$W$+jets cross section as a function of $y(\ell)\:+\:y({\rm
  first\:jet})$ for events with $\ge 1$~jets. Shown are predictions
  from {\sc Alpgen}, {\sc Sherpa}, and {\sc BlackHat-Sherpa}, and the
  ratio of theoretical predictions to data.}
  \label{fig:result-xsec-sylj}
\end{figure}

\begin{figure}[htb]
  \centering
  \includegraphics[width=0.9\linewidth]{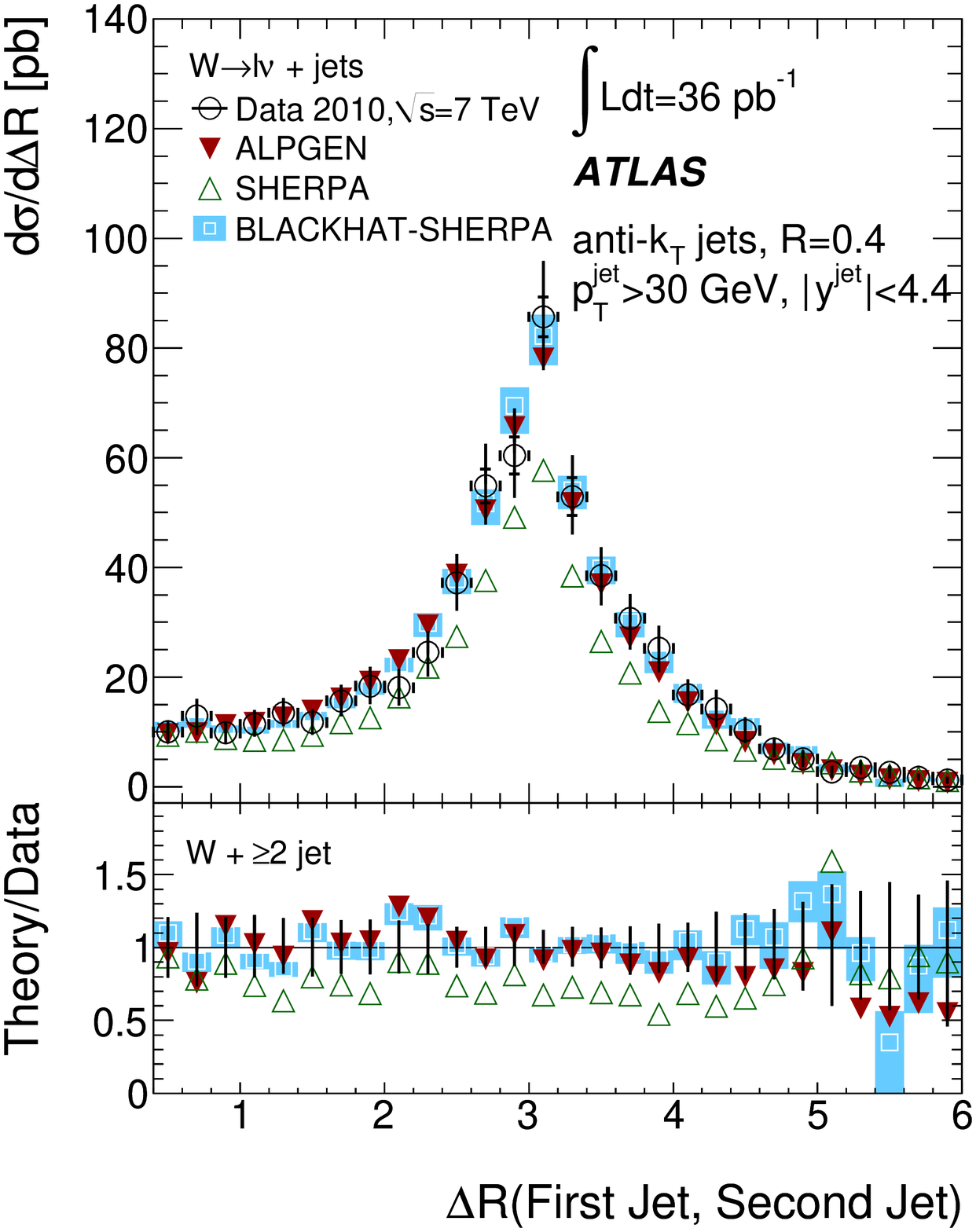}
  \caption{$W$+jets cross section as a function of $\Delta R({\rm
  first\:jet,\:second\:jet})$ for events with $\ge 2$~jets. Shown are
  predictions from {\sc Alpgen}, {\sc Sherpa}, and {\sc
  BlackHat-Sherpa}, and the ratio of theoretical predictions to data.}
  \label{fig:result-xsec-dRjj}
\end{figure}

\begin{figure}[!htb]
  \centering
  \includegraphics[width=0.9\linewidth]{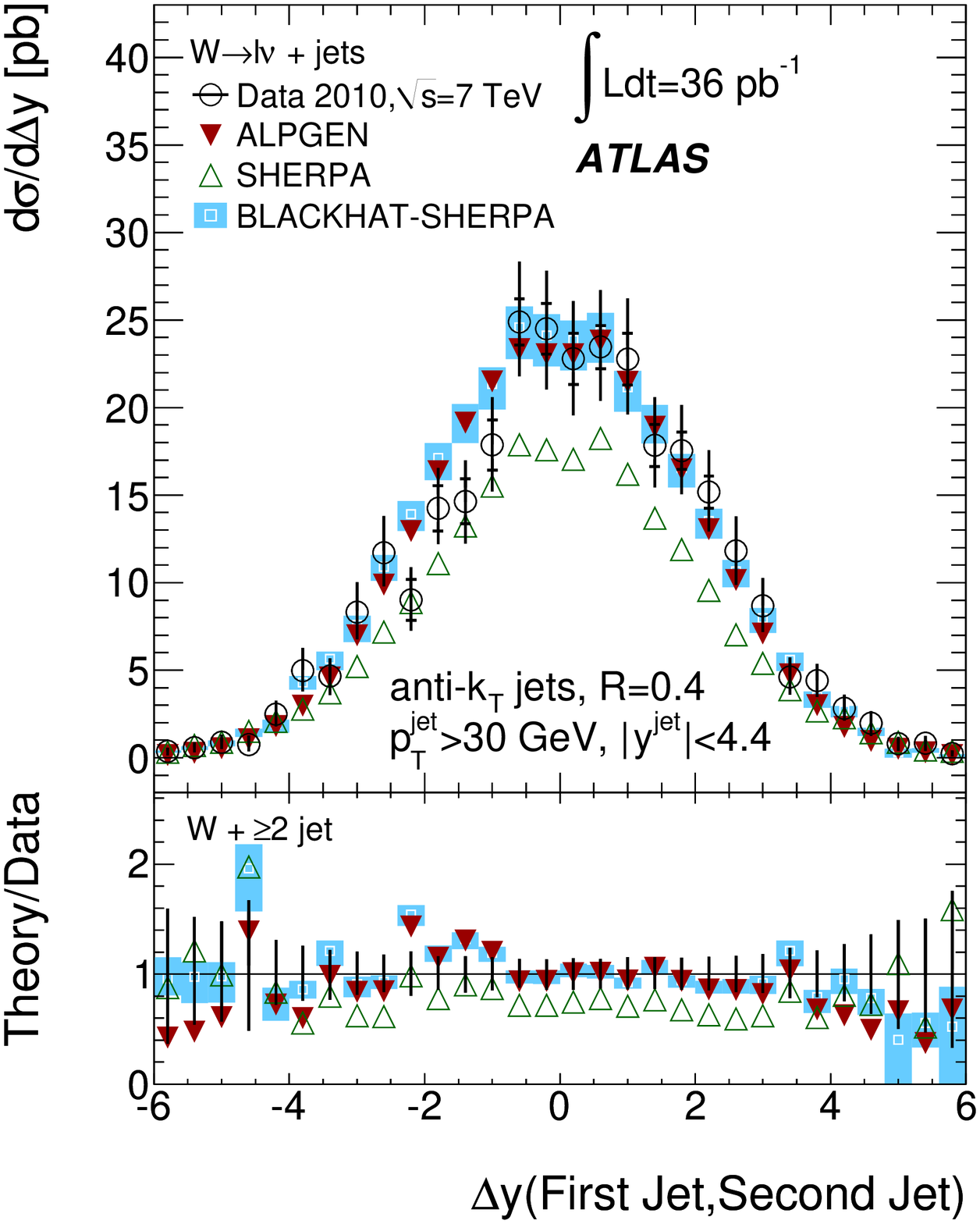}
  \caption{$W$+jets cross section as a function of $y({\rm
  first\:jet})\:-\:y({\rm second\:jet})$ for events with $\ge
  2$~jets. Shown are predictions from {\sc Alpgen}, {\sc Sherpa}, and
  {\sc BlackHat-Sherpa}, and the ratio of theoretical predictions to
  data.}
  \label{fig:result-xsec-dYjj}
\end{figure}

\begin{figure}[!htb]
  \centering
  \includegraphics[width=0.9\linewidth]{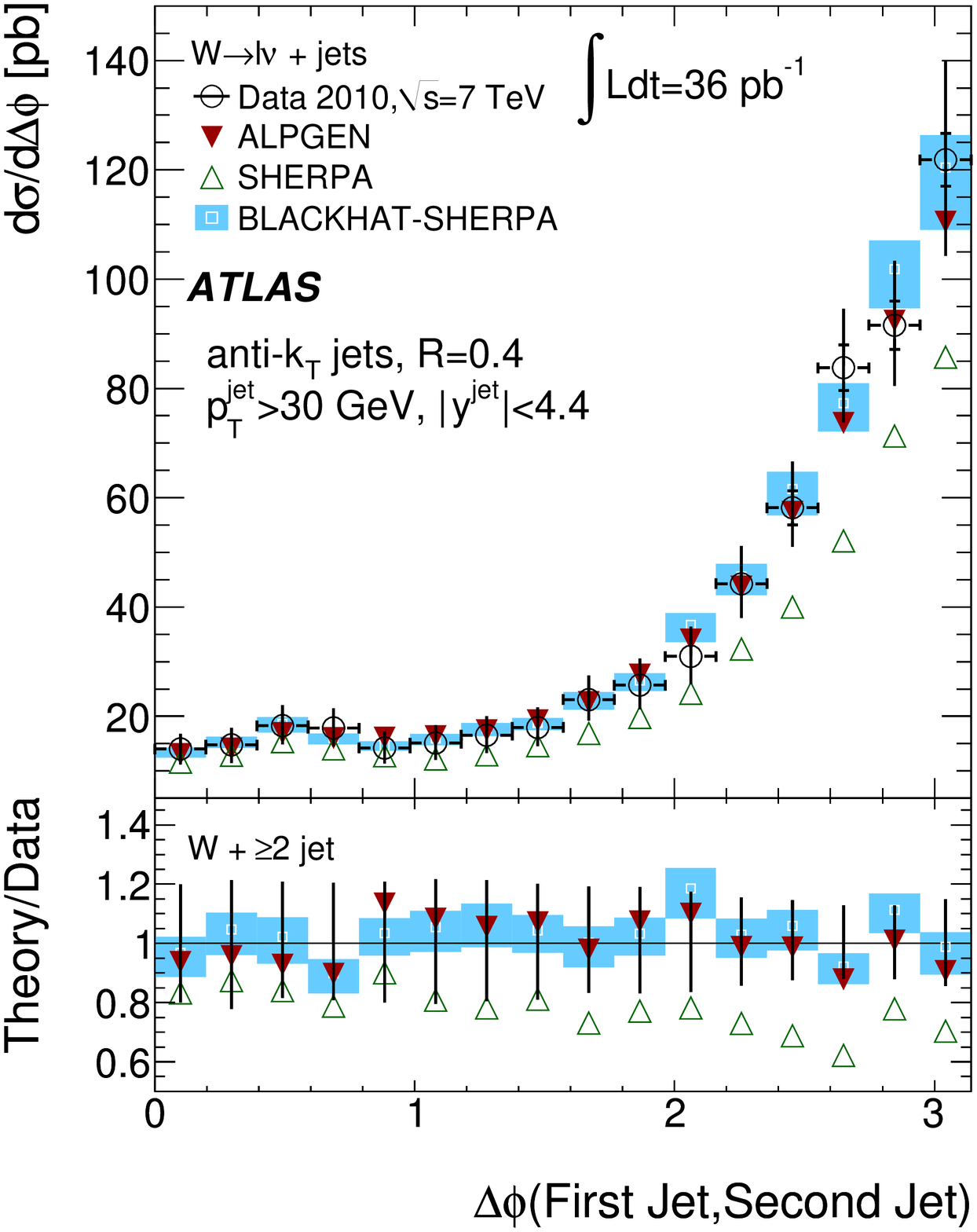}
  \caption{$W$+jets cross section as a function of $\Delta\phi({\rm
  first\:jet,\:second\:jet})$ for events with $\ge 2$~jets. Shown are
  predictions from {\sc Alpgen}, {\sc Sherpa}, and {\sc
  BlackHat-Sherpa}, and the ratio of theoretical predictions to data.}
  \label{fig:result-xsec-dPhijj}
\end{figure}

\section{Cross Section Results}
\label{sec:resuls}

\par The measured $W$+jets cross sections were calculated in the
limited kinematic region defined in Section~\ref{sec:unfold}. All
cross sections were multiplied by the leptonic branching ratio,
$Br(W\rightarrow\ell\nu)$.

\par The cross sections for the \Wen\ and \Wmn\ channels were
calculated separately and then compared. The two sets of cross
sections were found in good agreement within their uncorrelated
uncertainties. The systematic uncertainties specific to the individual
channels were considered fully uncorrelated and the common systematic
uncertainties fully correlated. Results for the electron and muon
channels were combined using three passes of the Best Linear Unbiased
Estimator (BLUE) technique~\cite{BLUE,BLUE1,AIBLUE}. Three iterations
were required to compute the upper systematic uncertainty, the central
value, and the lower systematic uncertainty. The combination improved
uncertainties and fluctuations in the tails of the measured
distributions.

\par Particle level expectations from {\sc Alpgen} and {\sc Sherpa}
simulations as well as a calculation using {\sc BlackHat-Sherpa} were
compared to the measured cross sections. {\sc Pythia} is shown only
for selected distributions that are given as a function of corrected
jet multiplicity. As {\sc Pythia} features LO matrix element accuracy
for events with up to one jet, it does not provide a good description
of the data for jet multiplicities greater than one. The {\sc Alpgen},
{\sc Pythia}, and {\sc Sherpa} predictions were normalized to the NNLO
inclusive $W$-boson production cross section. The version of {\sc
BlackHat-Sherpa} used here provides NLO predictions at parton level
for $W$-boson production with $N_{\rm{jet}} \le 4$.  No additional
normalization was applied to the {\sc BlackHat-Sherpa} predictions.

\par The measured $W$+jets cross sections and the cross section ratios
are shown as a function of the corrected jet multiplicity in
Figs.~\ref{fig:result-xsec} and~\ref{fig:result-xsecratio}. The cross
section is shown as a function of the \pT\ of the first jet for
$N_{\rm{jet}} \ge 1$ to $N_{\rm{jet}} \ge 4$ events separately in
Fig.~\ref{fig:result-xsec-jet1}, the second jet for $N_{\rm{jet}}\ge2$
to $N_{\rm{jet}}\ge4$ events separately in
Fig.~\ref{fig:result-xsec-jet2}, the third jet for $N_{\rm{jet}}\ge3$
and $N_{\rm{jet}} \ge 4$ events separately in
Fig.~\ref{fig:result-xsec-jet3}, the fourth jet for $N_{\rm{jet}}\ge4$
events in Fig.~\ref{fig:result-xsec-jet4}. The jets are ordered from
the highest to lowest \pT. The differential cross section as a
function of \HT\ is shown for $N_{\rm{jet}}\ge1$ to $N_{\rm{jet}}\ge4$
in Fig.~\ref{fig:result-xsec-ht}. Here \HT\ is defined as a scalar sum
over \pt\ of the lepton, neutrino (\MET), and all jets in the
event. \HT\ is often used to set the renormalization and factorization
scales in fixed-order calculations and is therefore an interesting
variable to compare between data and predictions.

The measured \HT\ distribution for events with one or more jets is not
well described by the {\sc BlackHat-Sherpa} prediction. The prediction
is calculated inclusively, at NLO, for events with a $W$ boson and one
or more jets: because of the limited order of the calculation, matrix
elements with three or more real emissions of final-state partons are
not included in the calculation. In contrast, {\sc Alpgen}, where LO
matrix-element terms with up to five final-state partons are utilized,
describes the data well. The data themselves are, as stated above,
inclusive of all higher jet multiplicities. A modified treatment of
{\sc BlackHat-Sherpa} prediction was introduced, where higher-order
NLO terms with two, three, and four real emissions were also added to
the $N_{\rm jet}\geq1$ distribution: this is shown in
Fig.~\ref{fig:result-xsec-ht-compar}. The higher-order terms were
combined by matching them exclusively in jet multiplicity by counting
parton jets with \pt~$>$~30~\GeV. The matching scheme is required to
reduce double-counting of cross sections. This case illustrates the
challenges of comparing NLO calculations to complex inclusive jet
variables like \HT. In Fig.~\ref{fig:result-xsec-mj} the cross
sections are shown as a function of the invariant mass, $m({\rm
jets})$, of the first two, three, and four jets for events with
$N_{\rm{jet}}\ge2$, $N_{\rm{jet}}\ge3$, and $N_{\rm{jet}}\ge4$,
respectively. The invariant mass of the multijet system is also
considered for the renormalization and factorization scales in
fixed-order pQCD calculations. Overall, these distributions constitute
a set of tests for factorization and renormalization scales used in
calculations of $\alpha_s$; the {\sc Alpgen} samples demonstrate a
better agreement with data than {\sc Sherpa} due to differences in the
scales and PDFs described in Section~\ref{sec:mc}.

\par Distributions dependent on rapidities of the leptons and the
first jet are shown in Figs.~\ref{fig:result-xsec-yj},
\ref{fig:result-xsec-dylj}, and~\ref{fig:result-xsec-sylj} for $y({\rm
first\:jet})$, $y(\ell)\:-\:y({\rm first\:jet})$, and
$y(\ell)\:+\:y({\rm first\:jet})$, respectively. These distributions
are sensitive to PDFs used for calculations of LO and NLO matrix
elements. Predictions from {\sc BlackHat-Sherpa} and {\sc Sherpa} were
produced with CTEQ6.6M, a NLO PDF, while {\sc Alpgen} used CTEQ6L1, a
LO PDF. The shape of the distributions from {\sc Sherpa} were found to
be similar to {\sc BlackHat-Sherpa}. {\sc Alpgen} gave a different
description of the $y(\ell)\:-\:y({\rm first\:jet})$ distribution. The
deviations observed between the data and {\sc BlackHat-Sherpa} at high
jet rapidities in Fig.~\ref{fig:result-xsec-yj} may be caused by
insufficient knowledge of the gluon PDFs at high $x$.

\par Lastly, distances between the first two jets are explored in
Figs.~\ref{fig:result-xsec-dRjj}, \ref{fig:result-xsec-dYjj},
and~\ref{fig:result-xsec-dPhijj} by defining the distance as $\Delta
R({\rm first\:jet,\:second\:jet})$, $y({\rm first\:jet})\:-\:y({\rm
second\:jet})$, and $\Delta\phi({\rm first\:jet,\:second\:jet})$,
respectively. This set of measurements offers a test of hard parton
radiation at large angles and of matrix element to parton shower
matching schemes. The majority of jets are modeled via the ME
calculation for the jet pairs with large angular separation, when
$\Delta R$ and $\Delta\phi$ are close to $\pi$. Collinear radiation at
small angular separation, when $\Delta R$ is small, is produced mainly
via the parton shower. Overall, {\sc Alpgen} and {\sc BlackHat-Sherpa}
demonstrate good agreement with the data while {\sc Sherpa} deviates
due to the differences in PDFs, $\alpha_s$, and factorization scales.

\par All distributions were also produced with the selection
requirement on $\pt^{\rm jet}$ reduced from 30~\GeV\ to 20~\GeV. The
results for the softer threshold are given in
Appendix~\ref{app:xsec_20GeV}. The softer threshold makes the cross
sections more sensitive to the non-pQCD and experimental effects,
especially for forward jets.

\par All these cross sections accompanied by the non-pQCD and QED
corrections are available in HEPDATA.

\section{Conclusions}

\par This paper presents a measurement of the $W$+jets cross section
as a function of jet multiplicity in $pp$ collisions at
$\sqrt{s}=7$~\TeV\ in both electron and muon decay modes of the $W$
boson, based on an integrated luminosity of 36~\ipb. The ratios of
cross sections $\sigma(W+\ge N_{\rm{jet}})/\sigma(W+\ge
N_{\rm{jet}}-1)$ have been calculated for inclusive jet
multiplicities, $N_{\rm{jet}}$, that range between $1-4$ for the
$\pt^{\rm jet}$~$>$~30~\GeV\ jet threshold and between $1-5$ for the
$\pt^{\rm jet}$~$>$~20~\GeV\ threshold. Measurements are also
presented of the \pT\ distribution of the first through fourth jets in
the event, of the invariant masses of two or more jets, of the
distances between the lepton and the first jet, of the distances
between the first two jets, and of the \HT\ distribution. The results
have been corrected for all detector effects and are quoted in an
ATLAS-specific range of jet and lepton kinematics. This range is
almost fully covered by the detector acceptance, so as to avoid
model-dependent extrapolations and to facilitate the comparison with
theoretical predictions. Good agreement is observed between the
predictions from the multi-parton matrix element generator {\sc
Alpgen} and the measured distributions. At the same time, {\sc Sherpa}
demonstrates a slightly worse agreement with the experimental results
than {\sc Alpgen}.  The paper features the first comparison between
the NLO predictions and the LHC data for events with a $W$ boson and
four jets. Calculations based on NLO matrix elements in MCFM
(available for jet multiplicities $N_{\rm{jet}}\le 2$) and in {\sc
BlackHat-Sherpa} (available for jet multiplicities $N_{\rm{jet}} \le
4$) are generally in good agreement with the data; deviations are
observed in the ${\rm d}\sigma(W+\geq{\rm\:jet})/{\rm d}\HT$
distribution at large \HT\ and in the tails of ${\rm d}\sigma/{\rm
d}y({\rm jet})$ and ${\rm d}\sigma/{\rm d}(y(\ell)-y({\rm jet}))$
distributions.

\begin{acknowledgments}
\par We are grateful to the {\sc BlackHat-Sherpa} collaboration and
Daniel Maitre for all their help.

\par We thank CERN for the very successful operation of the LHC, as
well as the support staff from our institutions without whom ATLAS
could not be operated efficiently.

\par We acknowledge the support of ANPCyT, Argentina; YerPhI, Armenia;
ARC, Australia; BMWF, Austria; ANAS, Azerbaijan; SSTC, Belarus; CNPq
and FAPESP, Brazil; NSERC, NRC and CFI, Canada; CERN; CONICYT, Chile;
CAS, MOST and NSFC, China; COLCIENCIAS, Colombia; MSMT CR, MPO CR and
VSC CR, Czech Republic; DNRF, DNSRC and Lundbeck Foundation, Denmark;
ARTEMIS, European Union; IN2P3-CNRS, CEA-DSM/IRFU, France; GNAS,
Georgia; BMBF, DFG, HGF, MPG and AvH Foundation, Germany; GSRT,
Greece; ISF, MINERVA, GIF, DIP and Benoziyo Center, Israel; INFN,
Italy; MEXT and JSPS, Japan; CNRST, Morocco; FOM and NWO, Netherlands;
RCN, Norway; MNiSW, Poland; GRICES and FCT, Portugal; MERYS (MECTS),
Romania; MES of Russia and ROSATOM, Russian Federation; JINR; MSTD,
Serbia; MSSR, Slovakia; ARRS and MVZT, Slovenia; DST/NRF, South
Africa; MICINN, Spain; SRC and Wallenberg Foundation, Sweden; SER,
SNSF and Cantons of Bern and Geneva, Switzerland; NSC, Taiwan; TAEK,
Turkey; STFC, the Royal Society and Leverhulme Trust, United Kingdom;
DOE and NSF, United States of America.

\par The crucial computing support from all WLCG partners is
acknowledged gratefully, in particular from CERN and the ATLAS Tier-1
facilities at TRIUMF (Canada), NDGF (Denmark, Norway, Sweden),
CC-IN2P3 (France), KIT/GridKA (Germany), INFN-CNAF (Italy), NL-T1
(Netherlands), PIC (Spain), ASGC (Taiwan), RAL (UK) and BNL (USA) and
in the Tier-2 facilities worldwide.
\end{acknowledgments}

\bibliography{wjets_pub}

\clearpage
\appendix
\section{Results for a jet threshold of $\pt > 20$ GeV}
\label{app:xsec_20GeV}

\par Here we present results for jets selected with a 20~\GeV\
threshold in \pt. The distributions are the same variables as for jets
with the 30~\GeV\ threshold shown in Section~\ref{sec:resuls} except
that the data with $N_{\rm{jet}}\geq5$ was used for physics
conclusions; the 20~\GeV\ threshold improved the signal-to-background
ratio and event count. The softer threshold makes the cross sections
more sensitive to the non-pQCD and experimental effects such as the
underlying event model, multiple parton interactions, parton
fragmentation, hadronization, and pile-up $pp$ interactions. The
corrections accounting for the non-pQCD effects, that were applied to
{\sc BlackHat-Sherpa} calculations, increased monotonically with the
absolute value of jet rapidity from $\sim$1.0 up to $\sim$2.4. The
uncertainties on the corrections are also larger in the forward
region.

\begin{figure}[htb]
  \centering
  \includegraphics[width=0.9\linewidth]{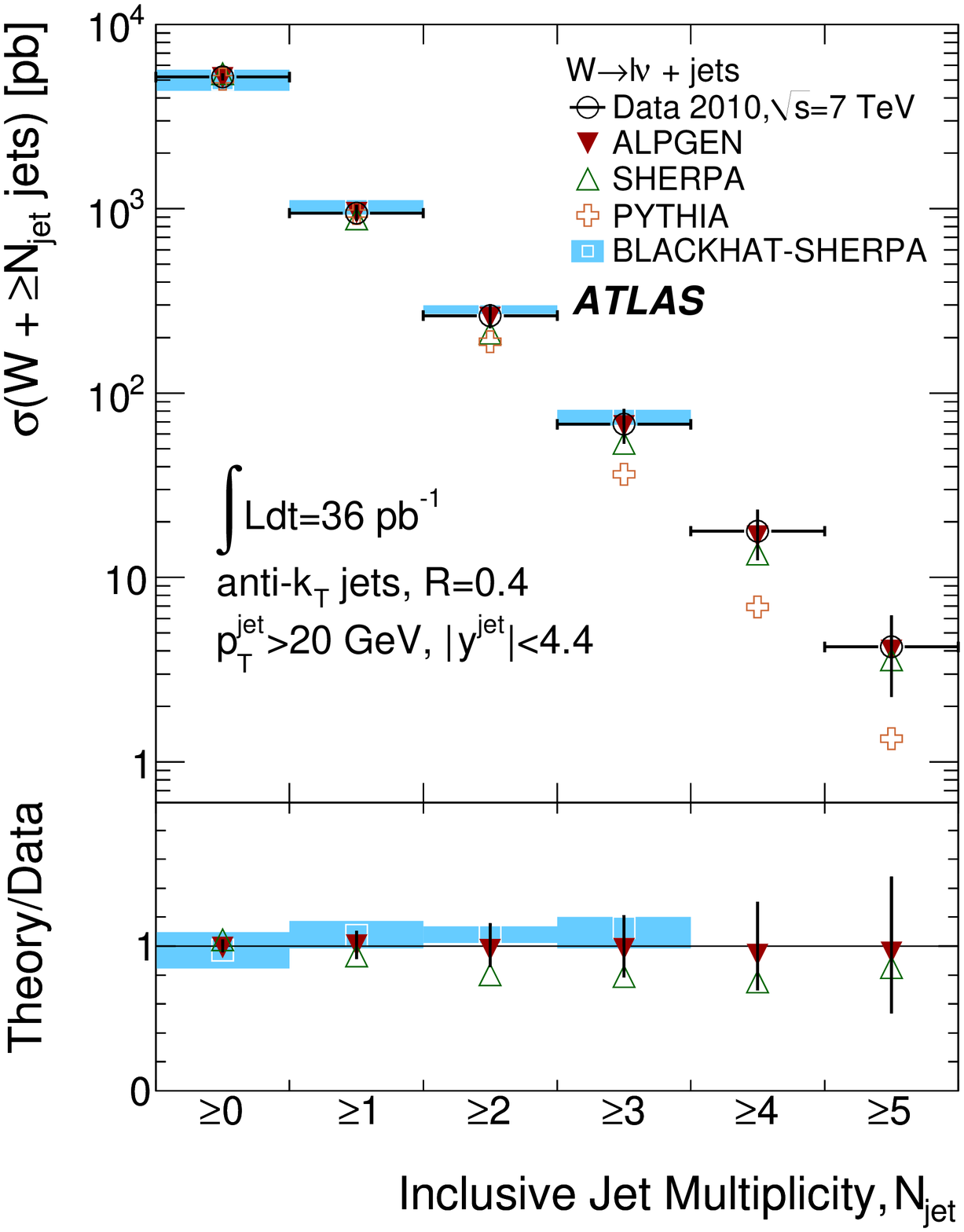}
  \caption{$W$+jets cross section results as a function of corrected
  jet multiplicity.  The following remarks apply to this and
  subsequent figures unless specific comments are given. The cross
  sections are quoted in the restricted kinematic region described in
  Section~\ref{sec:unfold}, except here
  $\pt^{\rm{jet}}$~$>$~20~\GeV. For the data, the statistical
  uncertainties are shown with a tick on the vertical bars, and the
  combined statistical and systematic uncertainties are shown with the
  full error bar. Also shown are predictions from {\sc Alpgen}, {\sc
  Sherpa}, {\sc Pythia} and {\sc BlackHat-Sherpa}, and the ratio of
  theoretical predictions to data ({\sc Pythia} is not shown in the
  ratio). The distributions from {\sc Sherpa}, {\sc Pythia} and {\sc
  Alpgen} were normalized to the NNLO total $W$-boson production cross
  section.}
  \label{fig20:result-xsec}
\end{figure}
\begin{figure}[htb]
  \centering
  \includegraphics[width=0.9\linewidth]{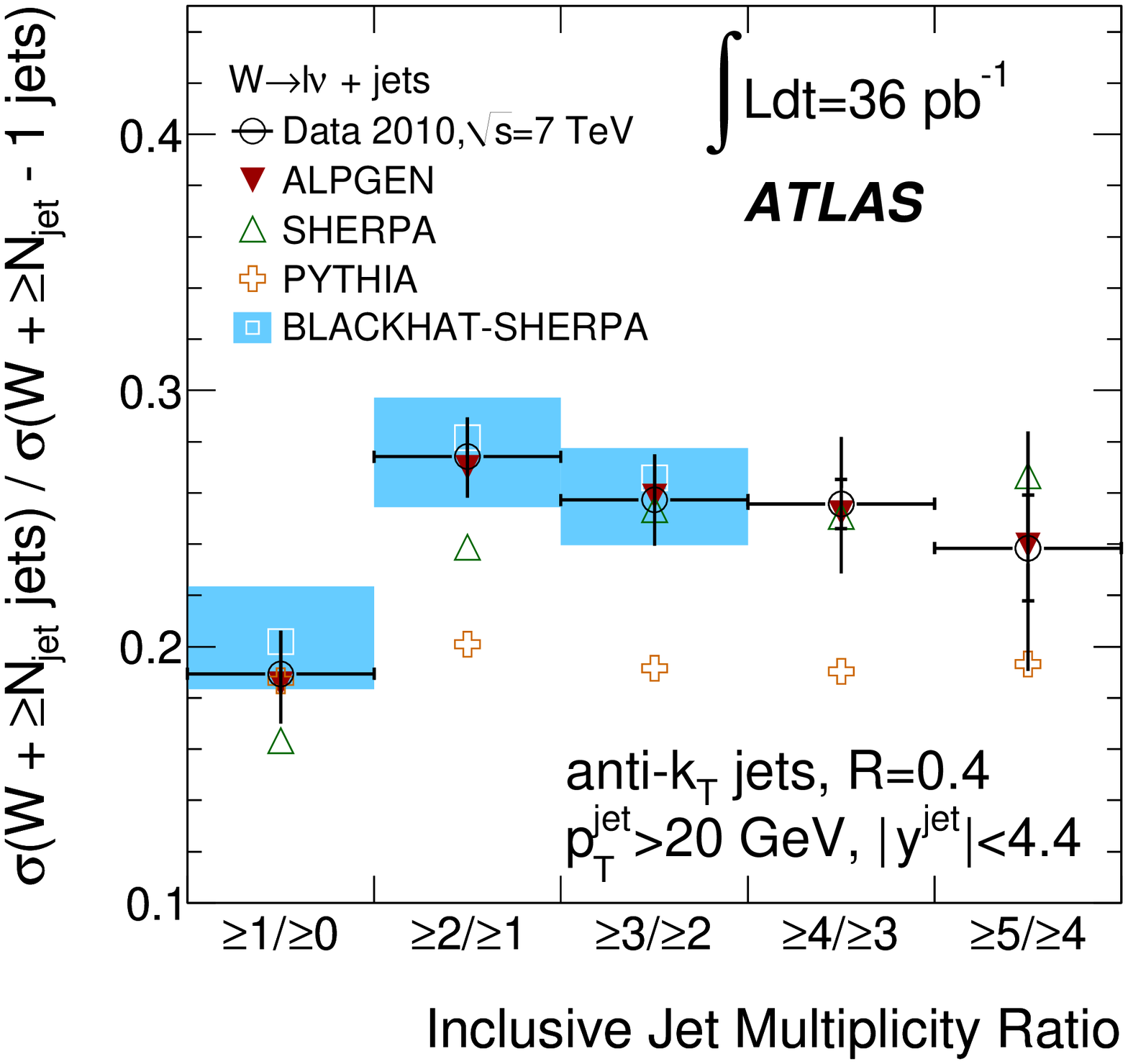}
  \caption{$W$+jets cross section ratio results as a function of
  corrected jet multiplicity.}
\label{fig20:result-xsecratio}
\end{figure}
\begin{figure}[htb]
  \centering
  \includegraphics[width=0.9\linewidth]{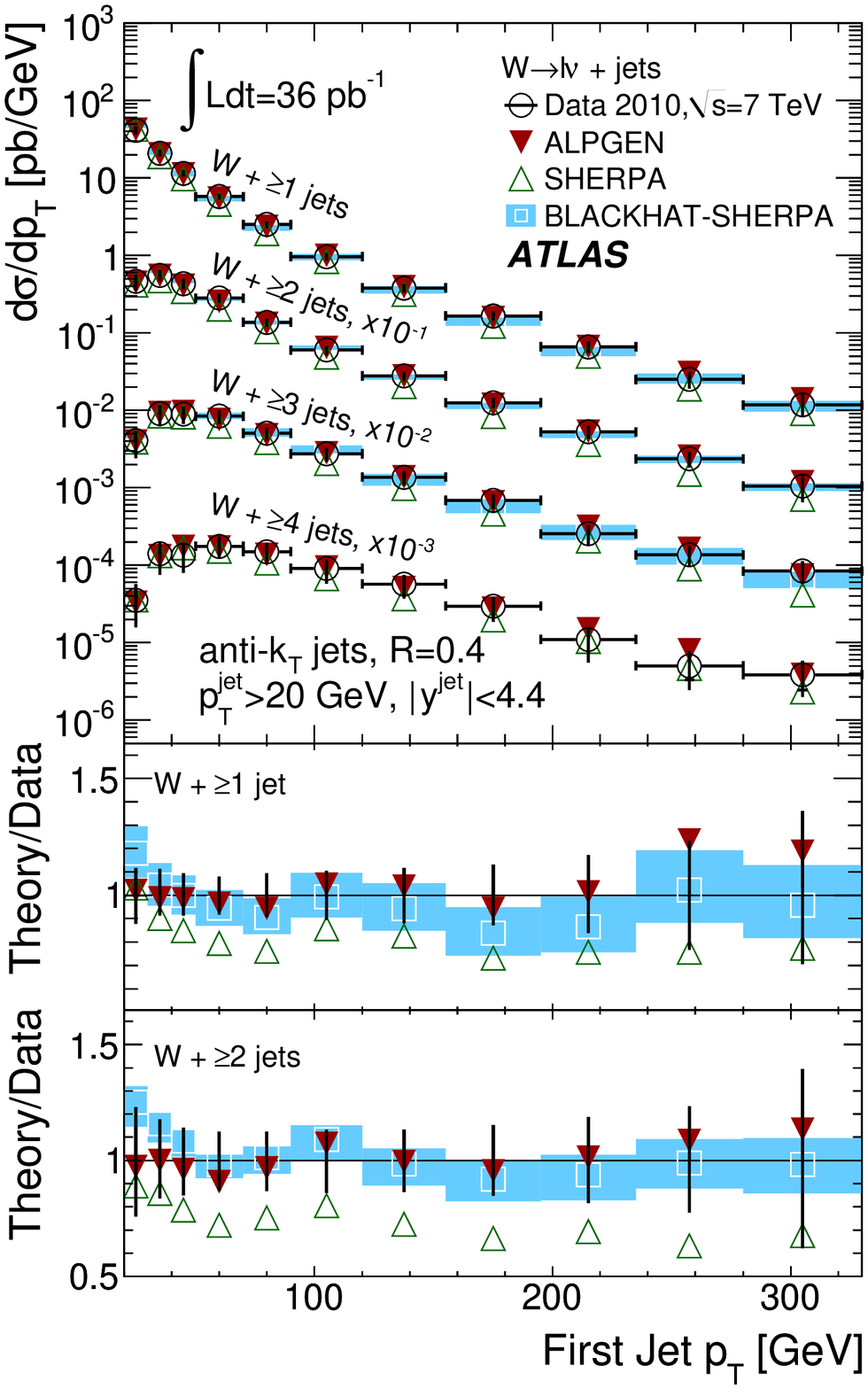}
  \caption{$W$+jets cross section as a function of the \pT\ of the
  first jet in the event. The \pT\ of the first jet is shown
  separately for events with $\ge1$~jet to $\ge4$~jet. The $\ge2$~jet,
  $\ge3$~jet, and $\ge4$~jet distributions have been scaled down by
  factors of 10, 100, and 1000 respectively. Shown are predictions
  from {\sc Alpgen}, {\sc Sherpa}, and {\sc BlackHat-Sherpa}, and the
  ratio of theoretical predictions to data for $\ge 1$~jet and $\ge
  2$~jet events.}
  \label{fig20:result-xsec-jet1}
\end{figure}
\begin{figure}[htb]
  \centering
  \includegraphics[width=0.9\linewidth]{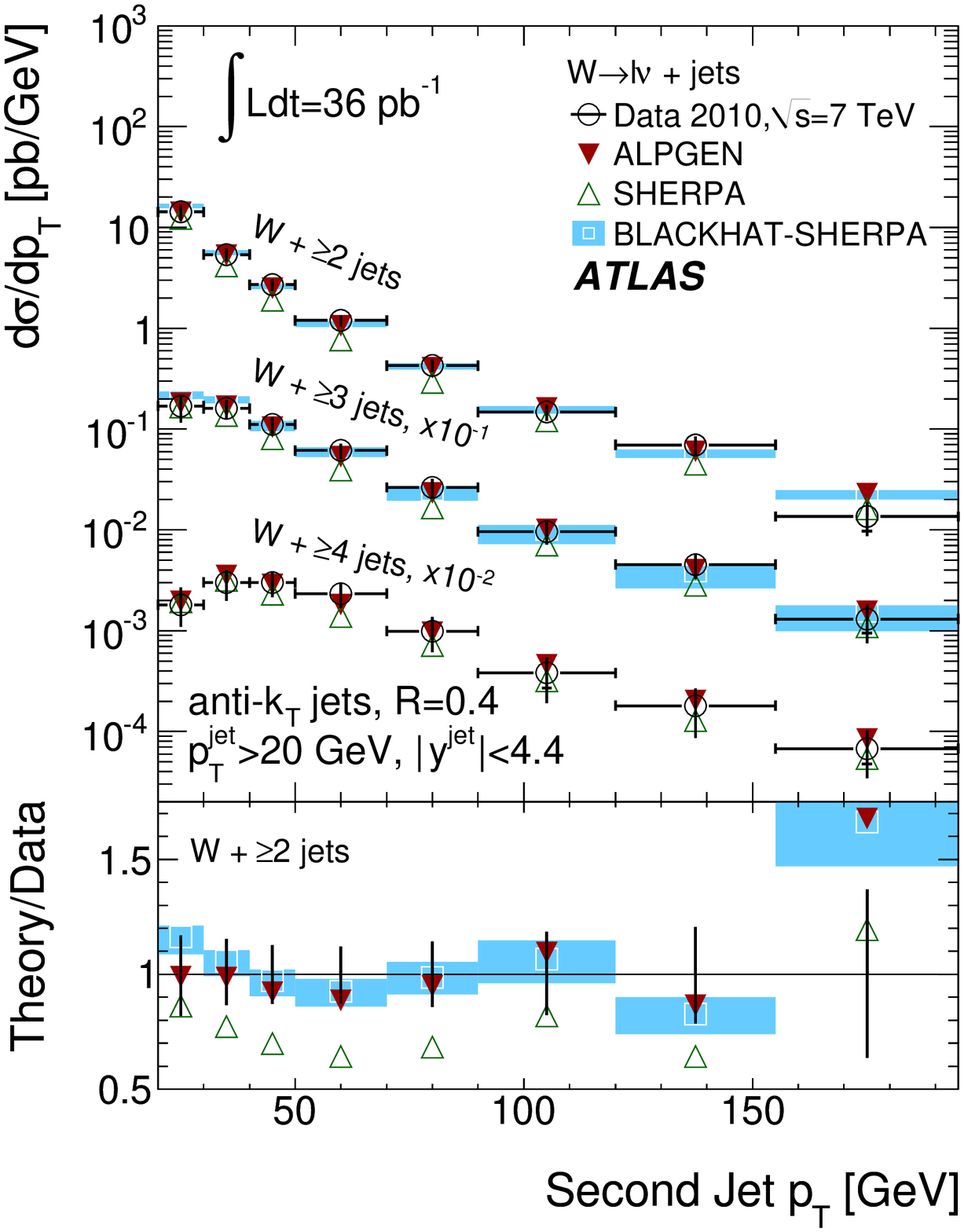}
  \caption{$W$+jets cross section as a function of the \pT\ of the
  second jet in the event.  The \pT\ of the second jet is shown
  separately for events with $\ge2$~jet to $\ge4$~jet. The $\ge3$~jet
  and $\ge4$~jet distributions have been scaled down by factors of 10
  and 100 respectively. Shown are predictions from {\sc Alpgen}, {\sc
  Sherpa}, and {\sc BlackHat-Sherpa}, and the ratio of theoretical
  predictions to data for $\ge2$~jet events.}
  \label{fig20:result-xsec-jet2}
\end{figure}
\begin{figure}[htb]
  \centering
  \includegraphics[width=0.9\linewidth]{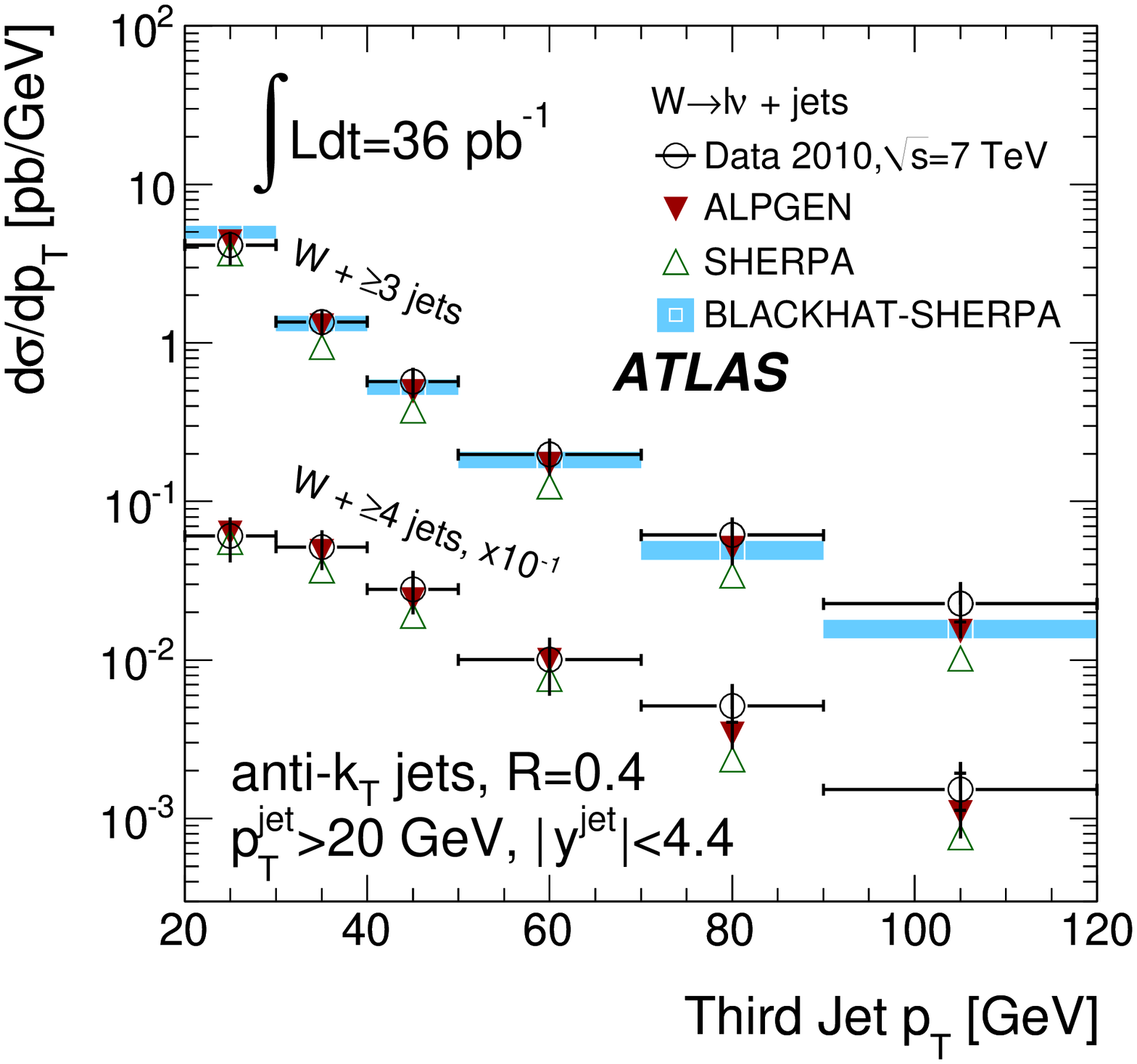}
  \caption{$W$+jets cross section as a function of the \pT\ of the
  third jet in the event. The \pT\ of the third jet is shown
  separately for events with $\ge3$~jet and $\ge4$~jet. The $\ge4$~jet
  distribution has been scaled down by a factor of 10. Shown are
  predictions from {\sc Alpgen}, {\sc Sherpa}, and {\sc
  BlackHat-Sherpa}.}
  \label{fig20:result-xsec-jet3}
\end{figure}
\begin{figure}[htb]
  \centering
  \includegraphics[width=0.9\linewidth]{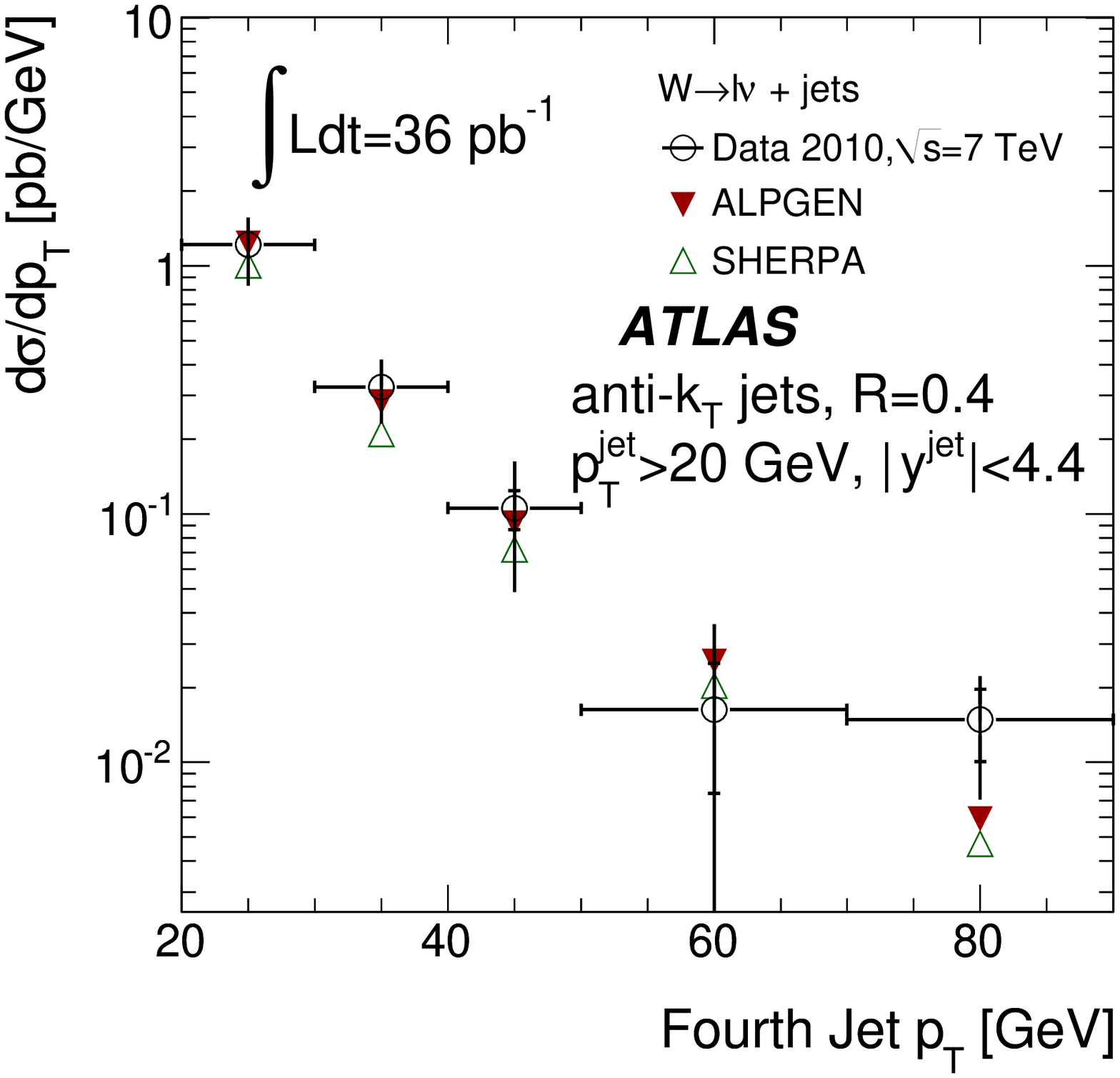}
  \caption{$W$+jets cross section as a function of the \pT\ of the
  fourth jet in the event. The distributions are for events with $\ge
  4$~jet. Shown are predictions from {\sc Alpgen} and {\sc Sherpa}.}
  \label{fig20:result-xsec-jet4}
\end{figure}
\begin{figure}[htb]
  \centering
  \includegraphics[width=0.9\linewidth]{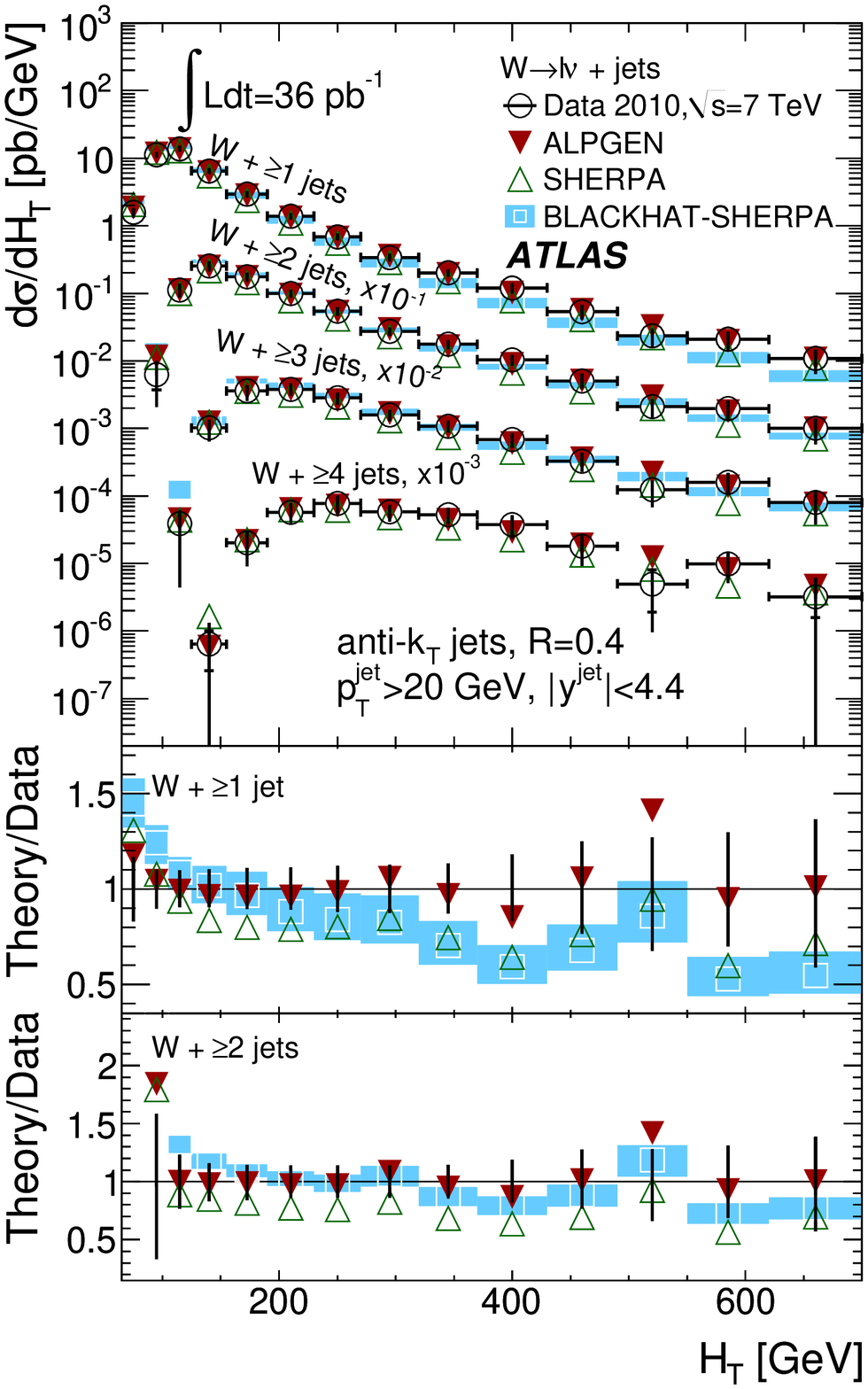}
  \caption{$W$+jets cross section as a function of \HT, shown
  separately for $\ge 1$~jets to $\ge 4$~jets. The $\ge 2$~jet, $\ge
  3$~jet, and $\ge 4$~jet distributions have been scaled down by
  factors of 10, 100, and 1000 respectively. Shown are predictions
  from {\sc Alpgen}, {\sc Sherpa}, and {\sc BlackHat-Sherpa}, and the
  ratio of theoretical predictions to data for $\ge 1$~jet and $\ge
  2$~jet events.}
  \label{fig20:result-xsec-ht}
\end{figure}
\begin{figure}[htb]
  \centering
  \includegraphics[width=0.9\linewidth]{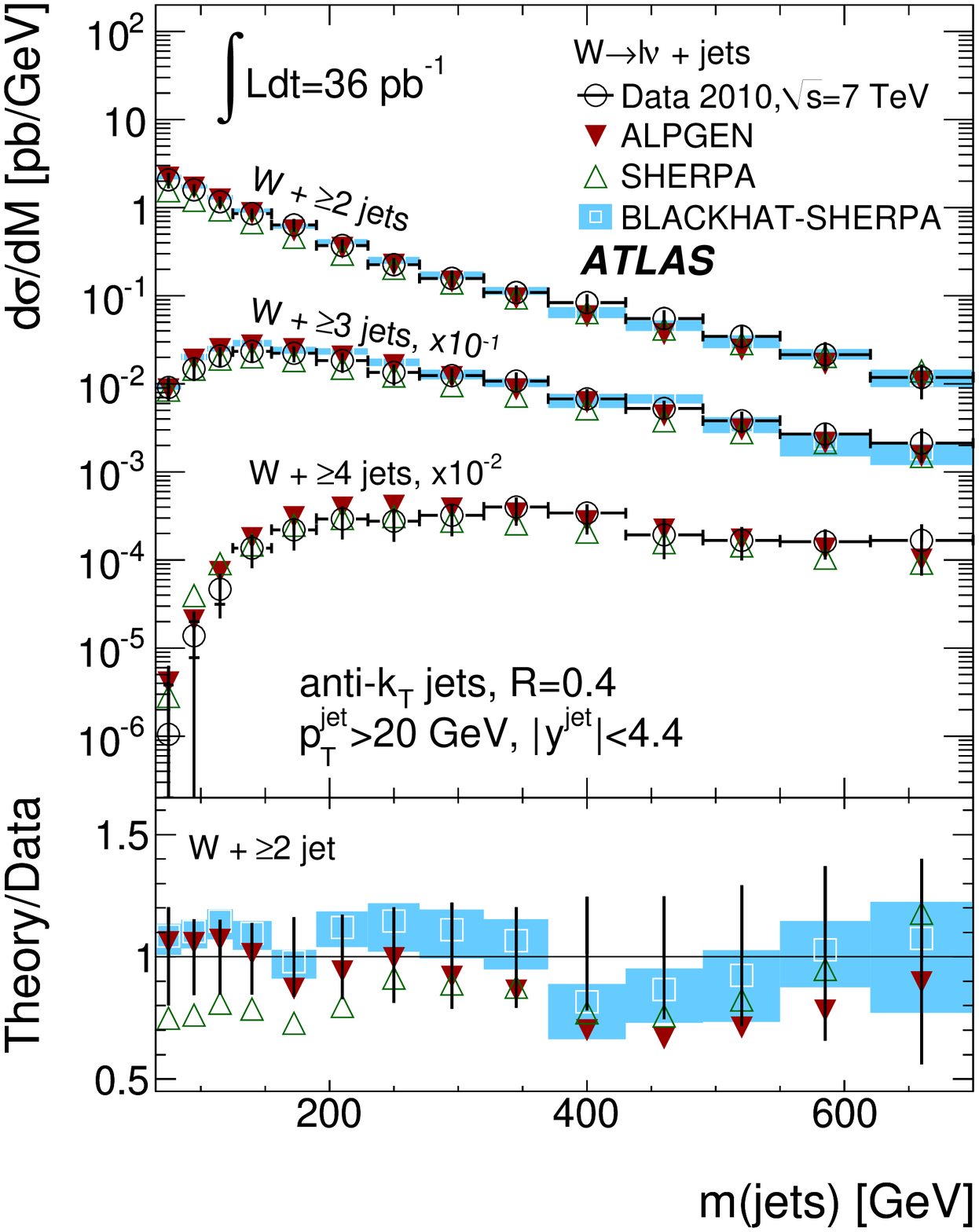}
  \caption{$W$+jets cross section as a function of $m({\rm jets})$,
  shown separately for $\ge 2$~jets to $\ge 4$~jets. The $\ge 3$~jet
  and $\ge 4$~jet distributions have been scaled down by factors of 10
  and 100, respectively. Shown are predictions from {\sc Alpgen}, {\sc
  Sherpa}, and {\sc BlackHat-Sherpa}, and the ratio of theoretical
  predictions to data for $\ge 2$~jet events.}
  \label{fig20:result-xsec-mj}
\end{figure}
\begin{figure}[htb]
  \centering
  \includegraphics[width=0.9\linewidth]{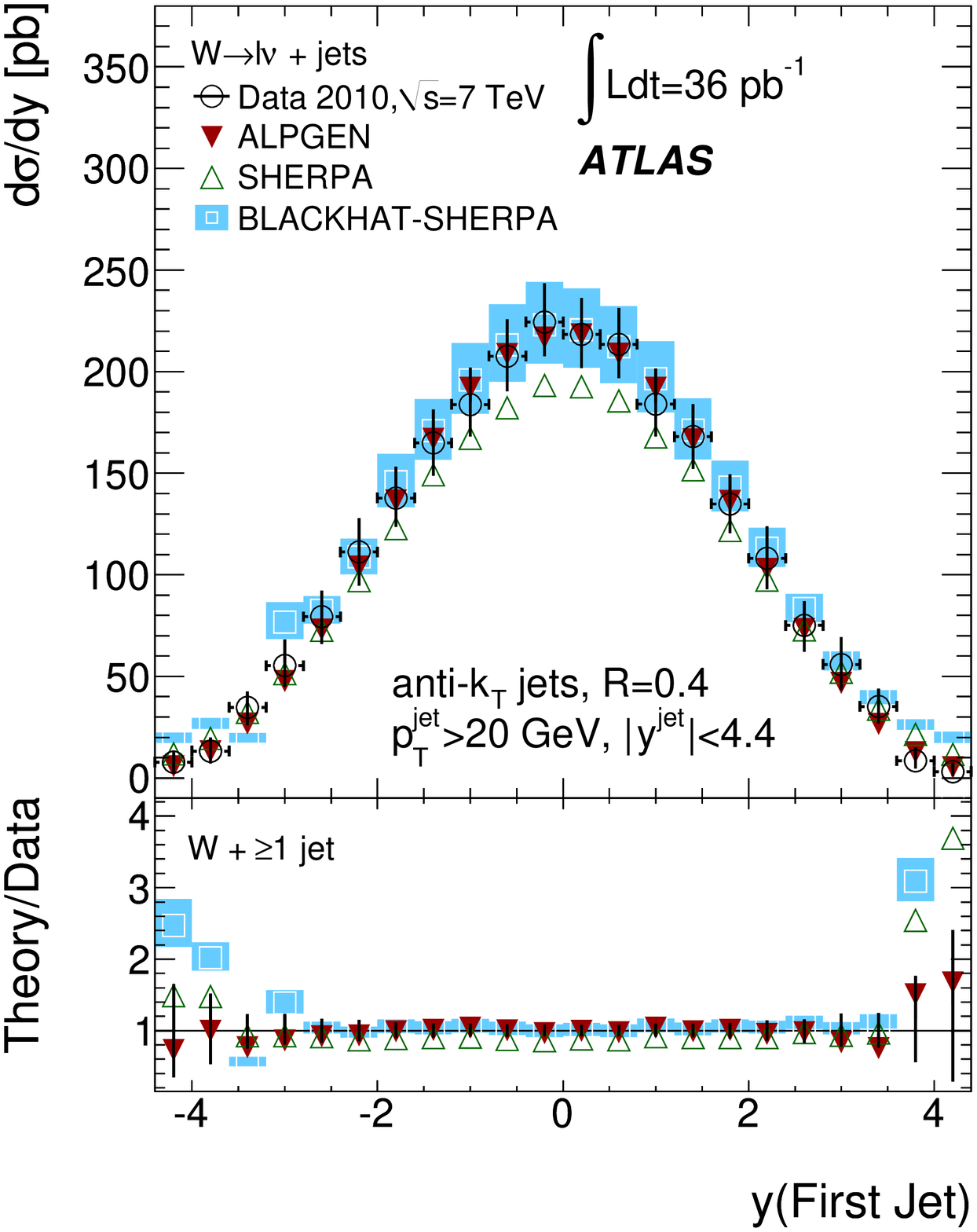}
  \caption{$W$+jets cross section as a function of $y({\rm
  first\:jet})$ for events with $\ge 1$~jets. Shown are predictions
  from {\sc Alpgen}, {\sc Sherpa}, and {\sc BlackHat-Sherpa}, and the
  ratio of theoretical predictions to data.}
  \label{fig20:result-xsec-yj}
\end{figure}
\begin{figure}[htb]
  \centering
  \includegraphics[width=0.9\linewidth]{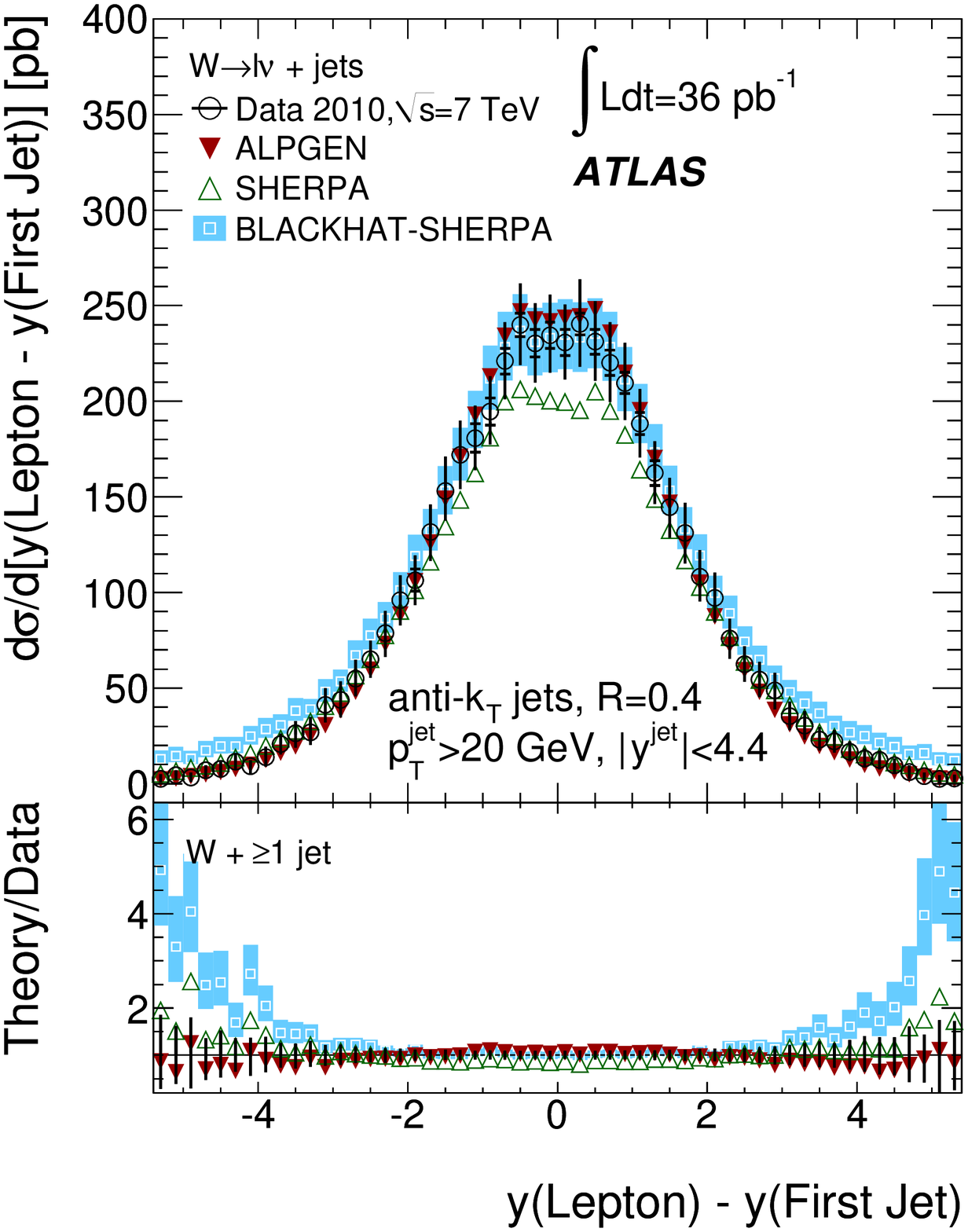}
  \caption{$W$+jets cross section as a function of $y(\ell)\:-\:y({\rm
  first\:jet})$ for events with $\ge 1$~jets. Shown are predictions
  from {\sc Alpgen}, {\sc Sherpa}, and {\sc BlackHat-Sherpa}, and the
  ratio of theoretical predictions to data.}
  \label{fig20:result-xsec-dylj}
\end{figure}
\begin{figure}[htb]
  \centering
  \includegraphics[width=0.9\linewidth]{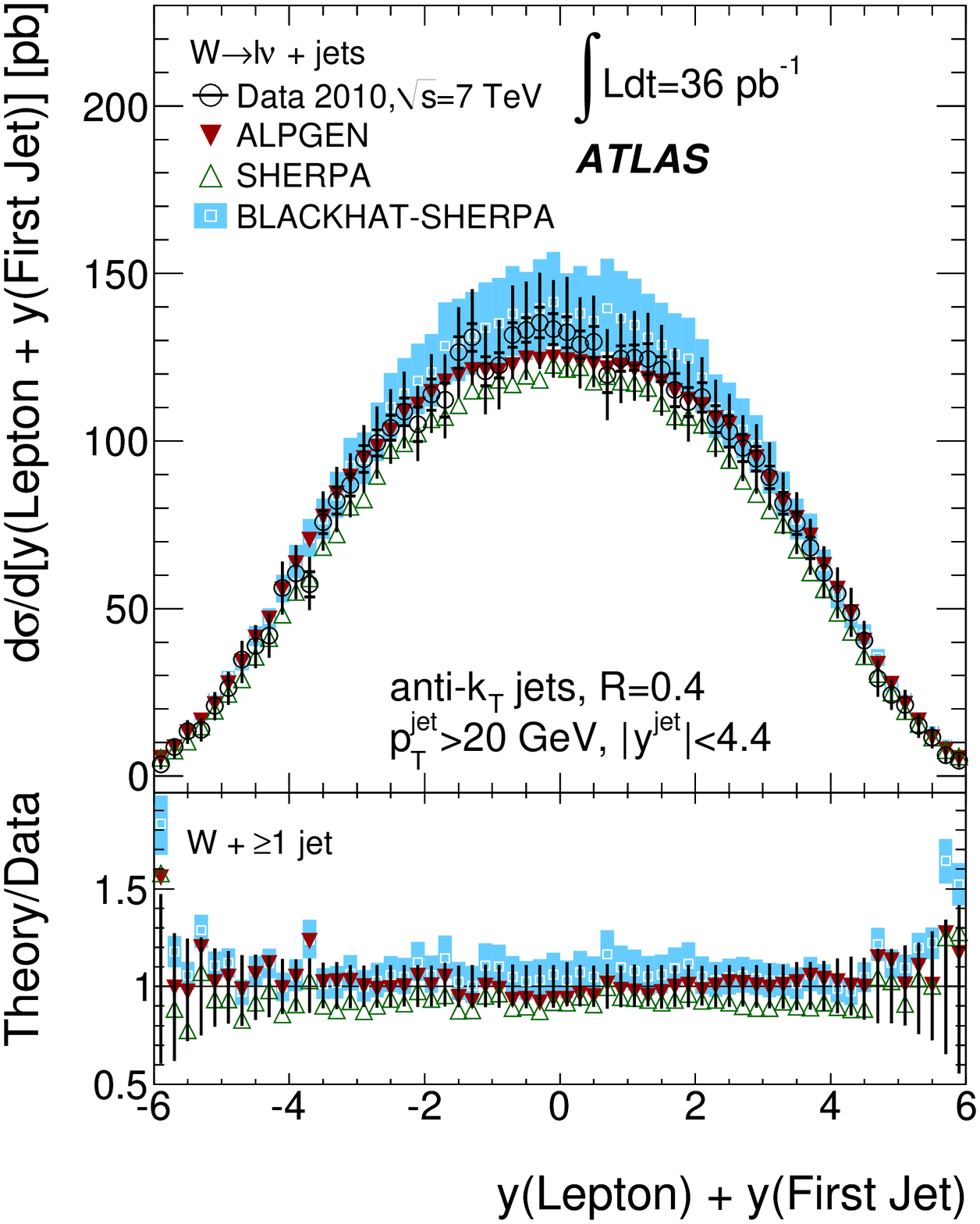}
  \caption{$W$+jets cross section as a function of $y(\ell)\:+\:y({\rm
  first\:jet})$ for events with $\ge 1$~jets. Shown are predictions
  from {\sc Alpgen}, {\sc Sherpa}, and {\sc BlackHat-Sherpa}, and the
  ratio of theoretical predictions to data.}
  \label{fig20:result-xsec-sylj}
\end{figure}
\begin{figure}[htb]
  \centering
  \includegraphics[width=0.9\linewidth]{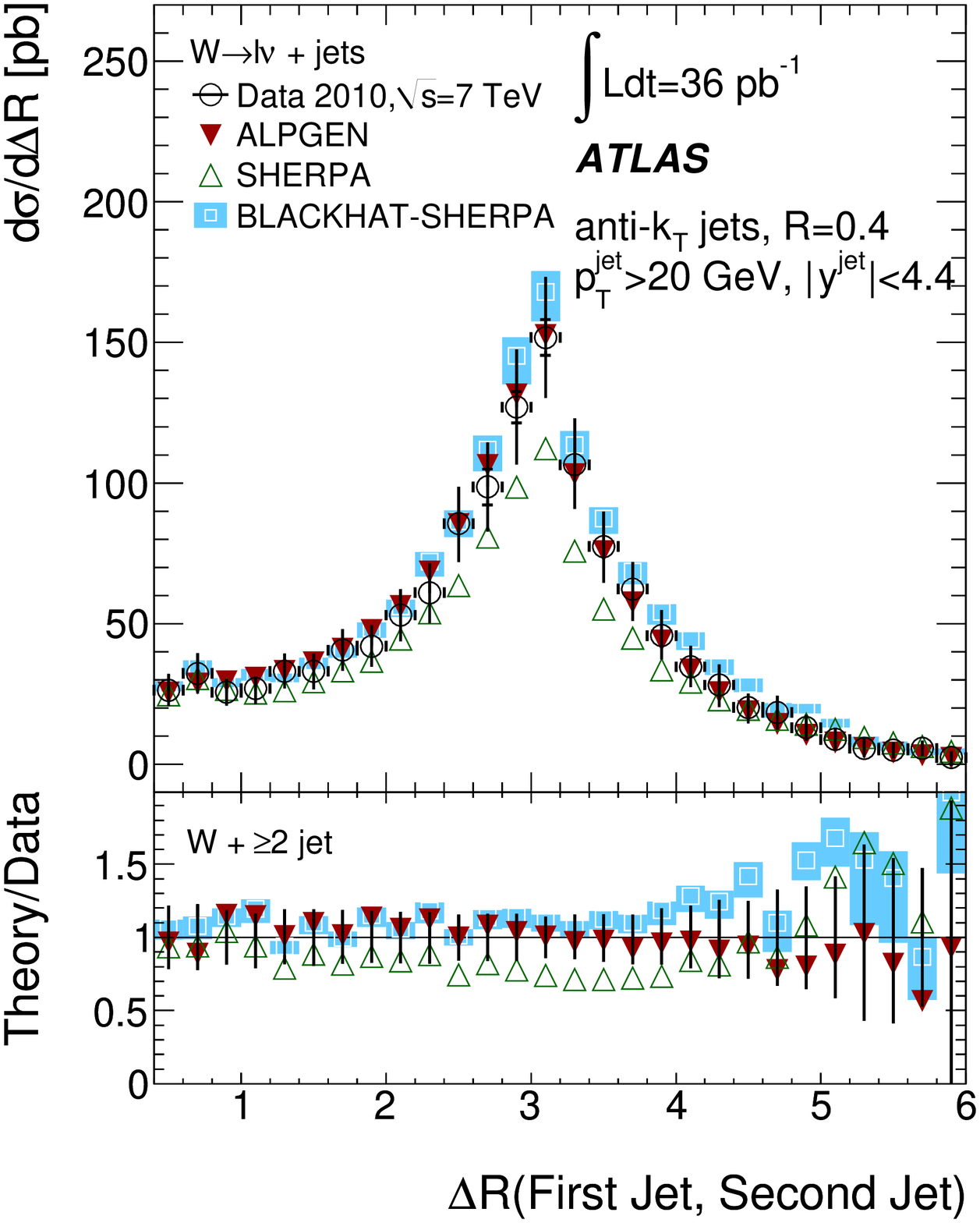}
  \caption{$W$+jets cross section as a function of $\Delta R({\rm
  first\:jet,\:second\:jet})$ for events with $\ge 2$~jets. Shown are
  predictions from {\sc Alpgen}, {\sc Sherpa}, and {\sc
  BlackHat-Sherpa}, and the ratio of theoretical predictions to data.}
  \label{fig20:result-xsec-dRjj}
\end{figure}
\begin{figure}[!htb]
  \centering
  \includegraphics[width=0.9\linewidth]{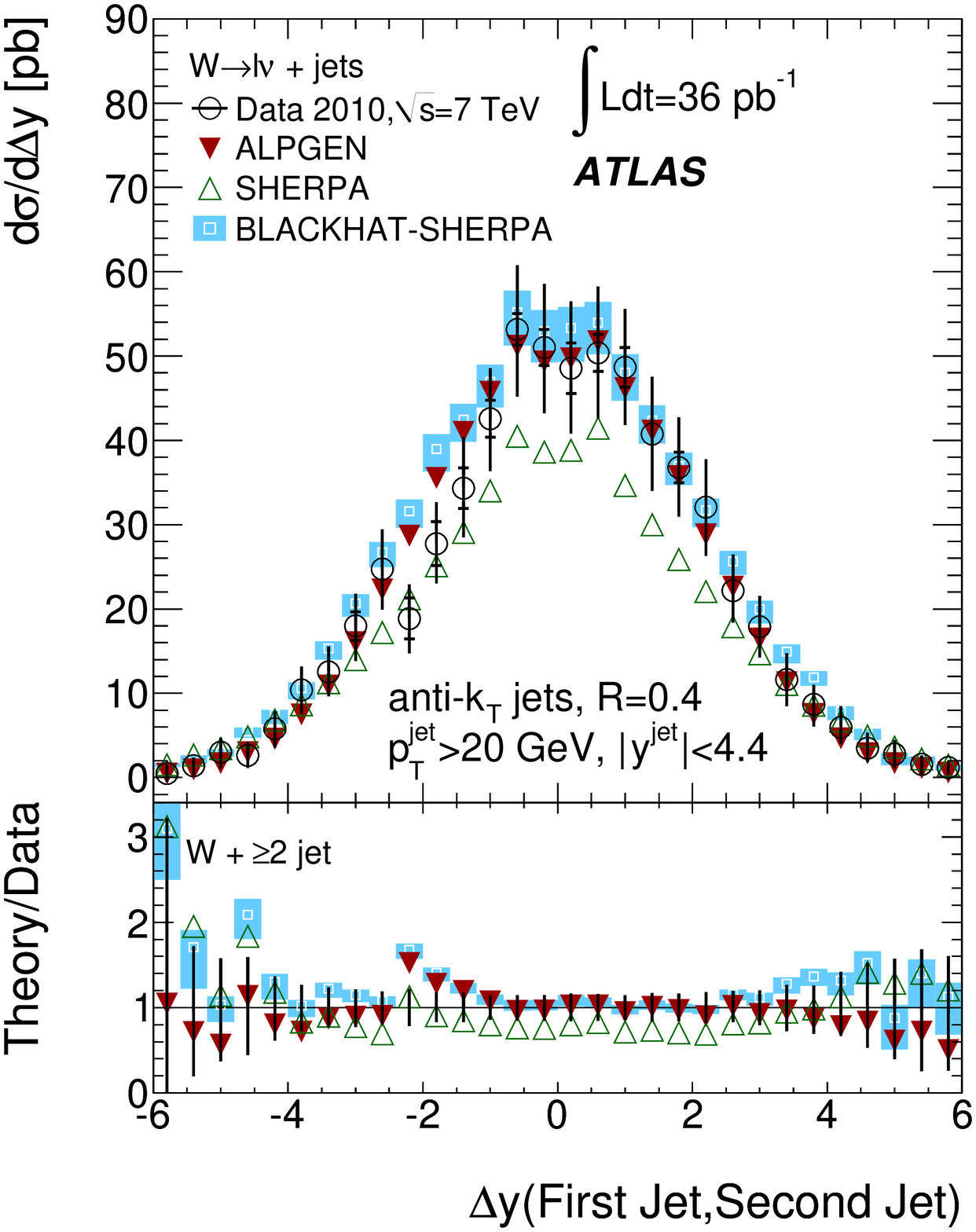}
  \caption{$W$+jets cross section as a function of $y({\rm
  first\:jet})\:-\:y({\rm second\:jet})$ for events with $\ge
  2$~jets. Shown are predictions from {\sc Alpgen}, {\sc Sherpa}, and
  {\sc BlackHat-Sherpa}, and the ratio of theoretical predictions to
  data.}
  \label{fig20:result-xsec-dYjj}
\end{figure}
\begin{figure}[!htb]
  \centering
  \includegraphics[width=0.9\linewidth]{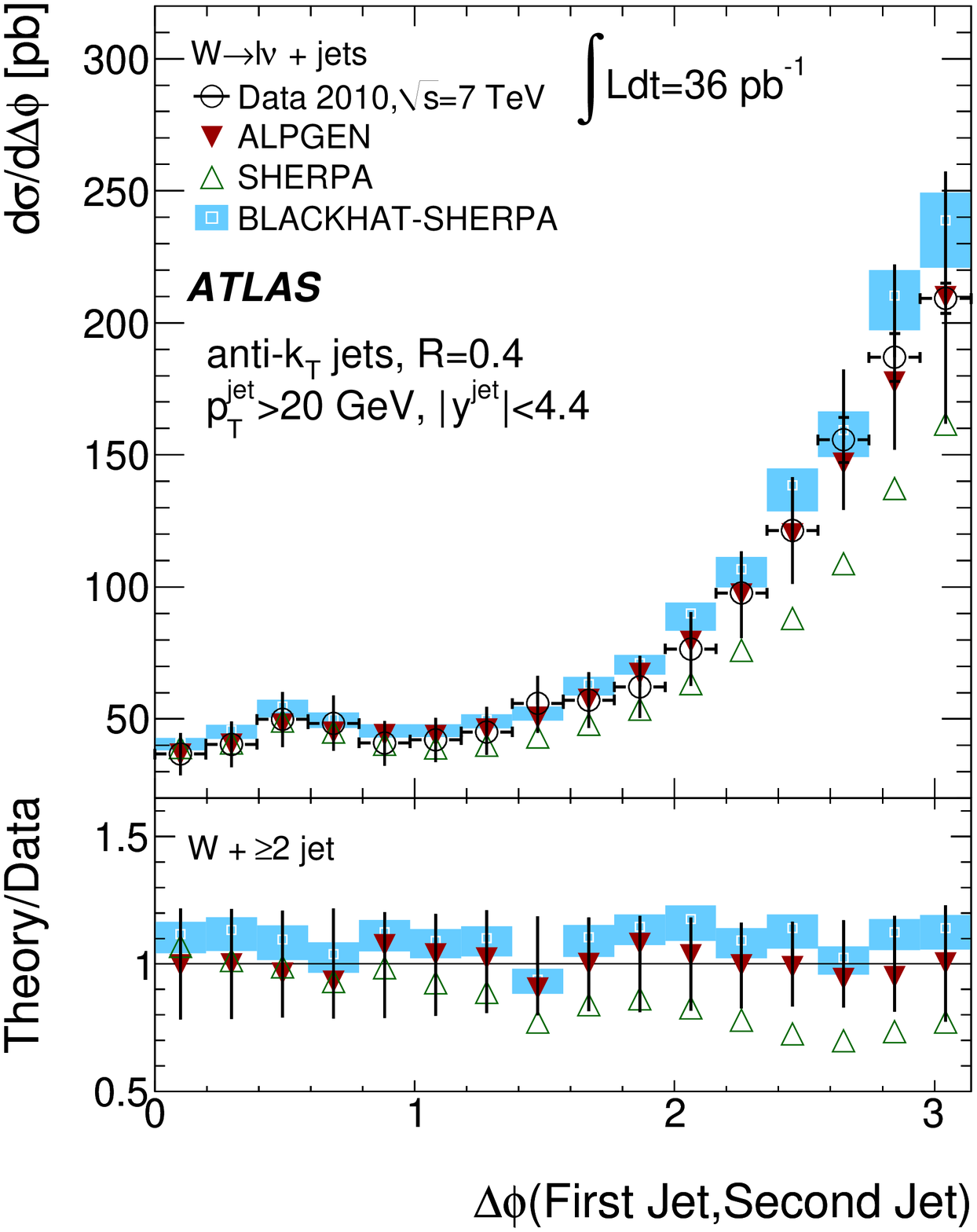}
  \caption{$W$+jets cross section as a function of $\Delta\phi({\rm
  first\:jet,\:second\:jet})$ for events with $\ge 2$~jets. Shown are
  predictions from {\sc Alpgen}, {\sc Sherpa}, and {\sc
  BlackHat-Sherpa}, and the ratio of theoretical predictions to data.}
  \label{fig20:result-xsec-dPhijj}
\end{figure}

\clearpage
\onecolumngrid
\begin{flushleft}
{\Large The ATLAS Collaboration}

\bigskip

G.~Aad$^{\rm 48}$,
B.~Abbott$^{\rm 110}$,
J.~Abdallah$^{\rm 11}$,
A.A.~Abdelalim$^{\rm 49}$,
A.~Abdesselam$^{\rm 117}$,
O.~Abdinov$^{\rm 10}$,
B.~Abi$^{\rm 111}$,
M.~Abolins$^{\rm 87}$,
H.~Abramowicz$^{\rm 152}$,
H.~Abreu$^{\rm 114}$
% ...
\bigskip

% ...
$^{10}$ Institute of Physics, Azerbaijan Academy of Sciences, Baku, Azerbaijan\\
$^{11}$ Institut de F\'isica d'Altes Energies and Departament de F\'isica de la Universitat Aut\`onoma  de Barcelona and ICREA, Barcelona, Spain\\
% ...
$^{48}$ Fakult\"{a}t f\"{u}r Mathematik und Physik, Albert-Ludwigs-Universit\"{a}t, Freiburg i.Br., Germany\\
$^{49}$ Section de Physique, Universit\'e de Gen\`eve, Geneva, Switzerland\\
% ...
$^{87}$ Department of Physics and Astronomy, Michigan State University, East Lansing MI, United States of America\\
% ...
$^{110}$ Homer L. Dodge Department of Physics and Astronomy, University of Oklahoma, Norman OK, United States of America\\
$^{111}$ Department of Physics, Oklahoma State University, Stillwater OK, United States of America\\
% ...
$^{114}$ LAL, Univ. Paris-Sud and CNRS/IN2P3, Orsay, France\\
% ...
$^{117}$ Department of Physics, Oxford University, Oxford, United Kingdom\\
% ...
$^{152}$ Raymond and Beverly Sackler School of Physics and Astronomy, Tel Aviv University, Tel Aviv, Israel\\
% ...
\end{flushleft}

\end{document}